\documentclass[fleqn,usenatbib,twocolumn]{mnras}%twocolumn in square brackets 
\usepackage{newtxtext,newtxmath}
\usepackage[T1]{fontenc}
\usepackage{ae,aecompl}
\usepackage[normalem]{ulem}
\usepackage{graphicx}	% Including figure files
\usepackage{amsmath}	% Advanced maths commands
\usepackage{array, booktabs} % Table
\usepackage{lipsum}
\usepackage{hyperref}
\usepackage{float}
\usepackage{url}
\date{}
\title[QPOML: Quasi-Periodic Oscillations and Machine Learning]{QPOML: A Machine Learning Approach to Detect and Characterize Quasi-Periodic Oscillations in X-ray Binaries}
\author[Kiker et al.]{Thaddaeus J. Kiker,$^{1,2,3}$\thanks{E-mail: thaddaeuskiker@protonmail.com}, James F. Steiner$^{4}$, Cecilia Garraffo$^{4}$, \newauthor Mariano M\'endez$^5$, and Liang Zhang $^{6}$\\
$^{1}$ Sunny Hills High School, 1801 Lancer Way, Fullerton, CA 92833, USA\\
$^{2}$ Department of Physics, Columbia University, New York, NY 10027, USA \\
$^{3}$ Earth Science Division, NASA Goddard Space Flight Center, Greenbelt, MD, USA  \\ 
$^{4}$ Center for Astrophysics | Harvard \& Smithsonian, 60 Garden St. Cambridge, MA 02138, USA\\
$^{5}$ Kapteyn Astronomical Institute, University of Groningen, P.O. BOX 800, 9700 AV Groningen, The Netherlands\\
$^{6}$ Key Laboratory for Particle Astrophysics, Institute of High Energy Physics, Chinese Academy of Sciences, Beijing 100049, People's Republic of China
} 
 
\date{Accepted XXX. Received YYY; in original form 2022 November 9}

\pubyear{2022}

\begin{document}
\label{firstpage}  
\pagerange{\pageref{firstpage}--\pageref{lastpage}}
\maketitle

\begin{abstract}
Astronomy is presently experiencing profound growth in the deployment of machine learning to explore large datasets. However, transient quasi-periodic oscillations (QPOs) which appear in power density spectra of many X-ray binary system observations are an intriguing phenomena heretofore not explored with machine learning. In light of this, we propose and experiment with novel methodologies for predicting the presence and properties of QPOs to make the first ever detections and characterizations of QPOs with machine learning models. We base our findings on raw energy spectra and processed features derived from energy spectra using an abundance of data from the \textit{NICER} and \textit{Rossi X-ray Timing Explorer} space telescope archives for two black hole low mass X-ray binary sources, GRS 1915+105 and MAXI J1535-571. We advance these non-traditional methods as a foundation for using machine learning to discover global inter-object generalizations between—and provide unique insights about—energy and timing phenomena to assist with the ongoing challenge of unambiguously understanding the nature and origin of QPOs. Additionally, we have developed a publicly available Python machine learning library, QPOML, to enable further Machine Learning aided investigations into QPOs.
\end{abstract}


\section{Introduction}

At the ends of their lives, massive stars ``do not go gentle into that good night'' \citep{thomas1952country}. Instead, if their initial mass exceeds $\sim8$ M$_\odot$, core collapse leads to spectacular Type II supernovae \citep{Schlegel1995}. If the compact remnant remains bound or becomes bound to a non-degenerate companion star, the result can be a neutron star (NS) or black hole (BH) remnant \citep{gilmore2004}. In special cases, this object maintains a non-degenerate partner, and together these may form an X-ray binary (XRB) system, in which the non-degenerate star engages in mass-exchange with its compact partner \citep{tauris2006}. Such systems are characterized by accretion from the donor star, through accretion disks \citep{SS73} and are the sources for jets \citep{gallo2005,neutronstarjet} and winds \citep{bhwinds,neutronStarWind}. Additional exotic phenomena like thermonuclear surface burning  \citep{thermonuclear} have also been observed in neutron star binaries. Both BH and NS systems are both observed to emit thermal X-ray radiation with temperatures $\sim1$ keV that is understood to arise from the conversion of gravitational potential to radiative energy. Neutron stars can produce thermal emission at their surfaces, and the optically thick, geometrically thin accretion disks around both NSs and BHs can produce strong thermal X-ray emission \citep{SS73}. Furthermore, BH and NS XRBs both also show hard X-ray flux coming from Compton up-scattering of thermal disk emission by a cloud of hot electrons around the compact source known as the corona \citep{corona1979,coronae1982}. Comptonized emission is commonly modeled by a power law relationship $N(E)\propto E^{-\Gamma}$, where $\Gamma$ is the photon index \citep{McClintockRemillard2006}. Strongly-Comptonized spectra commonly exhibit reflection features like a fluorescent, relativistically broadened 6.4 keV Fe K$\alpha$ line \citep{ironlines1989} and $\sim30$ keV Compton hump \citep{x-rayreflectionmodels2005}. These systems can be transient in activity and undergo evolution in spectral states \citep{gardenier2018}, ranging from hard, to intermediate, and to soft \citep{McClintockRemillard2006}, which are coupled with mass-accretion rate \citep{terrabytedone}, spectral hardness or thermal dominance, and thereby position on a hardness-intensity or color-color diagram track \citep{ingram2019}, and the presence/absence of quasi-periodic oscillations (QPO) of the observed X-ray radiation \citep{McClintockRemillard2006}. These QPOs are detected as narrow peaks in power-density spectra \citep{homanBelloniQPOstates}. In the past thirty years, numerous theories, including but not limited to relativistic precession \citep{lense-thirring-original}, precesssing inner flow \citep{precessionandlense}, corrugation modes \citep{corrugation}, accretion ejection instability \citep{accretion-ejection}, and propagating oscillatory shock \citep{propagatingoscillatoryshock} have been advanced to explain the occurrence of QPOs in black hole, as well as neutron star, XRB systems. Yet, there is not consensus as to which model is most plausible. In black-hole systems, most of the observed QPOs have been at low frequencies (LF) $\leq 30$ Hz \citep{belloni2020typeB}. Only a small subset has BHXRBs have exhibited high-frequency QPOs (HFQPO). LF QPOs are further subdivided canonically into three classes \citep{casellaABC}: Type-A QPOs are the rarest, sometimes appearing in the intermediate or soft state as broad, low amplitude features centered between 6-9 Hz and usually lacking harmonic companions \citep{motta2011}. Type-B QPOs are more common, and can be seen during the short soft intermediate state and have shown some connection with jet behavior \citep{globaltypeB,garciaMendez2021}. Finally, type C QPOs are the most common, and can be detected  as narrow features in the low-hard and hard-intermediate states with harmonic companions  \citep{typeCmodel2016}. Their fundamental frequencies range from $\sim$ 0.1-30 Hz depending on state, and almost always correlate strongly with spectral features like $\Gamma$ and luminosity \citep{motta2015LotsofQPOs}. As for HFQPOs, we recommend readers to \cite{motta2011}, \cite{MendezHFQPOs2013}, and \cite{stellaVietriISCO}. QPOs are also observed in neutron star systems \citep{Belloni2002,wang2016QPOreview}. We focus on LFQPOS from BHXRBs in this paper and recommend \cite{vanDerKlis2006} and \cite{wang2016QPOreview} for reviews of neutron star specific QPOs and \cite{ingram2019}, \cite{oneHZqpo}, \cite{hectohertzKato2005}, \cite{Revnivtsev2001}, and \cite{mendezBelloni2021} of QPOs in XRBs in general. All in all, hundreds of XRBs have been observed since the discovery of Sco X-1 \citep{scox-1,lmxbCatalog,blackcat} and a large fraction show some type of QPO. 

Machine learning is a revolutionary subfield of artificial intelligence in which models teach themselves patterns in data rather than operating by externally supplied hard-coded rules \citep{goodfellow2016deep}. With data available to astronomers approaching the petabyte domain \citep{2014MLinAstro}, this aspect of machine learning has helped it supplement traditional methods in addressing the ever growing volume and increasing complexity of astronomical data, while also providing new perspectives on old phenomena \citep{bigUNIbigDATA,astroMLarticlenotPython}. Consequently, machine learning has been used prolifically to classify variable stars \citep{mlVariableStars2011}, search for exoplanets \citep{mlExos}, detect pulsars \citep{detectPulsars}, predict solar flares \citep{solarflareML}, classify and even discover galaxies \citep{classifyGalaxies,discoverGalaxies}. However, although machine learning techniques has been applied to a number of problems related XRBs as well, e.g, to classify and identify X-ray binaries \citep{grsML2017,arnason2020,Sreehari2021,deBeurs2022,grsML2022,yang2022}, predict compact object identity \citep{Pattnaik2021}, and study gravitational waves \citep{MLgravWave}, this subfield contains tens of thousands of observations that have never been explored with machine learning to detect QPOs themselves. For the first time, in this work we seek to develop a methodology for using machine learning to detect QPOs, because we believe that our theoretical understanding of QPOs and their exotic progenitor systems would benefit from insights this approach could provide \citep{mlandtheory}. Our approach is unique, because although the externally determined presence of QPOs has been used as a binary input parameter in accretion state classifiers such as those in \cite{Sreehari2021}, QPOs have never before been the output of machine learning prediction themselves. The rest of this paper is structured as follows: in Section \ref{sec:obs} we describe the observations upon which we base our work. Following this, in Section \ref{sec:data_analysis} we describe the energy and spectral fitting procedures we employ to produce input/output data from these observations for the machine learning models and methods which we detail in Section \ref{sec:methods}. We present our results in Section \ref{sec:results}, and we discuss these results contextually in Section \ref{sec:discussion}. Finally, we conclude in Section \ref{sec:conclusion}. Additional work concerning demonstrating \texttt{QPOML} and model comparison are presented in following appendices. 

\section{Observations}\label{sec:obs}

%\begin{table}
%    \centering\label{tab:sources}
%    \caption{Description of Sources %Included in this Study}
%%    \begin{tabular}{l c c c}
%    \hline 
%    \hline
%    Source & Class & Instrument & Number %of Observations  \\
%    \hline
%    MAXI J1535-571 & BH & \textit{NICER} & 270 () \\
%    GRS 1915+105 & BH & \textit{RXTE} & 620 \\
%   \hline
%    Total & & & 1278 \\
%    \hline 
%    \end{tabular}
%\end{table}

\subsection{GRS 1915+105}
% https://academic.oup.com/mnras/article/349/2/393/956946

GRS 1915+105 is a well studied galactic low mass XRB system composed of a $12.4^{+2.0}_{-1.8}$ M$_{\odot}$ primary and a $1.2$ M$_{\odot}$ K III secondary \citep{GRSCompanion,Greiner2003} on a $34$ d period located at a distance of $8.6^{+2.0}_{-1.6}$ kpc from the Earth \citep{GRSDISTANCE}. The secondary star in this system overflows its Roche lobe. GRS 1915+105 was one of the first microquasar jet systems, with (apparent) superluminal motion detected from a ballistic jet launched with an inclination $70\pm2\deg$ \citep{grsjetinclination}. Since its discovery in 1992 \citep{grs-discovery}, this somewhat peculiar source has displayed unique timing and spectral patterns which have been organized into 14 separate variability classifications depending on its variability state \citep{belloni2000,Hannikainen2005}. With its 16-year archive of observations of this source we considered all data from the Rossi X-ray Timing Explorer (RXTE) Proportional Counter Array (PCA; $2-60$ keV) that are also included in \cite{GRSDATAPAPER}, \cite{mendez2022couplingNATURE}, and \cite{garciaGRS2022MNRAS}. These include a great number of detections of type C QPOs between 1996 and 2012. Energy and power-density spectra (PDS) have been derived from binned, event, and GoodXenon data as described in \cite{GRSDATAPAPER}. Briefly, PDS have been constructed by averaging 128 s long intervals at 1/128 s time resolution, normalized according to \cite{leahy1983norm}, and Poisson noise subtracted \citep{zhangPDSsubtraction}. Of the 625 timing observations in \cite{GRSDATAPAPER}, we have 554 matching energy spectra.  

\subsection{MAXI J1535-571}

MAXI J1535-571 was discovered by the MAXI/GSC nova alert system as a hard X-ray transient system undergoing outburst in 2017 by \cite{maxiATELdiscovery}, and it was first suggested to be black hole system by \cite{MAXIisBHATel}. Since discovery, it has been suggested as a $\sim 10.39M_{\odot}$ BH, $\sim 5$ kpc distant \citep{MAXI-distance}. MAXI J1535-571 has displayed state transitions \citep{Nakahira2018}, reflaring events \citep{cuneo2020}, and hysteresis during its main outburst \citep{hysteresis-MAXI}. Furthermore, it has been determined to possess a near-maximal dimensionless spin parameter of $a = \frac{cJ}{GM^2}>0.99$ \citep{miller2018,HXMT-MAXI-SPIN}. To study this source we use data from the International Space Station mounted, soft X-ray (0.5-12 keV) observatory Neutron star Interior Composition ExploreR (NICER) \citep{NICER} which has unequaled spectral-timing capabilities in soft X-rays.  

We have filtered our \textit{NICER} data following standard practices, excluding South Atlantic Anomaly passages in order to identify continuous good time intervals (GTIs) which are extracted and analyzed individually.  Data from detectors 14, 34, and 54 have been excised owing to a propensity for elevated noise or spurious events in those detectors.  Additionally, for each GTI, the average event rates of overshoot, undershoot, and X-ray events are compared amongst the detector ensemble, and any detector which has a median absolute deviation (MAD) $>15$ is also excised for that GTI\footnote{The MAD is a robust statistic which is insensitive to outliers. 15 MAD corresponds to approximately 10$\sigma$ for a Gaussian-distribution.}).  All spectra have been corrected for deadtime (generally $<1\%$). {\em NICER} backgrounds have been computed using the 3C50 background model \citep{Remillard_3C50}, as well as using a proprietary and similar background model which replaces the 3C50's ``hrej'' and ``ibg'' indexing with cutoff-rigidity ``COR\_Sax'' and overshoot-rate indexing. We have removed any data with a background count rate $\geq5$ counts/s, exclude observations for which the source-to-background count ratio is $<10$, and reject observations with exposure times $t\simeq60$ s. Additionally, we require at least 5000 net source counts to ensure reliable energy and power-density spectral results, and we consider the remaining data sufficiently bright and insensitive to the selection between these similar background models. Energy spectra have been rebinned from the 10 eV PI channels by a factor ranging from 2--6 in order to oversample {\em NICER}'s energy resolution by a factor $\gtrsim 2$, while also requiring a minimum of 5 counts per bin. From $1-4096$ Hz, PDS are computed using events in the energy range from $0.2-12$ keV, for a light-curve sampling at 2$^{-13}$s ($\approx 122\mu$s). PDS are computed individually and averaged together using 4s segments for $t<160$s and 16s segments for $t\geq160$s. Below 1 Hz, PDS are computed by averaging together results for 128s segments for $t\geq 128s$ 64s segments for $64\leq t<128s$ and 4s segments for $t<64s$.  The resulting PDS is then logarithmically rebinned in  $\sim3$\% frequency intervals, the Poisson noise subtracted, and the rms$^2$ Hz$^{-1}$ normalization adopted.

Although we have less MAXI J1535-571 observations with QPOs for analysis (in large part due to the source's transient nature), one benefit of using \textit{NICER} over \textit{RXTE} data for this source (if we could have used \textit{RXTE} data) is that \textit{NICER} spectral channels do not suffer from gain drift over epochs like \textit{RXTE} PCA (which affected energy-channel conversions), and thus we can use the \textit{NICER} energy spectra as raw inputs to our regression and classifier models, in addition to the engineered features discussed in Section \ref{sec:data_analysis} and Section \ref{sec:feature_engineering}.

Overall, we selected these two sources for this initial evaluation of our methodology because they represent two very different types of LMXRBs. On one hand, GRS 1915+105 has long been known as a markedly unusual source in terms of its outburst behaviors and states (e.g. its very abnormal, three-decade long transient outburst, regular/irregular bursts, dips, etc., behaviors influenced by GRS 1915+105's orbital period and accretion disk size, the longest and largest respectively known among LMXRBs), wheres on the other hand, MAXI J1535-571 is, in comparison to GRS 1915+105, a far more typical source in terms of outburst states, QPO-spectral parameter associations, and tracks through the hardness-intensity diagram \citep{Taam1996,Truss2006,Nakahira2018,Bhargava2019,cuneo2020,Koljonen2021,garciaGRS2022MNRAS}. Hence, between these two sources we aim to evaluate our methods across a spectrum of typical to challenging spectral-timing relationships. Furthermore, in choosing objects observed with different instruments, we aim to take advantage of the different strengths of each instrument, such as the plethora of \textit{RXTE'} observations and the high spectral resolution of \textit{NICER}  \citep{NICER}.

% Originally focused on providing rotation-resolved spectroscopy of neutron stars with unprecedented sensitivity, \textit{NICER} has also been utilized extensively to study black holes in XRBs, e.g. \cite{cuneo2020} \textbf{add more}

\begin{figure*}
    \centering
    \includegraphics[width=0.48\textwidth]{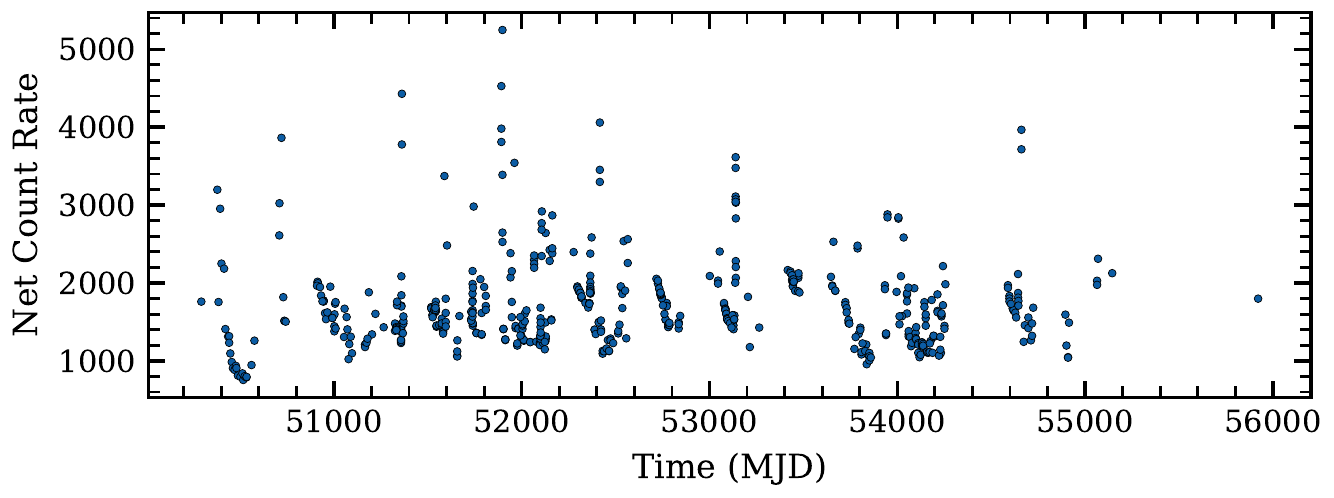}
    \includegraphics[width=0.48\textwidth]{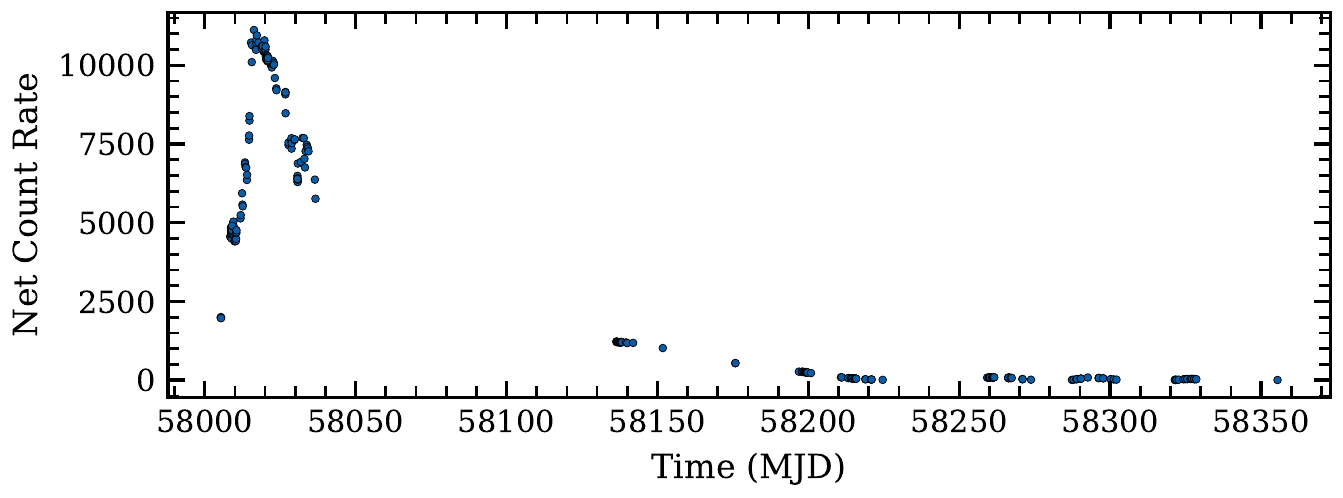}
    \caption{Light curves of GRS 1915+105 (left) and MAXI J1535-571 (right) for the observations used in this work. Net count rates are calculated as the sum of the background subtracted counts divided by observation time for every observation of each source. Note the persistent nature of GRS 1915+105 versus the transient flare of MAXI J1535-571 (reflaring epochs of MAXI J1535-571 are not included given the lack of QPOs detected there in previous works).}
    \label{fig:lightcurve}
\end{figure*}

\begin{figure*}
    \centering

    % MAXI % 
    \includegraphics[width=0.24\textwidth]{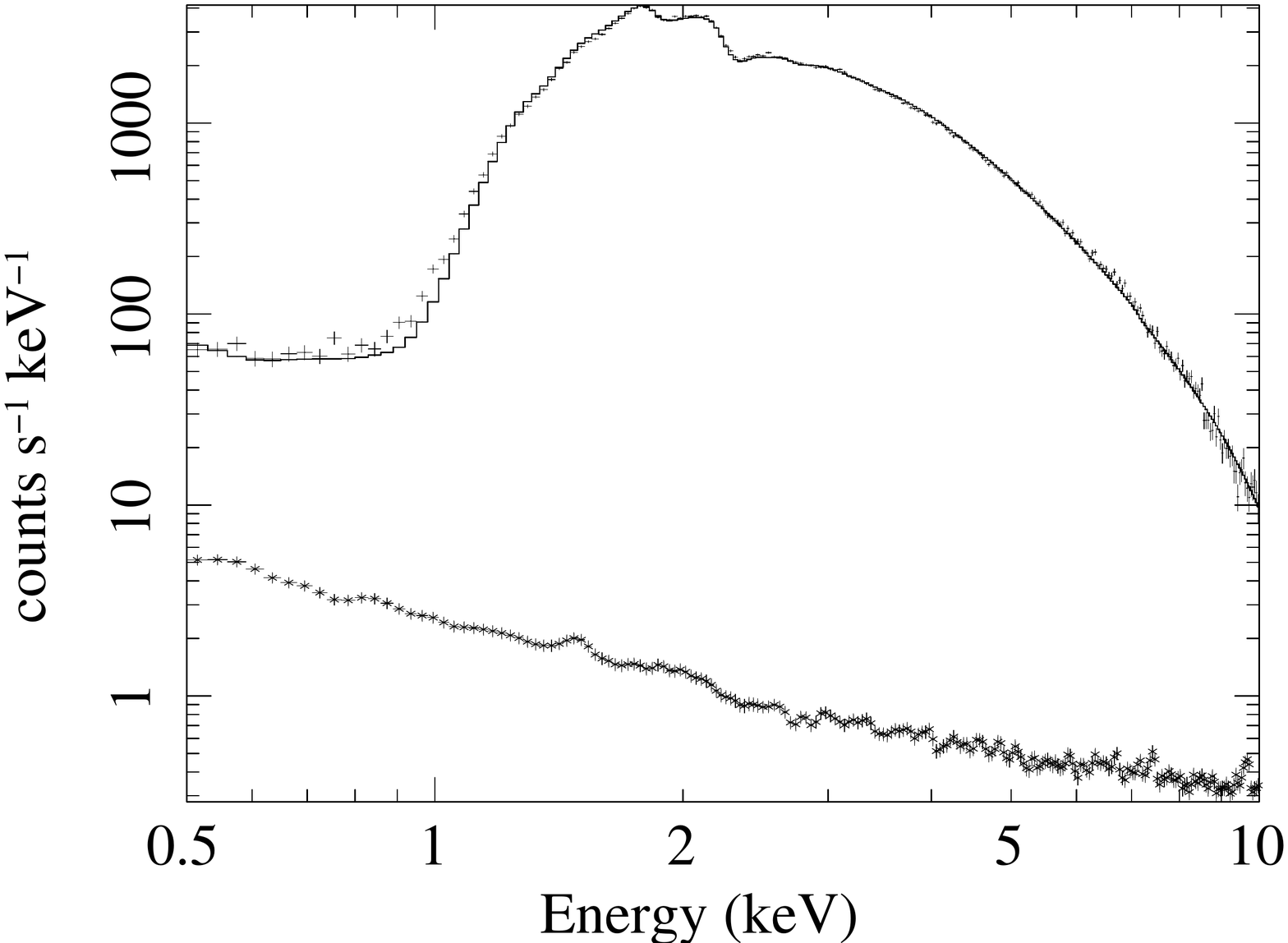}
    \includegraphics[width=0.24\textwidth]{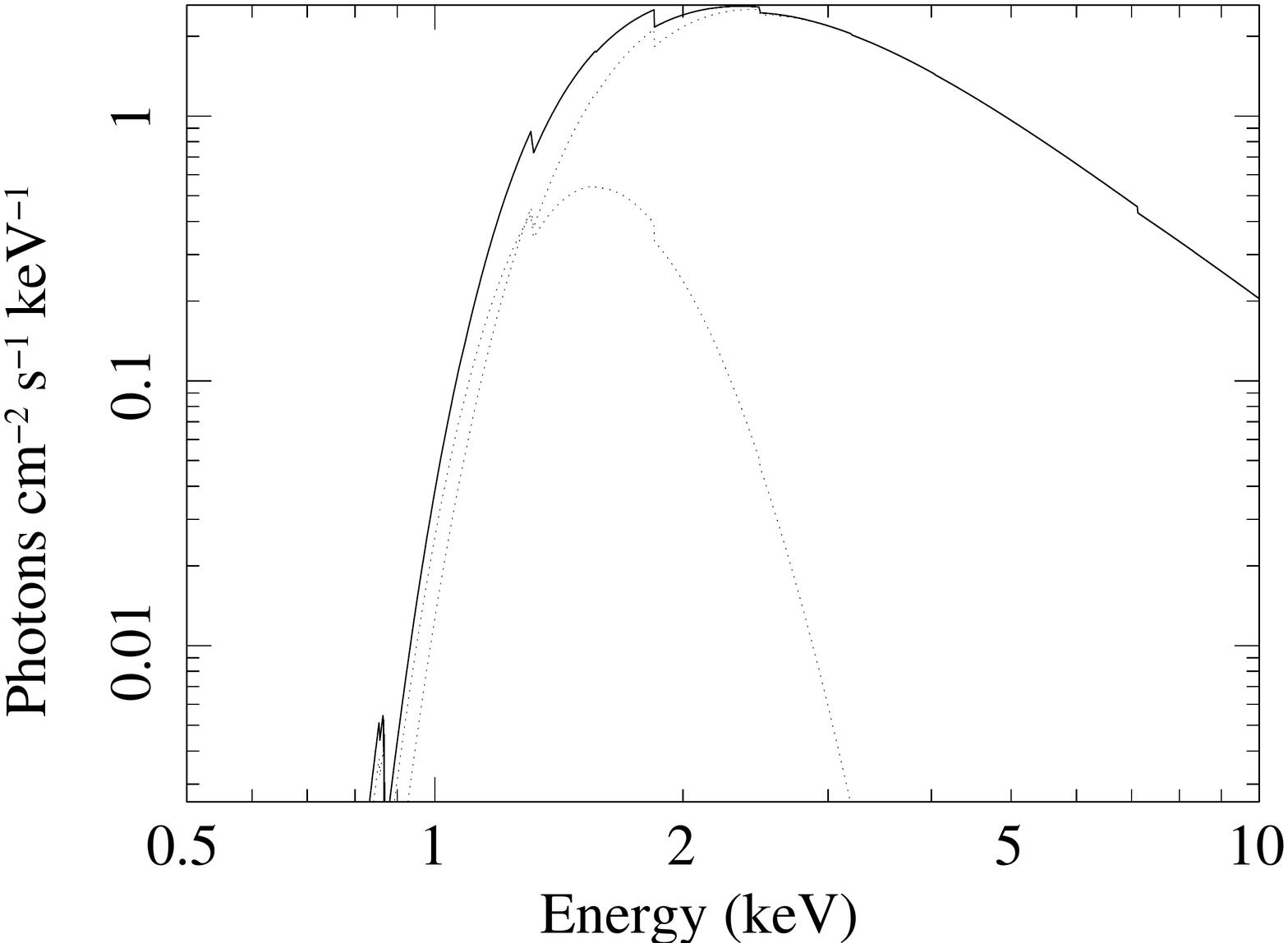}
    \includegraphics[width=0.24\textwidth]{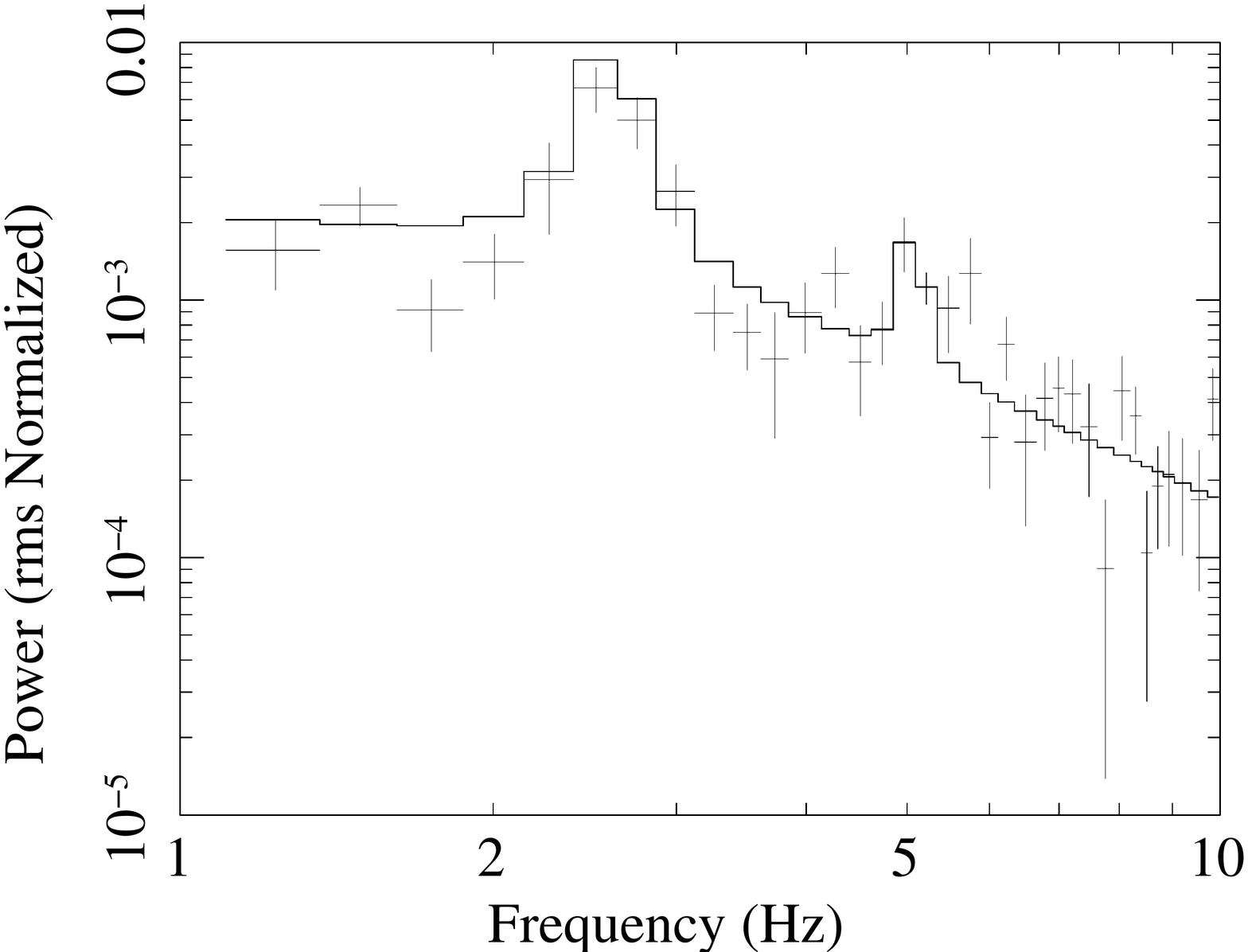}
    \includegraphics[width=0.24\textwidth]{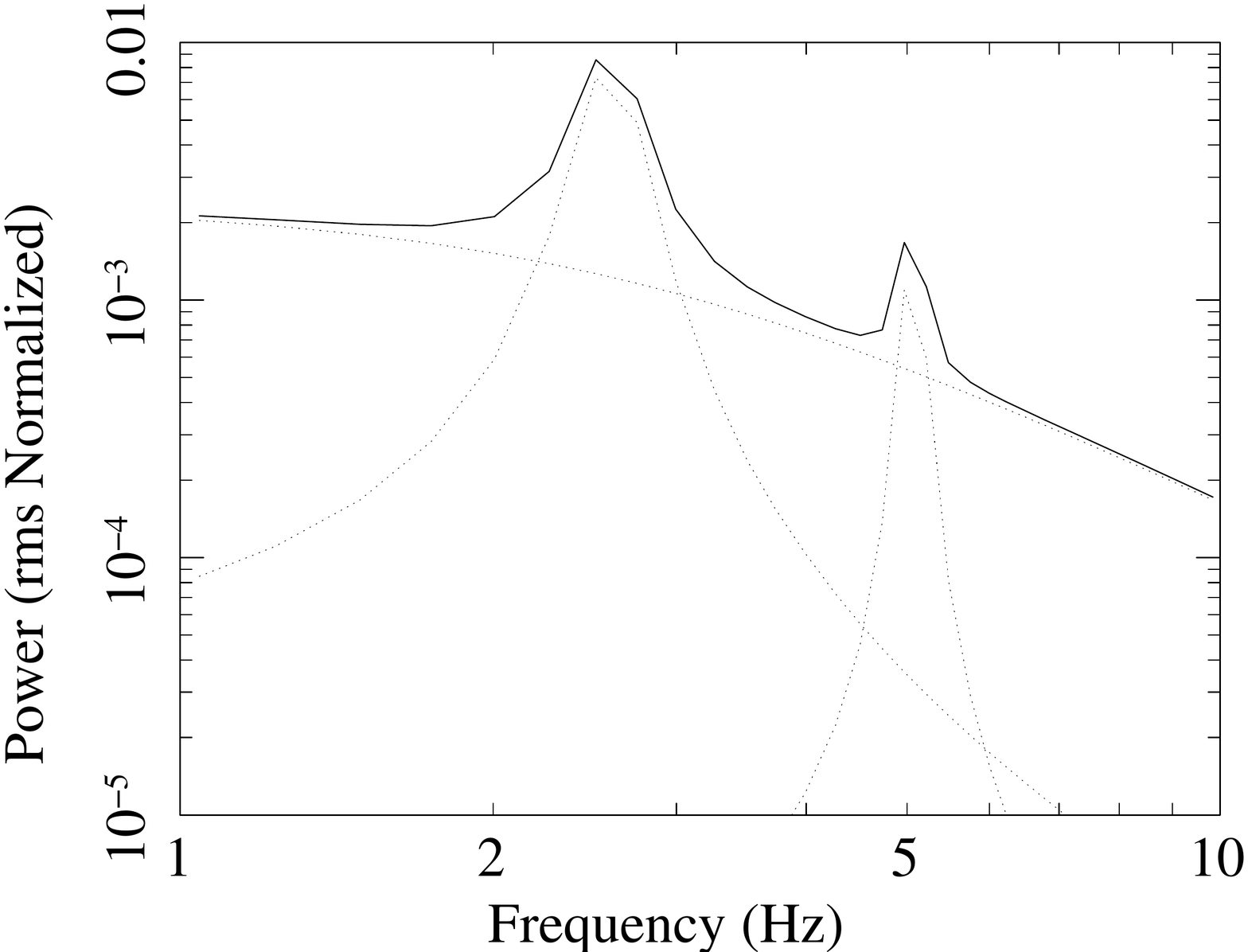}
    
    % GRS %
    \includegraphics[width=0.24\textwidth]{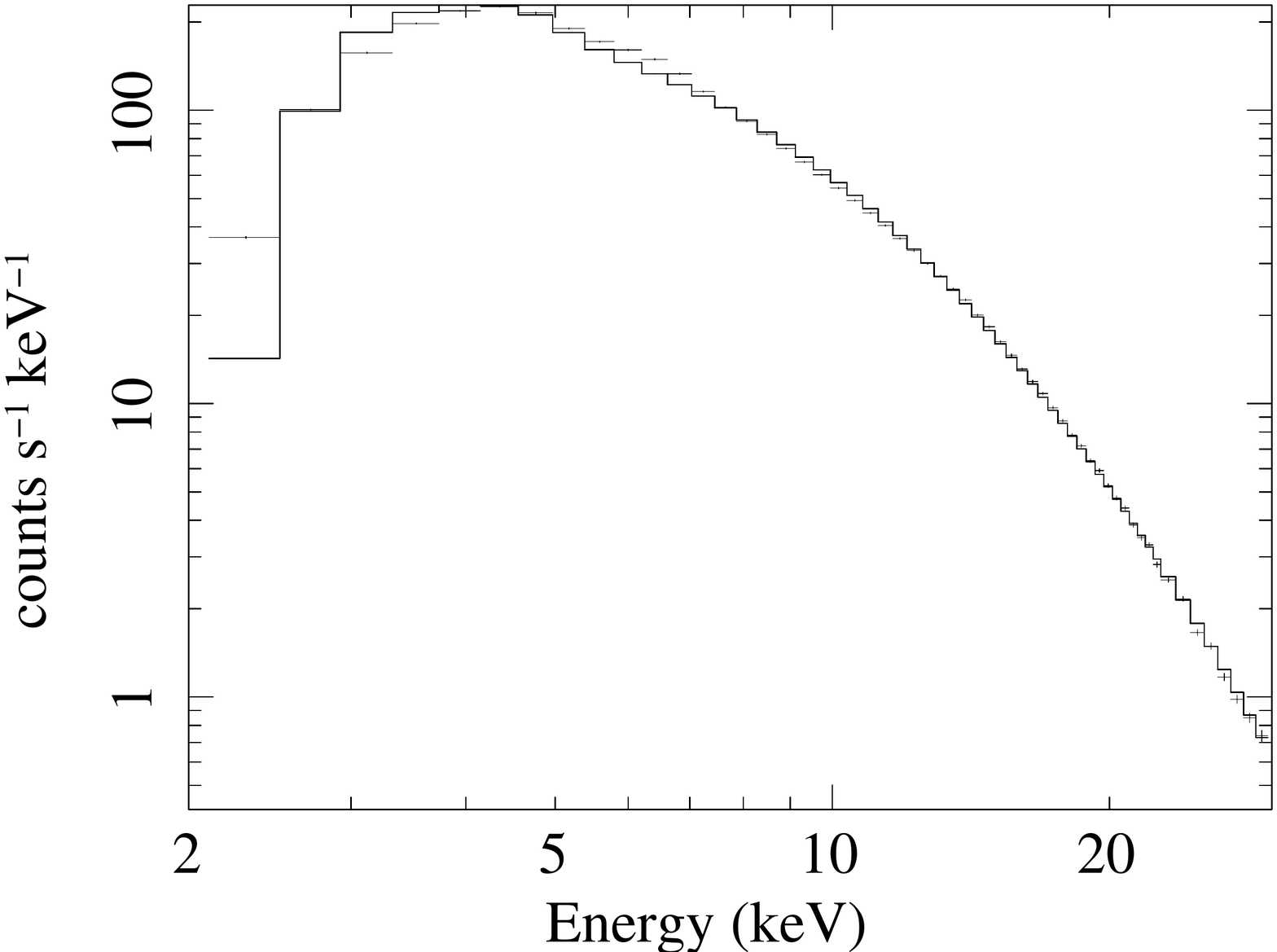}
    \includegraphics[width=0.24\textwidth]{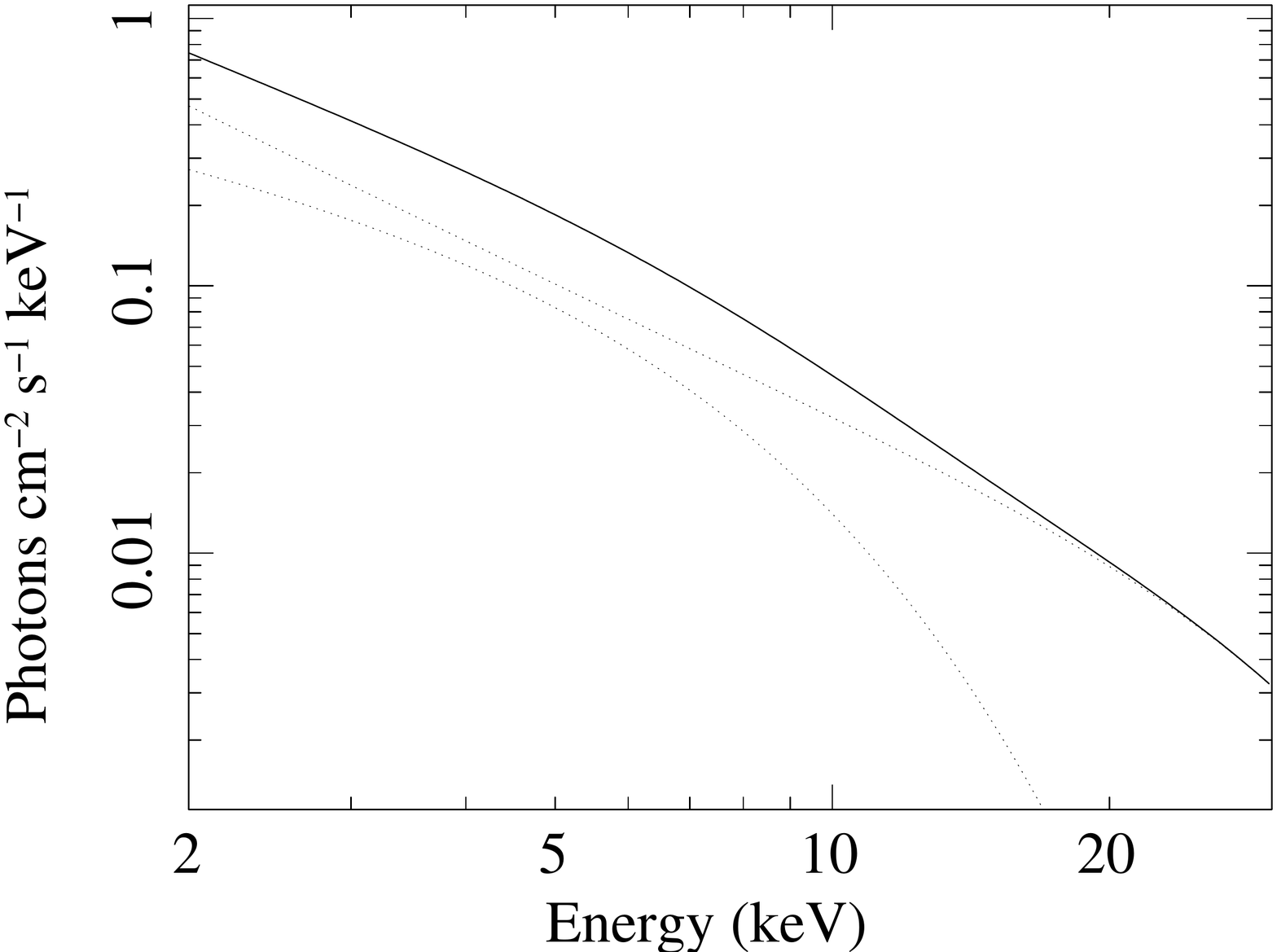}
    \includegraphics[width=0.24\textwidth]{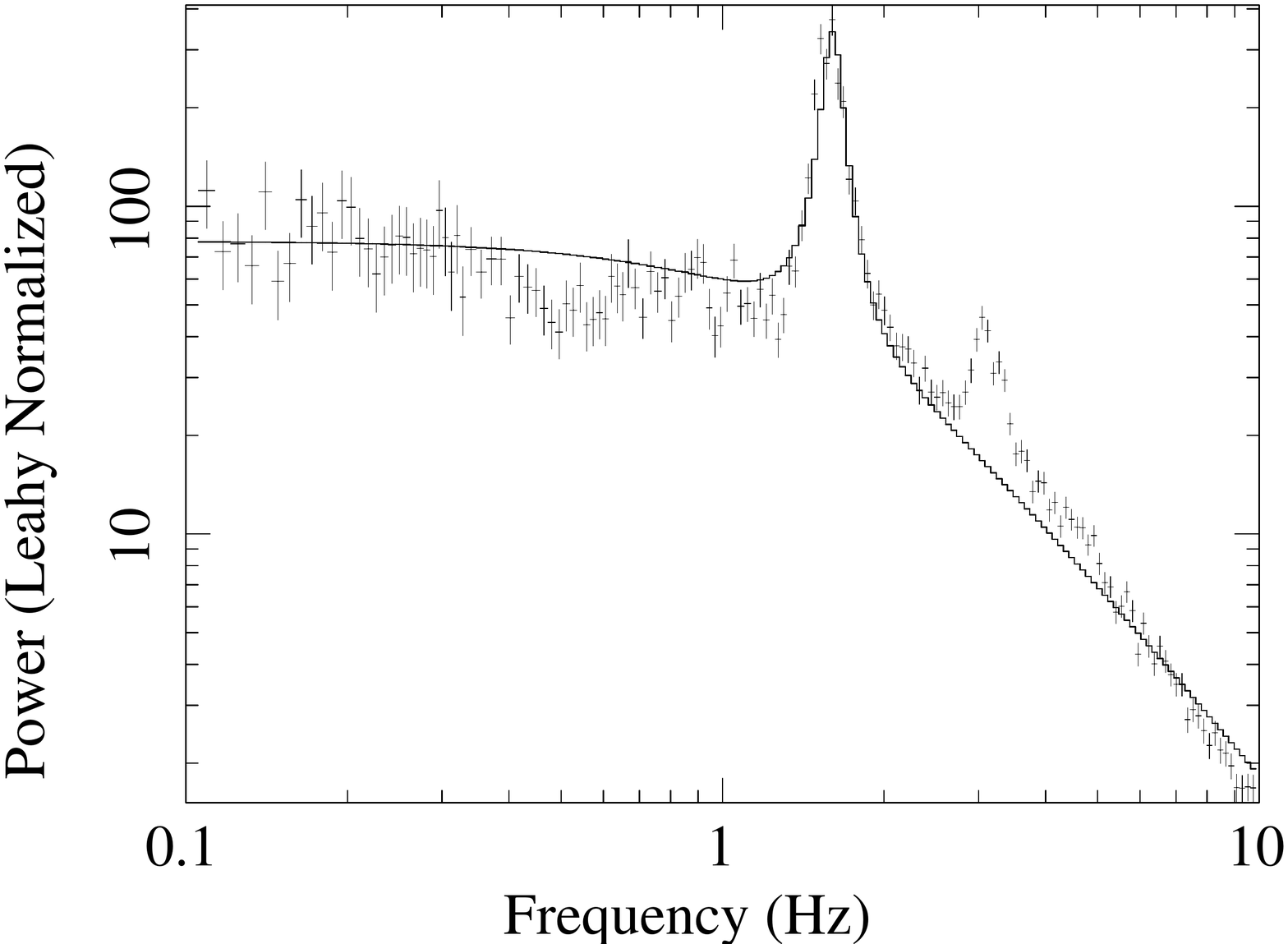}
    \includegraphics[width=0.24\textwidth]{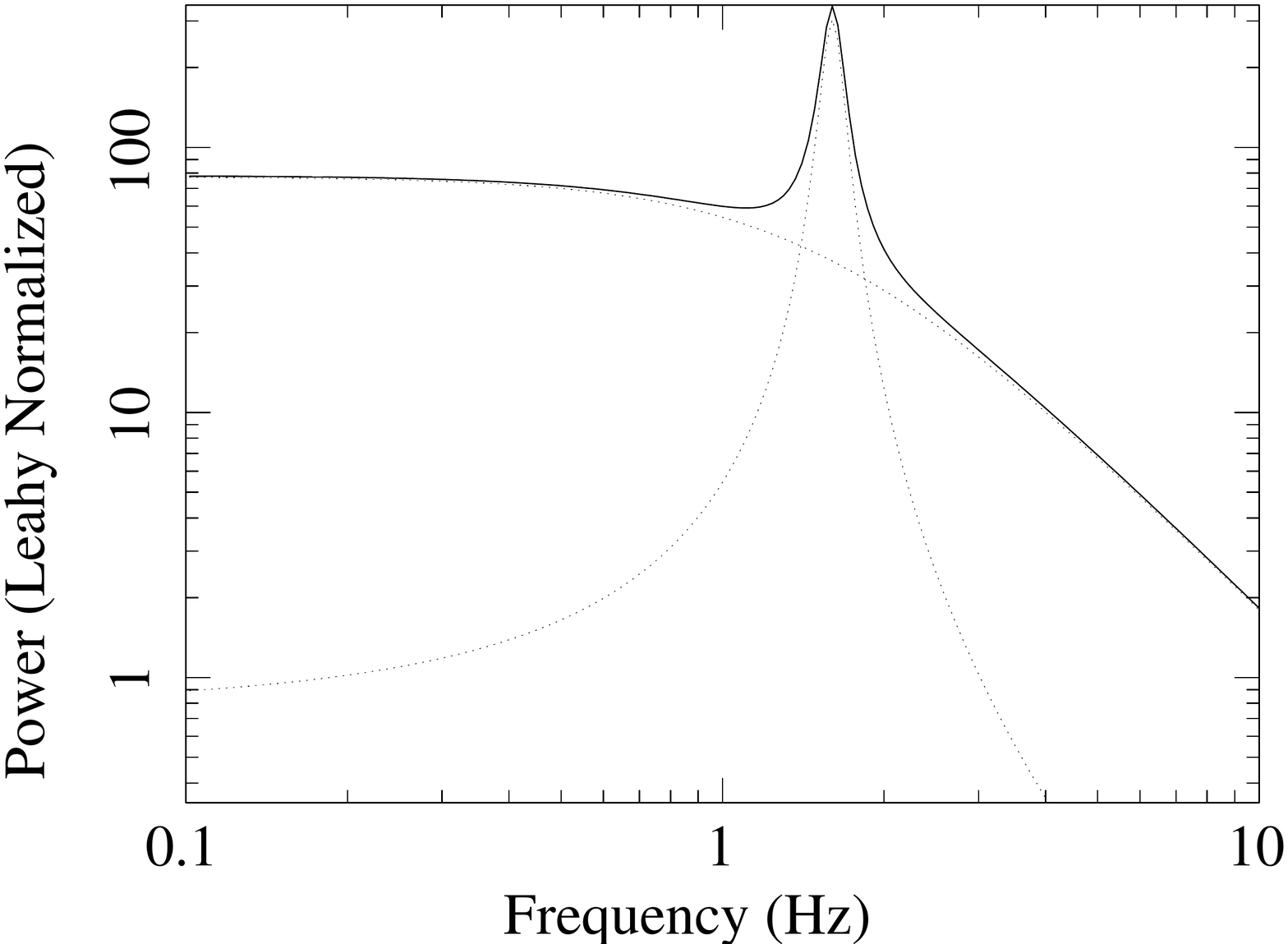}
    
    \caption{Example energy and power density spectra and models for MAXI J1535 observation 1050360105-21 on the top and the same for GRS 1915+105 observation 40116-01-01-07 on the bottom. For each row, from left to right, the first plot shows the energy spectrum and folded \texttt{tbabs*(nthcomp+diskbb)} model, the second shows energy spectrum model alone, the third shows the power density spectrum in the relevant frequency range, and the fourth shows the best fit Lorentzian PDS model alone. Best fit QPO features have been superimposed over zero centered Lorentzians used to model the power-density continuum. Only the fundamental (i.e. first harmonic) is fit for the GRS 1915+105 QPO (as discussed in Section \ref{sec:data_analysis}, this was an intentional choice to see how the models fair with seemingly simpler outputs).} 
    \label{fig:HID}
\end{figure*}

%\begin{figure*}
%    \centering
%    \includegraphics[width=0.4\textwidth]{figures/figure_3/fig3_10.pdf}
%    \includegraphics[width=0.42\textwidth]{figures/figure_3/fig3_11.pdf}
%    \caption{QPO fundamental frequency as a function of hardness ratio (defined in Section \ref{sec:data_analysis}) and net count rate as well as net count rate as %a function of QPO fundamental frequency and hardness ratio for observations of the black hole source GRS 1915+105 (see also \cite{evolvingPropertiesGRSCorona}). %The relative  degeneracies between parameters in these plots demonstrate the benefits from making predictions based on higher dimensional input spaces.}
%    \label{fig:HID}
%\end{figure*}

\section{Data Analysis}\label{sec:data_analysis}

\subsection{Energy Spectra}

%\textbf{Notes}
%\begin{itemize}
%    \item Both had disk temperature freely ranging from 0.2 to 3.0 keV, with nthcomp gamma %from 1.1 to soft upper limit of 3.5 and hard upper limit of 4.0. They both had kTin set to %diskbb disk temp. however, they are different in the ranges that their high temperature %rollover was allowed to traverse: for MAXI it was allowed to traverse 4-250 keV, whereas %for GRS it was allowed to traverse 4-40 keV (there are papers I based this off of, but %ignore why I didn't include it. not worth it)
%\end{itemize}

As previously mentioned and discussed in more detail in Section \ref{sec:feature_engineering}, we base our detection of QPOs on energy spectra and processed features from the energy spectra. Thus, to generate the processed spectral features we fit the energy spectra for both sources with \texttt{XSPEC} version $12.12.0$ \citep{XSPEC1999} using the three component model \texttt{tbabs*(diskbb+nthcomp)}, which represents a Tuebingen-Boulder absorbed multi-temperature blackbody and thermally Comptonized continuum \citep{Mitsuda1984,Zdziarski1996,Kubota1998,Zycki1999}. We fixed the equivalent hydrogen column densities to canonical values of $6\times10^{22}\;\mathrm{atoms}\;\mathrm{cm}^{-2}$ for GRS 1915+105 and $3.2\times10^{22}\;\mathrm{atoms}\;\mathrm{cm}^{-2}$ for MAXI J1535-571 based on \cite{astrosatviewofGRS} and \cite{cuneo2020}, respectively, with solar abundances in accordance with \cite{Wilms2000} and \cite{VernerCrossSections} cross-sections. We tied the \texttt{nthcomp} seed photon temperature to $\mathrm{T}_{\mathrm{in}}$ of \texttt{diskbb} for both sources, and let high energy rollover (electron temperature) freely vary between $4-40$ keV for GRS 1915+105 and $4-250$ keV during fitting for MAXI J1535-571, basing these ranges on \cite{zhangGRS2022} and \cite{Dong2022MAXIkTe}, respectively. For GRS 1915+105, we ignore channels $<2.5$ keV or $>25$ keV during fitting, calculate net count rate from the resulting range, and compute hardness as the sum of the ratio of the background subtracted channel net count rates for the ranges in \cite{zhangGRS2022}, except as a proportion rather than a ratio, i.e. $\frac{[13-60]\;\mathrm{keV}}{[2-7]+[13-60]\;\mathrm{keV}}$. Regarding MAXI J1535-571, we note the presence of instrumental residuals in the $1.7-2.3$ keV \textit{NICER} range, likely related to \textit{NICER}'s Au mirror coating and residual in the Si K $\alpha$ fluorescence peak, and following \cite{miller2018}, we address these by excluding the $1.7-2.3$ keV energy band from the spectral fitting process, and otherwise fit the range $0.5-10.0$ keV. We compute net count rate normalized to the number of \textit{NICER} detectors, and hardness ratios for MAXI J1535-571 observations as the proportion of the total net count rate contributed by the $3.0-10.0$ keV range, i.e. $\frac{[3.0-10.0]\;\mathrm{keV}}{[0.5-1.7]+[2.3-3.0]+[3.0-10.0]\;\mathrm{keV}}$. Altogether, for both sources we use the net count rate, hardness ratio, asymptotic power-law photon index, \texttt{nthcomp} normalization, inner disk temperature, and \texttt{diskbb} normalization for input parameters, which we discuss in more detail in Section \ref{sec:feature_engineering}. 

\subsection{Power Density Spectra}

Throughout this work, all QPOs for both sources are parameterized as Lorentzian distributions given by Equation \ref{eq:lorentz},

\begin{equation}
    A(f)=\frac{K(\frac{\sigma}{2\pi})}{(f-f_0)^2+(\frac{\sigma}{2})^2}
    \label{eq:lorentz}
\end{equation}

\noindent where $f$ is frequency in Hertz, $\sigma$ is full width at half maximum (FWHM), and $K$ is the normalization, as per \cite{XSPEC1999}. In the case of GRS 1915+105, QPO properties are obtained by fits to PDS following \cite{GRSDATAPAPER}. A QPO is considered significant when the ratio of the QPO power integral divided by its $1\sigma$ error $>3$ or quality factor $Q=\frac{v_0}{\sigma}$) $>2$ \citep{quality1999}, provided their frequency does not change significantly in an observation. Our primary use for this GRS 1915+105 data is to train machine learning regression models to predict the properties of the fundamental QPO feature, since \textit{only} data with matching QPO detections are used in our GRS 1915+105 machine-learning analysis. In all, this corresponds to $554$ QPOs. In contrast to this approach of fitting individual QPOs solely for regression, we use the energy and timing data from MAXI J1535-571 to explore both classification of observations into binary states of QPO presence/absence as well as multiclass QPO cardinality states \footnote{Also called multinomial classification \citep{bouveyron2019model}, when number of classes totals to $\geq 3$} based on binned raw energy spectra and processed features. Additionally, for MAXI J1535-571 we predict the properties for both the fundamental and frequently appearing harmonic in the PDS based on binned energy spectra and spectral parameterizations derived from energy spectra. Our QPO detection method for MAXI J1535-571 is slightly different than that of GRS 1915+105. Specifically, we determine the presence and properties of QPOs in PDS from MAXI J1535-571 by first fitting two zero-centered Lorentzian functions to PDS and then iteratively fitting a third Lorentzian over a logarithmically sampled set of $268$ frequencies $f$  between $1$ and $20$ Hz, where width is kept $\sigma<\frac{f}{10}$ for an initial fit, and then freed for a subsequent refined fitting step. A peak of qualifying distance ($\Delta \chi^2$ distance to neighboring samples) and threshold (horizontal distance between samples) is identified with the \texttt{scipy} function \texttt{find\_peaks} \citep{scikit-learn} in the resulting distribution of $-1\cdot\chi^2$ fit-statistic with peak height greater than the $\Delta10$ Akaike Information Criterion \citep{Akaike1998}. Finally, a visual inspection is required to accept a QPO candidate detection (to avoid potential spurious detections, e.g., at the frequency boundary). In $68$ of observations the fundamental is accompanied by the second harmonic (the fundamental itself is called the first harmonic), in $14$ observations it is alone, and in $188$ observations no QPO is detected. 

\section{Machine Learning Methods}\label{sec:methods}

\subsection{Model Selection}

In machine learning, models can be broadly divided by two sets of classification: (i) whether they operate in a supervised or unsupervised manner; and (ii) whether they are built for classification or regression \citep{bruce2017practical}. Since we are providing our models with explicit targets for loss minimization, our approach falls under the umbrella of supervised learning \citep{supervisedreview}, and as we are attempting to connect spectral information about XRBs with real-valued output vectors that describe QPOs in their power-density spectra, we also fall under (multi-output) regression \citep{multioutputreview}. In selecting our machine learning models for regression, we seek those that natively support multi-output regression, incorporate capabilities for mitigating overfitting, have precedents of working successfully with medium to small sized data sets, and natively communicate feature importances. Additionally, we seek to evaluate a collection of models against each other in light of the No-Free-Lunch-Theorem \citep{NoFreeLunch,avoidMLpitfalls}.  
    
Based on these criteria, we settle on a set of tree-based models and their descendants, specifically decision trees \citep{breiman1984original}, random forests \citep{breiman2001random}, and Extremely randomized trees \citep{extratrees}. Here we provide a brief summary of these models for context. Decision trees are the original tree-based regression model which operate by inferring discriminative splits in data and making predictions via a series of ``if-then-else'' decisions \citep{breiman1984original}. Random forests are more powerful derivatives of decision trees, and are based on an ensemble of decision trees trained via bootstrap aggregation \citep{breiman1996bagging,breiman2001random}. By incorporating predictions from such an ensemble, random forests reduce prediction variance while increasing overall accuracy when compared to a single decision tree \citep{lakshminarayanan2016decision}. Finally, Extremely randomized trees (also known as extra trees) are similar to random forests in this respect but operate with more randomization during the training process, as instead of employing the most discriminative thresholds within feature spaces for splits, extra trees select the best performing randomly drawn thresholds for splitting rules \citep{extratrees,scikit-learn}. Details on training and optimization are given in Section \ref{subsec:train_validate_tune}, where we also discuss our steps to avoid overfitting \citep{bruce2017practical}.

Together, these represent some of the most powerful yet lightweight machine learning models available, and meet our criteria for multi-output regression \citep{multioutputreview}, robustness to overfitting \citep{metarandomforests,evalTree-BasedEnsembles}, success with small/medium sized datasets \citep{floares2017smallest}, and feature importances \citep{featureimportancestrees}. An additional benefit of these models is that they are natively supported by the \texttt{TreeExplainer} method in the \texttt{SHAP} Python package \citep{SHAP2017}, which frees us from common pitfalls related to impurity and permutation based feature importances, which we discuss in more detail in Section \ref{sec:discussion}. Overall, we explore all the above models in addition to ordinary linear regression (to provide a base performance comparison) for the regression cases, but focus on random forest and logistic regression \citep{original-logistic} for classification cases.  

\subsection{Feature Engineering}\label{sec:feature_engineering}

As \cite{casari2018feature} detail, feature engineering is the process of transforming raw data to maximize predictive performance. After experimenting with different formats, we settled on the following in order to use derived features from spectral fits or raw spectral data as predictors and timing features as outcomes. We will hereafter refer to and experiment with two types of input data for our models: the first are rebinned net energy spectra, which we discuss below and will simply call ``energy spectra.'' The second type is the combination of \texttt{XSPEC} model-fit parameters and spectrum derived features like net count rate and hardness which we will designate the ``engineered features'' input type. When using engineered features for inputs, we format our input data as a matrix composed of vectors containing the net count rate, hardness ratio, asymptotic power-law photon index, \texttt{nthcomp} normalization, inner-disk temperature, and \texttt{diskbb} normalization for every observation. Hereafter, we refer to and present these values by the letters $\{A,B,C,D,E,F,G\}$ as shorthand. This input structure is visualized in Equation \ref{mat:context_matrix} as follows,

\begin{equation}\label{mat:context_matrix}
    \mathrm{IN}_{m\times7} =
    \begin{bmatrix}
    A_1 & B_1 & C_1 & D_1 & E_1 & F_1 & G_1 \\
    \vdots & \vdots & \vdots & \vdots & \vdots & \vdots & \vdots\\
    A_m & B_m & C_m & D_m & E_m & F_m & G_m \\
    \end{bmatrix}
\end{equation}

\noindent where $m$ is the number of observations. This format can be extended to any $n-$dimensional number of features, which we take advantage of when using raw energy spectra as input data. For the case of MAXI J1535-571, we compare the predictive performance of the models and provide different insights by using raw spectral data in the form of count rate values from $19$ channels, 0.5 keV wide apiece spanning the energy range $[0.5-10.0]$ directly as the input vectors within the input matrix, similar to \cite{Pattnaik2020}. This coarse spectral input strikes a balance between sparsity and precision, allowing us to determine importances for specific $0.5$ keV ranges while not overwhelming the models with too many input features given the overall sample size \citep{smallsample1991,van2020small}. With regards to regression, our QPO output matrix is similarly formatted as a vector matrix, with rows that match by index to vectors in the input matrix, but with an important addition regarding ordering (detailed below). A significant challenge relates to the prediction of not only the presence versus absence of QPOs in a given PDS, as well as (for present cases) the specific number of QPOs and the physical parameters of each QPO present. Over the course of an outburst, the number of QPOs present can change, as these are transient phenomena \citep{Remillard2006,ingram2019}. We account for this challenge of variable output cardinality by first identifying all QPO occurrences associated with an observation. Then, we order these occurrences and their features in a vector of length $L=N_f\times \mathrm{max}(N_s)$, where $N_f$ is the number of features describing every QPO (e.g. $N_f=3$ for frequency, width, and amplitude), and $N_s$ is the maximum number of simultaneous QPOs observed in any particular PDS in a data set. We then structure each output vector as a repeating subset of features for every QPO contained, and order these internal QPO parameterizations by frequency. If one or more of these occurrences are not detected in a PDS, their feature spaces in the vector are populated with zeros. This allows us to circumvent the aforementioned difficulty with variable output cardinality, because the models will learn during training to associate indices populated with zeros as QPO non-detections \citep{deepLearningPython}. As with input features, Equation \ref{mat:qpo_matrix} provides a visualization of the general QPO matrix output returned by our model, where each row corresponds to one observation matched with a row in the input matrix (both out of $m$ total observations). 

\begin{equation}\label{mat:qpo_matrix}
    \mathrm{OUT}_{m\times n} =
    \begin{bmatrix}
    f_{1,1} & \sigma_{1,1} &  K_{1,1} & \hdots & f_{1,n} &  \sigma_{1,n} &  K_{1,n}\\
    \vdots & \vdots & \vdots & \ddots & \vdots & \vdots & \vdots \\
    f_{n,1} & \sigma_{m,1} & K_{m,1} & \hdots & f_{m,n} & \sigma_{m,n} & K_{m,n} \\
    \end{bmatrix}
\end{equation}

In the case of MAXI J1535-571, the maximum number of QPOs simultaneously observed in a PDS is two, and each QPO is described in terms of its frequency, width, and amplitude, so the output matrix takes the shape $\mathrm{OUT}=m\times6$. Since we only regress for the fundamental in the GRS 1915+105 PDS, its output matrix takes the form $\mathrm{OUT}=m\times3$. Prior to reformatting the data in this manner, we applied a columnar min-max standardization to the \texttt{XSPEC}, and hardness input features, as well as the QPO Lorentzian output features, which linearly transformed each distribution into a $[\mathrm{max}(x'),\mathrm{min}(x')]=[0.1,1]$ range (as opposed to the traditional $[0-1]$ range given our decision to denote QPO non-detections with zero values) while preserving their shapes, according to Equation \ref{eq:minMax} \citep{Kandanaarachchi2019}.

\begin{equation}
    x' = \frac{x-\mathrm{min}(x)}{\mathrm{max}(x)-\mathrm{min}(x)} \times \frac{\mathrm{max}(x')-\mathrm{min}(x')}{\mathrm{min}(x')} \label{eq:minMax}
\end{equation}

This step is necessary to prevent features with relatively larger absolute amplitudes receiving undue weight, and it also frees the models from dependency on measurement units \citep{Akanbi2015, Han2012}. We did not apply this standardization step to channel count and net count rate input features, however, as the imposition of \textit{a priori} theoretical limits to these features is not as readily justifiable \citep{Pattnaik2020}.  \footnote{Standardization prior to splitting data into train and validation sets does not impair our model's predictive validity when input features are derived from 
\texttt{XSPEC} because its pre-adjusted inputs will always be constrained within the theoretical bounds applied during standardization for each feature (e.g. $\Gamma$ will always initially range between $x-y$ for a source, where $x$ can be a hard lower limit like $\Gamma=1.1$ and $y$ can be the corresponding hard upper limit during fitting, such as $\Gamma=5$).}

\subsection{Training, Validation, and Hyperparameter Tuning}\label{subsec:train_validate_tune}

To better understand our models in different data combinations and minimize statistical noise, while guaranteeing every observation gets included in a training, as well as at a separate time, validation instance, we employ a repeated $k$-fold cross-validation strategy \citep{olson2008advanced,repeatedkfold} for model evaluation (as opposed to solely using a default proportion-based train-test split). According to this procedure, our data is first split into a 90\% training and validation set, and then a 10\% held out test set. Before evaluating the models on this test set, the training and validation set is randomly split into $k=10$ folds. Given the relative class imbalance in the MAXI J1535-571 data in favor of observations without QPOs, for MAXI J1535-571, the folds for both regression and classification cases are also stratified during splitting, which means each fold maintains the same proportion of observations with QPOs \citep{ma2013imbalanced}. Then, every model is evaluated on each unique fold after being trained on the remaining folds, with the individual $k$-fold performance taken as the mean of these evaluations across the ten folds. We repeat this process five times (randomly shuffling the data between each iteration), and the final score for each model is calculated as the mean performance across the ten $k$-fold instances, either as the $f-$score for classification cases (a harmonic mean of the precision and recall), or the median absolute error for regression \citep{scikit-learn,kuhn2019applied}. Random initialization is kept the same between models to make sure each model is trained/tested on the same data within each fold, and to ensure fair comparison between these models, each was subject to automatic and individualized hyperparameter tuning via grid search prior during this evaluation \citep{dangeti2017statisticsML}. The specific hyperparameter values from which combinations were derived and evaluated for each model are presented in Table \ref{tab:hyper-tuning}. 

%In the case of XGBoost, for example, this included modulation of learning rate $\eta$, $\ell_{1}$ regularization, number of ensemble tree estimators, and maximum tree depth to minimize overfitting while maximizing predictive performance. For purposes of concise presentation, we mainly present figures generated from the tenth fold, which forms a validation set for discussion (e.g. Figure \ref{fig:grs_average_performances}, Figure \ref{fig:results_regression_grs}, Figure \ref{fig:results_regression_maxi}, etc.). Beyond these, there are some results shown like the receiver operator characteristic (ROC) curves in Figure \ref{fig:cm_and_roc_binary} that depict averaged results across repetitions and folds to give an aggregated understanding of referenced concepts that takes inter-fold variance into account.% The remainder of plots for all other models and folds are made available online in a supplementary figure set. 

\begin{table}\label{tab:hyper-tuning}
    \caption{Feature spaces for model hyperparameter tuning}
    \label{tab:pairwise}
    \centering

    \begin{tabular}{llll}
\toprule
 & Decision Tree & Random Forest & Extra Trees \\
\texttt{min\_samples\_leaf} & \{1,3\} & \{1,3\} & \{1,3\} \\ 
\texttt{min\_samples\_split} & \{2,4,6,8\} & \{2,4,6,8\} & \{2,4,6,8\} \\ 
\texttt{n\_estimators} & & \{50,100,150, & \{50,100,150, \\ 
& & 200,250,500\} & 200,250,500\} \\ 
\texttt{warm\_start} & & \{\texttt{True},\texttt{False}\} & \\
\midrule
\bottomrule
\end{tabular}
\end{table}

\subsection{Feature Selection}\label{sec:feature_selection}

Through feature selection, it is generally important to deal with potential multicollinearity by calculating Variance Inflation Factors (VIF) and removing features with VIF values $\gtrsim5$ \citep{kline1998principles, Sheather2008-mc}. However, we have chosen not to remove potentially collinear features prior to regression for the following reasons: first, the tree based models like random forest that we focus on are by design robust from the effects of multicollinearity \citep{Strobl2008,2021arXiv211102513C}. Second, since multicollinearity only affects the estimated coefficients of linear models, but not their predictive ability, applying a linear model to potentially collinear data is perfectly reasonable in our case, as we are using the linear model solely as a baseline against which we will compare the predictive capabilities of the more complicated random forests model; i.e., as we are applying the linear model, we are not interested in its components \citep{multicollinearityclass,multicollinearityregression}. We will, however, revisit multicollinearity when we interpret feature importances in Section \ref{sec:results}.

\section{Results} \label{sec:results}

\subsection{Regression}

\begin{figure}
    \centering
    \includegraphics[width=0.4\textwidth]{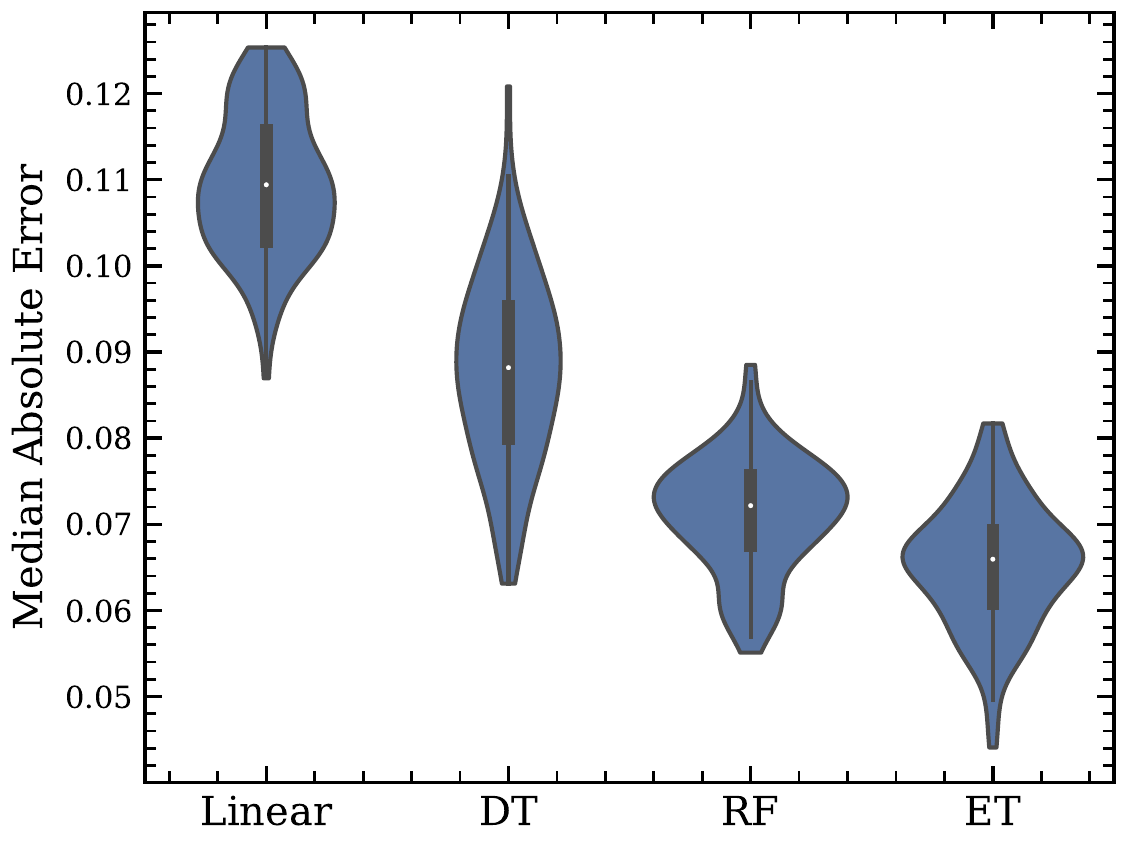}
    \includegraphics[width=0.4\textwidth]{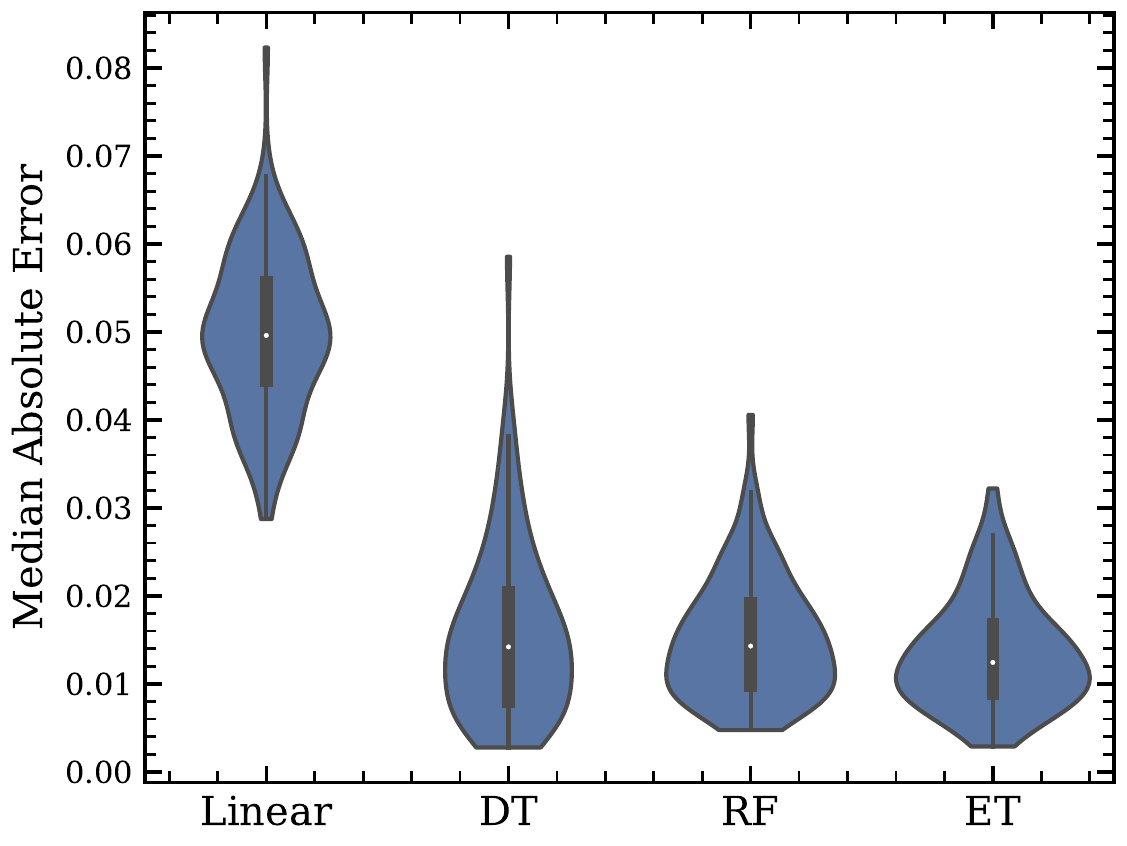}
    \caption{Gaussian kernel density estimate violin plot representations of aggregated median absolute error for each tested model across $k=10$ validation folds repeated $r=5$ times on GRS 1915+105 (feature input) data (top) and MAXI J1535-571 (feature input) data (bottom). The abbreviations DT, RF, and ET stand for the decision tree, random forest, and extra tree models, respectively. As further discussed in Section \ref{sec:results}, linear regression is outperformed by the classical machine learning models models across folds for each repitition round. Furthermore, the two ensemble tree based models clearly outperform the single decision tree model, which is to be expected.}
    \label{fig:grs_average_performances}
\end{figure}

As demonstrated in Figure \ref{fig:grs_average_performances}, on average our tree-based models outperform linear regression in every regression case, regardless of source or input feature type. Interestingly, as shown in Figure \ref{fig:results_regression_maxi} and Figure \ref{fig:results_regression_maxi_spectrum}, linear regression also seriously struggles to correctly assign $0$ values to observations lacking QPOs for both processed and rebinned energy spectra input data, a problem not faced by the other models (except random forest with rebinned energy spectra to a lesser degree). Furthermore, linear regression always has higher dispersion in the relationship between actual and predicted QPO frequency. Yet, despite their unified superiority versus linear regression, the machine learning models do differ significantly within fold amongst themselves, as shown in Figure \ref{fig:grs_average_performances}, Figure \ref{fig:results_regression_grs}, Figure \ref{fig:results_regression_maxi_features}, and \ref{fig:results_regression_maxi_spectrum}. Specifically, although decision tree provides a notable improvement in dispersion between true and predicted values, as well as a slope between these closer to unity, it is by far bested by random forest, and extra trees. Two additional interesting divergences in model performance occur between the sources, as well as between their input types. Regarding the former, all models trained and evaluated on GRS 1915+105 data have more overall dispersion and slopes tending further away from unity in their mapping between true and predicted frequency when compared to the same models for MAXI J1535-571 QPOs with processed input features. This can be clearly seen when comparing Figure \ref{fig:results_regression_grs} with Figure \ref{fig:results_regression_maxi_features}. The superior performance of the algorithms on MAXI J1535-571 are surprising for several reasons: first, with GRS 1915+105 the models never face the problem of false negatives or false positives because there are no QPO-absent data in this set. In contrast, MAXI J1535-571 observations are of varying composition, imbalanced in favor of QPO absence. Second, GRS 1915+105 has around two times more total observations, and around six times more observations with QPOs than MAXI J1535-571; in most cases training models on more data leads to corresponding increases in accuracy \citep{kalinin2020handbook,brefeld2020machine}. However, this assumption may not hold in instances like this, where models are being tested on different objects, as there may exist fundamentally stronger/more pronounced associations between spectral and QPO in one of the systems. The most likely reason for the inferior performance on GRS 1915+105 QPOs is that the underlying relationships between the input and output QPO features are likely more convoluted for GRS 1915+105, which is understandable given GRS 1915+105 has long been known to have complex variability states, and is in fact a bit of an oddball among black-hole systems. Additionally, potential confusion could arise because the models fitted on fundamental QPOs only in GRS 1915+105 intentionally lack the freedom to predict aspects about harmonics, which could lead to these models to potentially confuse signals for harmonics with fundamentals (this is an unexpected insight from our initial decision to only predict for the fundamental in GRS 1915+105 in an effort to explore how the models behave with simpler output space). Finally, to evaluate the performance of the multioutput aspect of the regression, we carry out pairwise nonparametric  two-sided goodness-of-fit Kolmogorov-Smirnov (KS) tests on permutations of QPO parameter residual arrays \citep{KS-test,KS-test-review}, and fail in all instances to reject the hypothesis that any pair of distributions of residual arrays between actual and predicted QPO parameters are not drawn from the same distribution ($p>0.76$ for all GRS 1915+105 and $p>0.99$ for all MAXI J1535-571 residual pair permutations, regardless of input type). This shows that the the models do not favor any particular QPO parameter in their regression and instead regress for each with statistically insignificant differences in accuracy (i.e. accuracy is not different for QPO features, both for the fundamental, as well as the harmonic when present). As for the second interesting divergence in model performance (by input type), surprisingly there is a pronounced difference in model performance when these regression models are trained on processed features as opposed to rebinned energy spectra: in all model cases, dispersion and slope both drastically worsen when models rely on the rebinned energy spectra directly. This is shown for MAXI J1535-571 regression between Figure \ref{fig:results_regression_maxi_features} and Figure \ref{fig:results_regression_maxi_spectrum} demonstrates that although the models could hypothetically learn some lower level representation of the concepts of hardness, overall net count rate, etc. from the data and not require the engineered features, with the amount of data provided  engineered features provide significant additional insight for the models to base decisions on that, exceeding what is provided by energy spectra alone. This would be an interesting idea to investigate with deep learning methods, which would far exceed these classical models' ability to learn abstractions in the data through automated feature extraction \citep{nadeauandbengio}. 

\subsection{Classification}

At least for MAXI J1535-571, binary classification of QPO absence/presence appears to be a fairly trivial task, as shown by the confusion matrices of the first repetition tenth folds in Figure \ref{fig:cm_and_roc_binary}. Additionally, as Figure \ref{fig:cm_and_roc_binary} also shows, our logistic regression classifier corollary to linear regression performs just as well as random forest in terms of accuracy and other classification metrics when trained on processed input data, with negligible difference for rebinned energy spectra as well. This is corroborated by the corresponding ROC curves also shown in Figure \ref{fig:cm_and_roc_binary}. The ROC curves show how a model has optimized between specificity (on the abscissa) and recall (also known as sensitivity; on the ordinate), with the ideal model displaying an ROC curve enclosing an area under curve (AUC) of $1$ \citep{bruce2017practical}. The curves in Figure \ref{fig:cm_and_roc_binary} represent the average ROC and AUC values with $1\pm\sigma$ deviations across all folds and repetitions evaluated. Both logistic regression and random forest decrease in average AUC when trained on rebinned energy spectra, but the decrease is most dramatic for logistic regression. We also present multiclass classification results for multinomial logistic regression and random forest based on processed and rebinned energy spectra input data in Figure \ref{fig:cm_multiclass}. In the case of processed input data, random forest clearly outperforms logistic regression, but both models actually experience noted decreases in accuracy when tasked with predicting multiple outputs corresponding to the actual number of QPOs in a MAXI J1535-571 observation based on rebinned energy spectra input. In fact, in the case of energy spectra inputs, random forest actually performs worse than logistic regression. Overall, the decreased performance of both models here is likely do to the class imbalance in the data set (as mentioned in Section \ref{sec:data_analysis}), which gives the models very few single QPO observations to use as training data per round. 

%\begin{equation}
%    \mathrm{F}_{\beta} = %(1+\beta^2)\cdot\frac{\mathrm{precision}\cdo%t\mathrm{recall}}{(\beta^2\cdot\mathrm{preci%sion})+\mathrm{recall}}
%\end{equation}
\begin{figure*}
    \centering
    \includegraphics[width=0.32\textwidth]{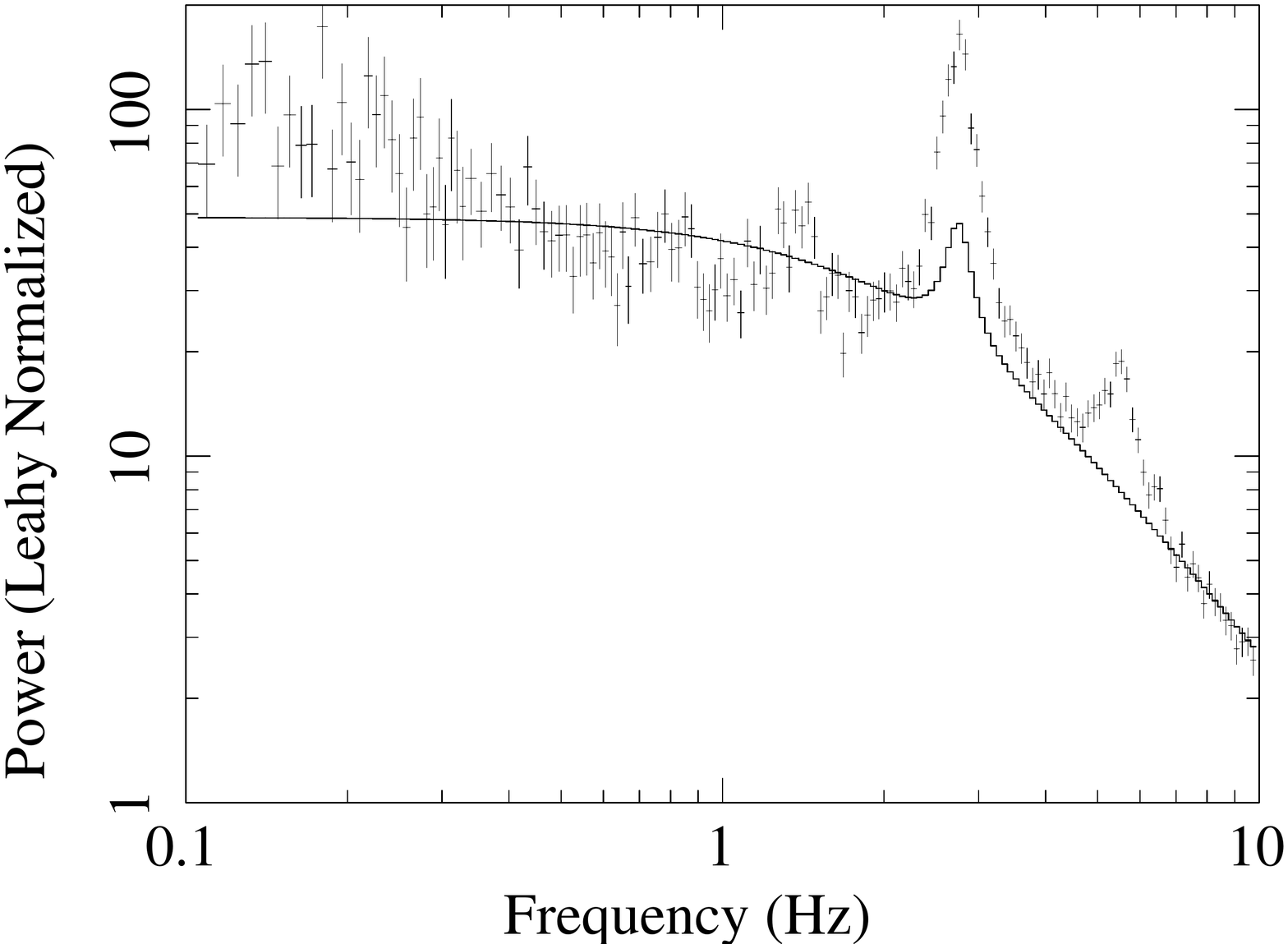}
    \includegraphics[width=0.32\textwidth]{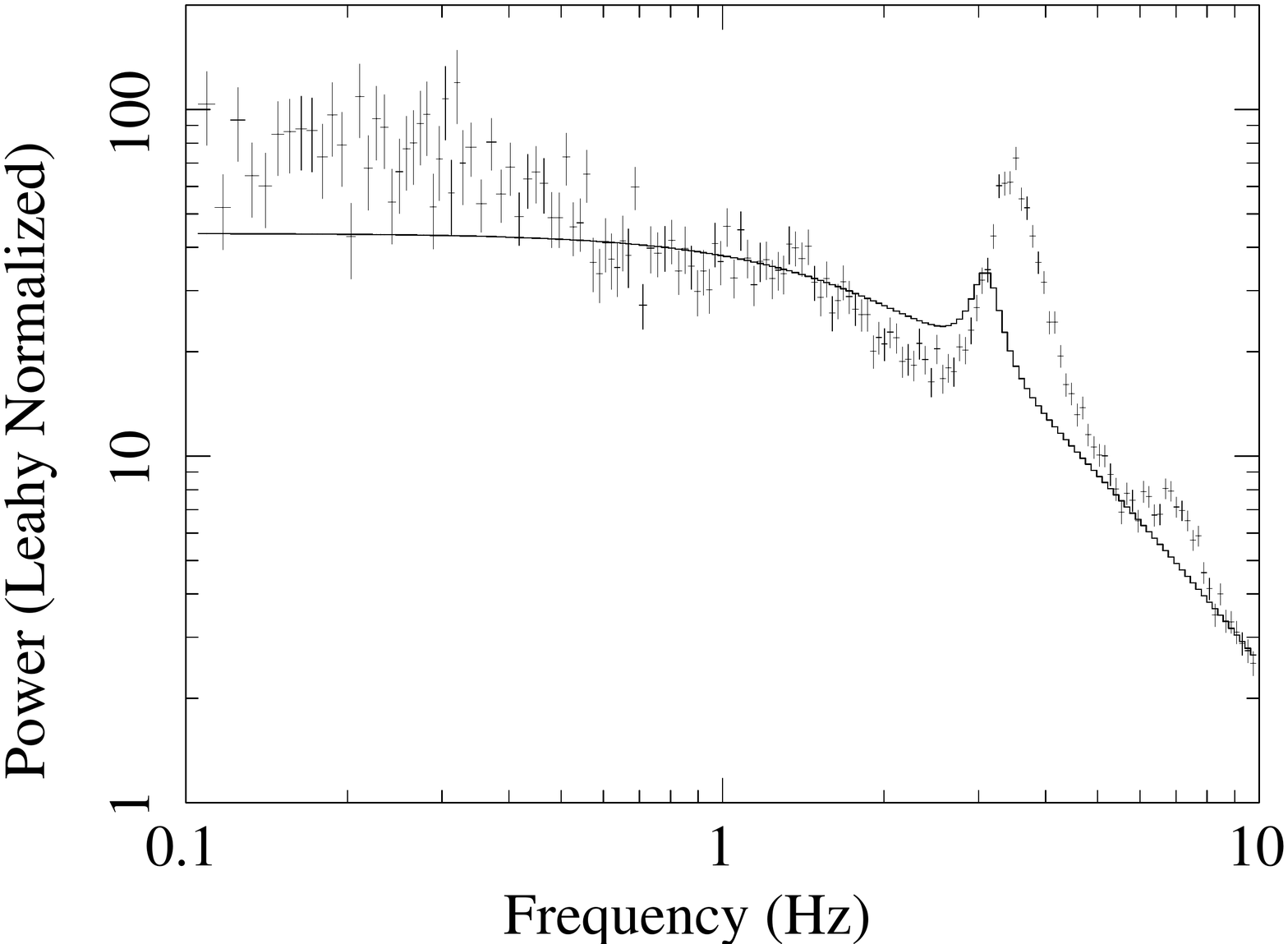}
    \includegraphics[width=0.32\textwidth]{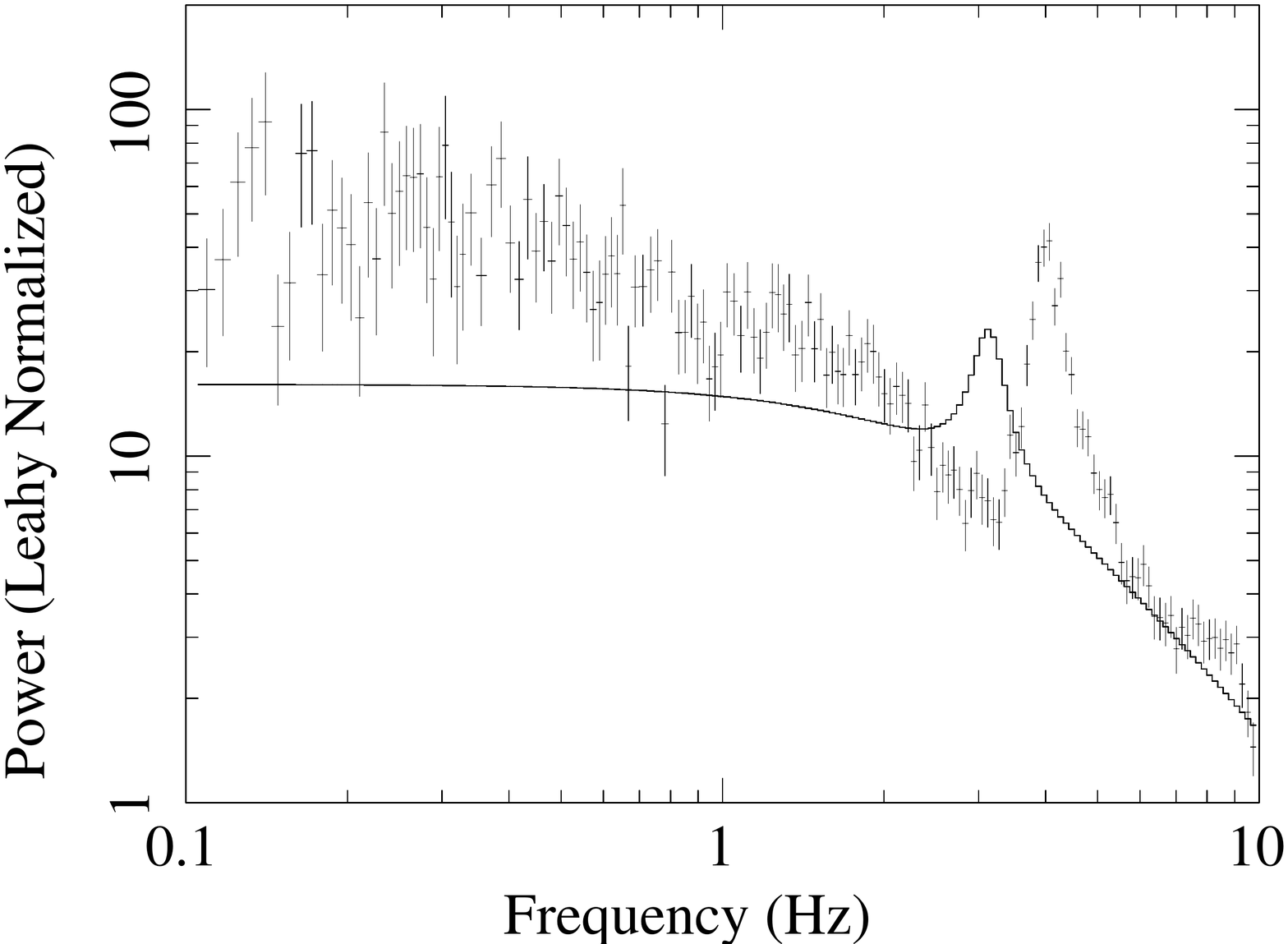}
    \includegraphics[width=0.32\textwidth]{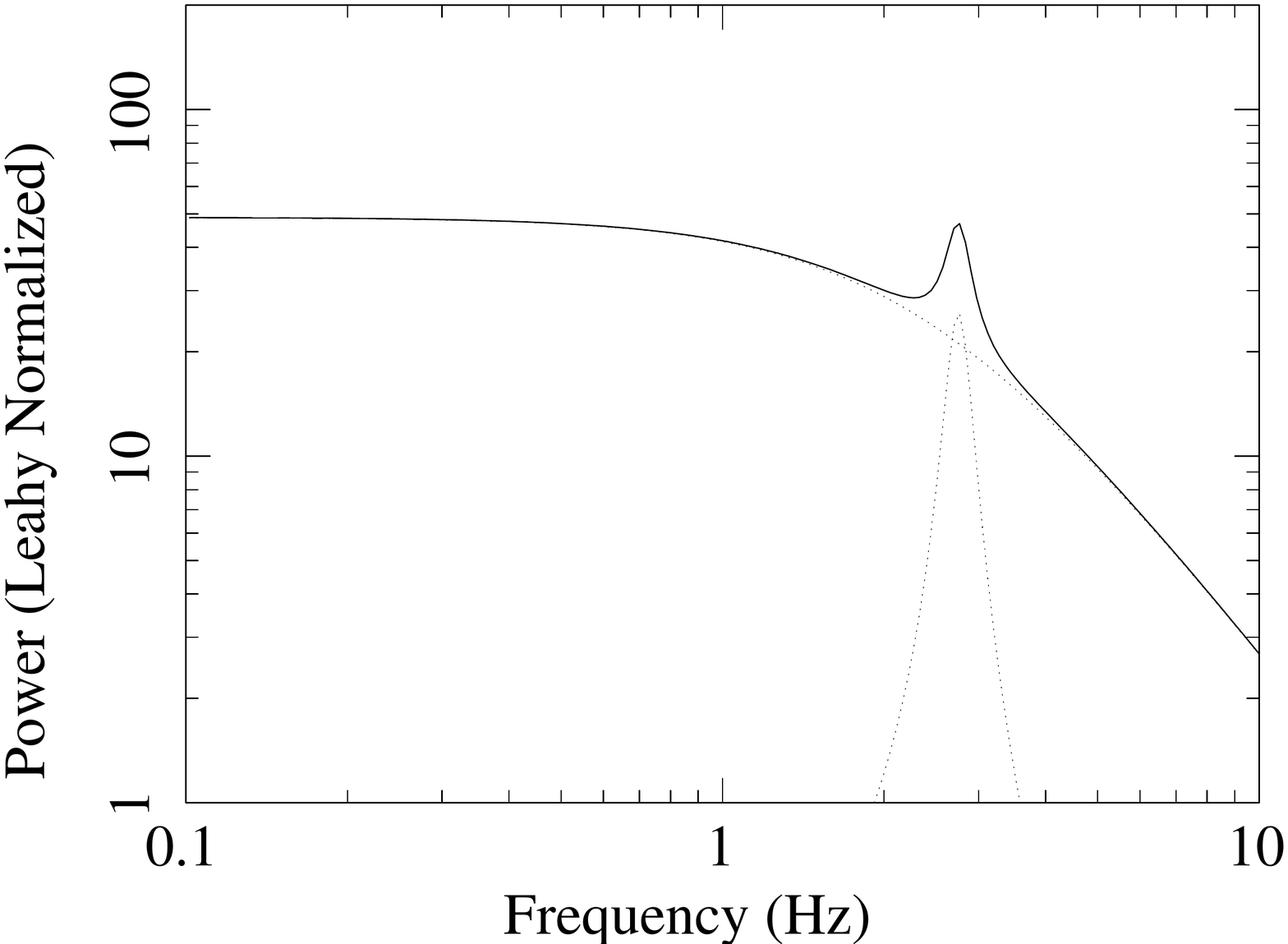}
    \includegraphics[width=0.32\textwidth]{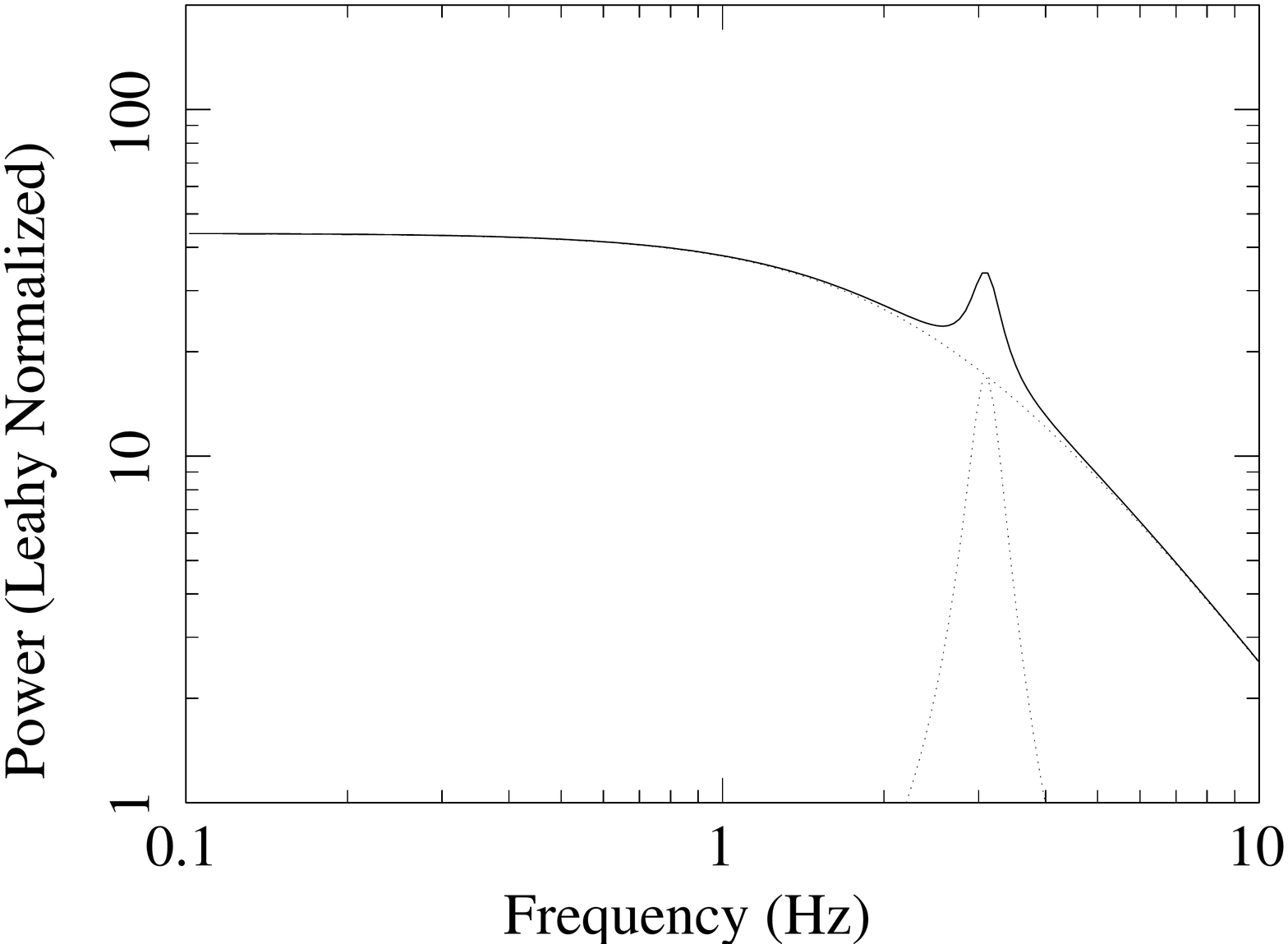}
    \includegraphics[width=0.32\textwidth]{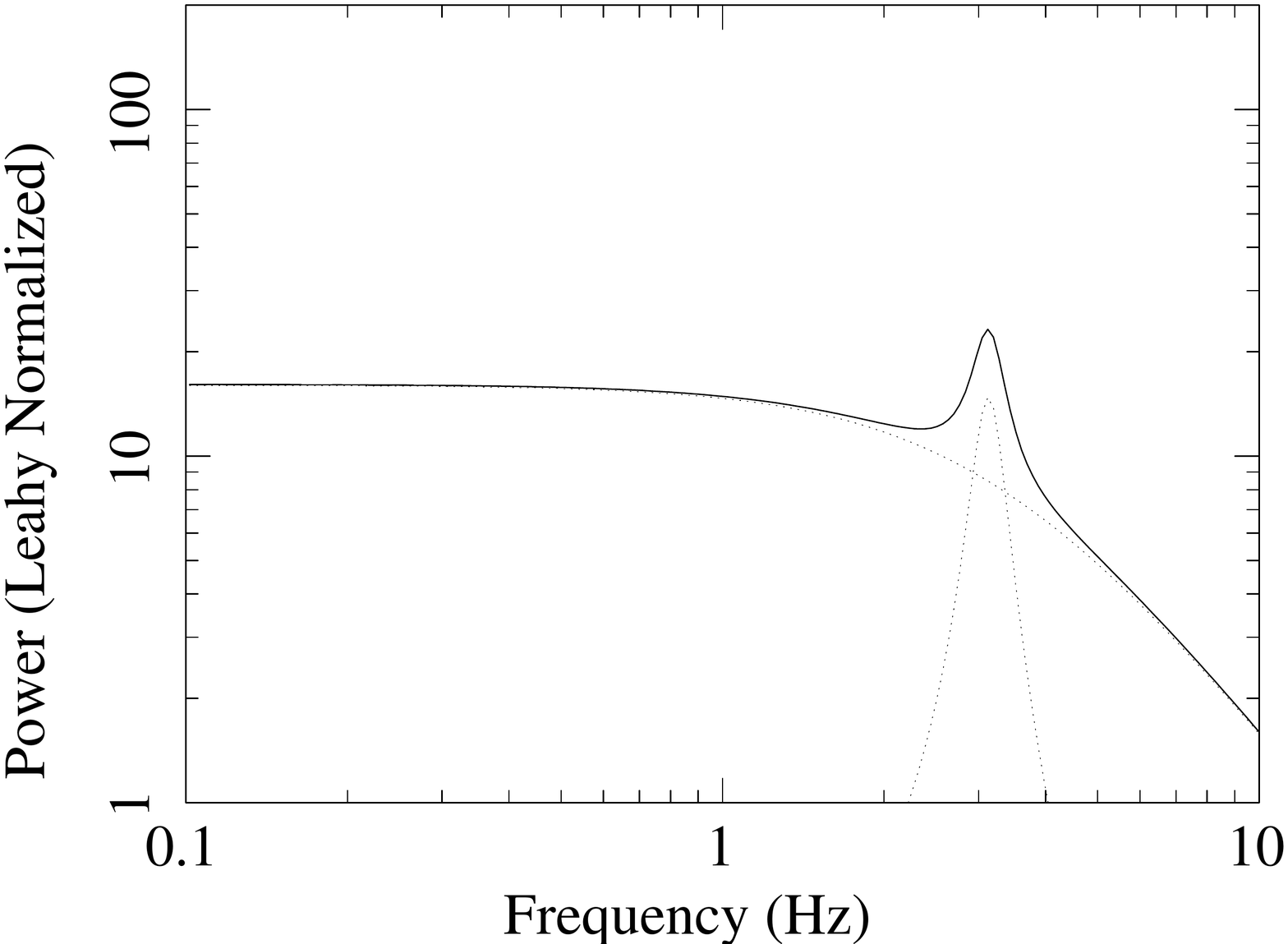}
    \caption{Example PDS with over plotted QPO predictions for the GRS 1915+105 observations 80701-01-54-02, 50703-01-28-01, 50703-01-24-01 ordered by column left to right from least (best-fitting) median to greatest (worst) Pythagorean sum of normalized errors on the three predicted QPO Lorentzian parameters (with corresponding models alone in bottom row). Note that the seemingly diminished height of the predicted QPOs is actually a consequence of how they were determined in the processing procedure, and in the case of the best observation 80701-01-54-02, the amplitude only differs by less than $0.3\%$ from the ``true'' amplitude value it was predicting, as the derived amplitudes had reduced amplitudes originally.}
    \label{fig:example_pds}
\end{figure*}

\begin{figure*}
    \centering
    \includegraphics[width=0.3\textwidth]{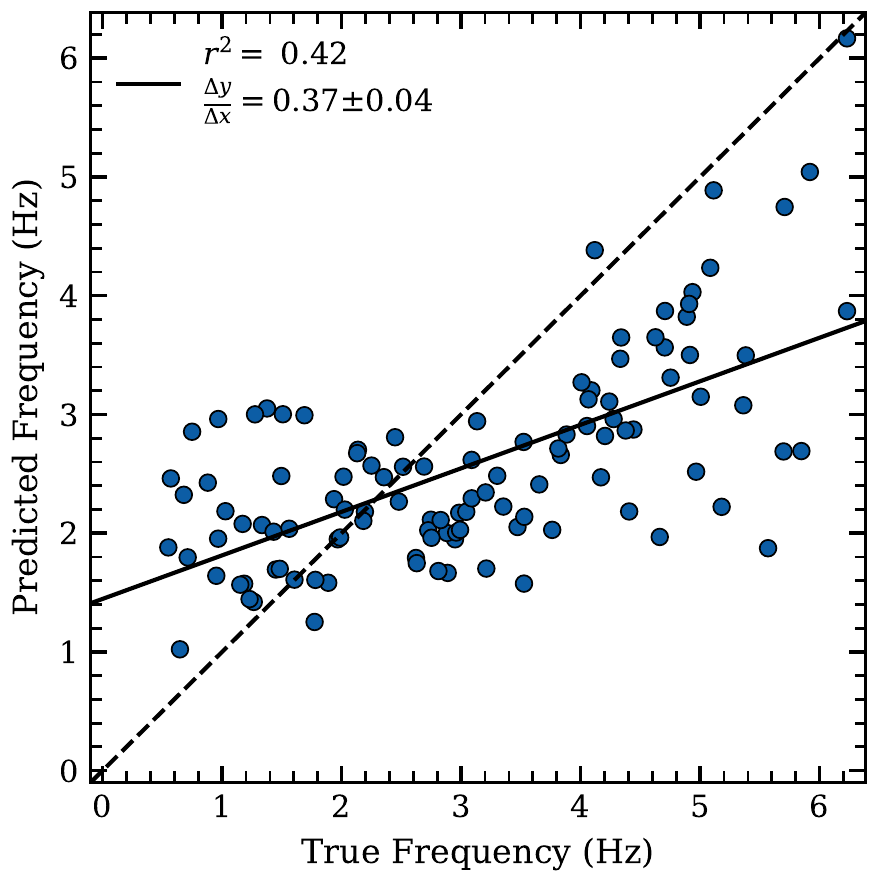}
    \includegraphics[width=0.3\textwidth]{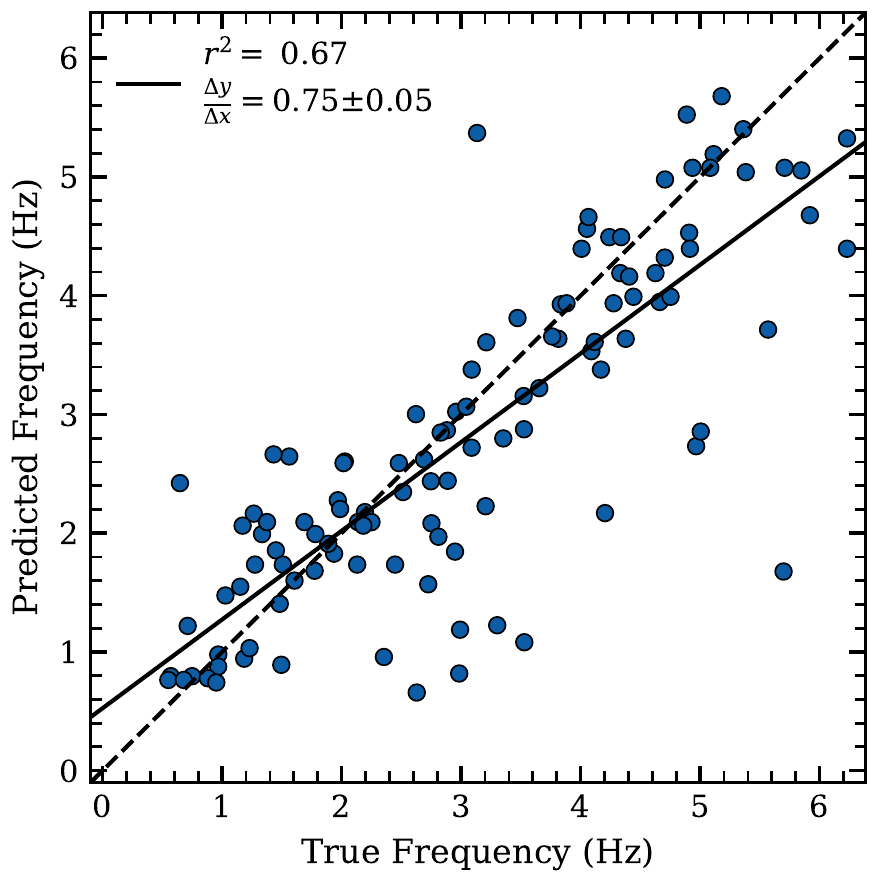}
    \includegraphics[width=0.3\textwidth]{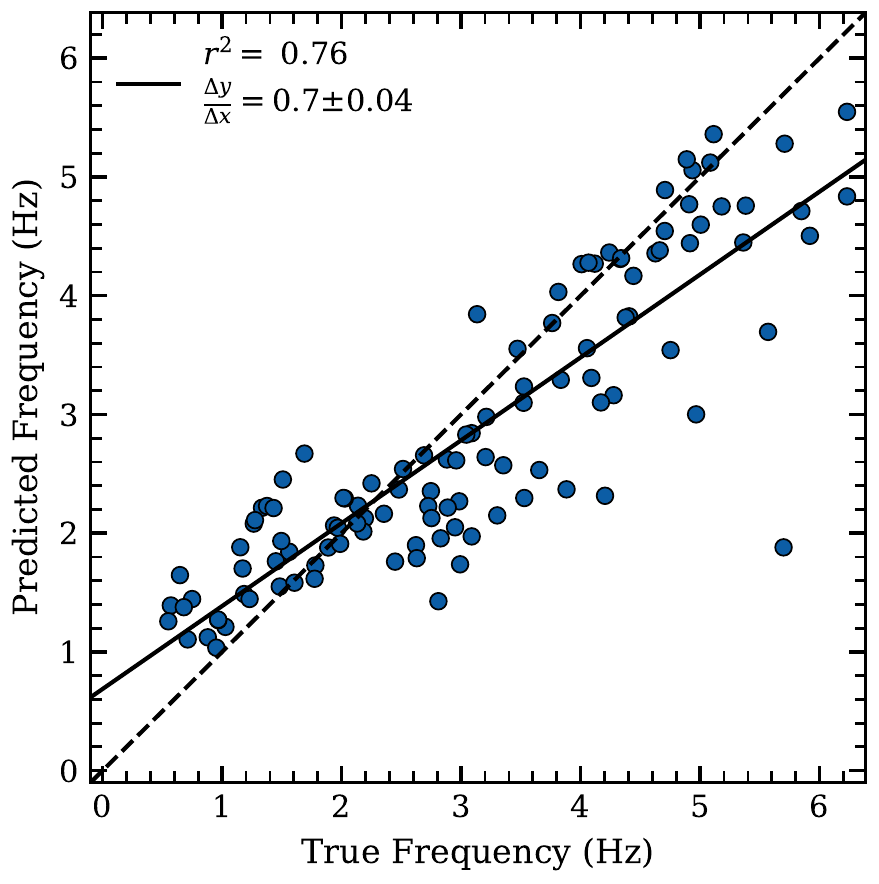}
    \includegraphics[width=0.3\textwidth]{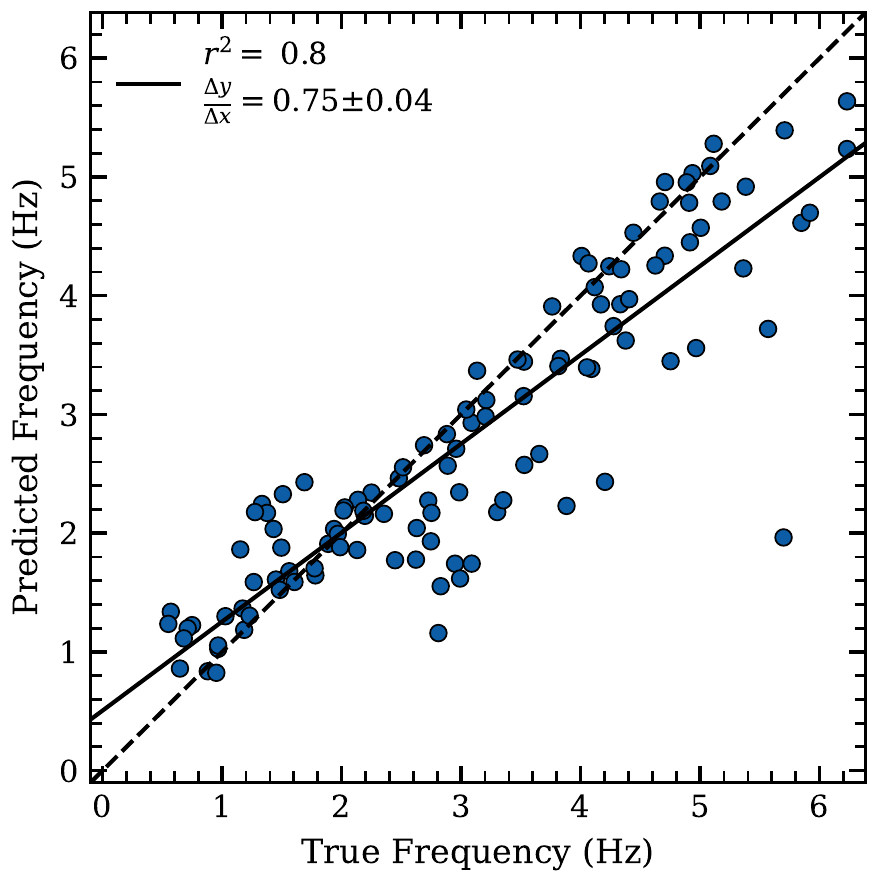}
    \caption{A results regression plot for all QPOs predicted from the test set for the source GRS 1915+105 as returned (from left to right) by linear regression, decision tree, random forest, and extra trees. The best models$-$random forest and extra trees$-$both minimize dispersion between true and predicted values (as quantified by $r^2$), while simultaneously producing the most 1:1 relationships between them (as quantified by best fit slope).}
    \label{fig:results_regression_grs}
\end{figure*}

\begin{figure*}
    \centering
    \includegraphics[width=0.3\textwidth]{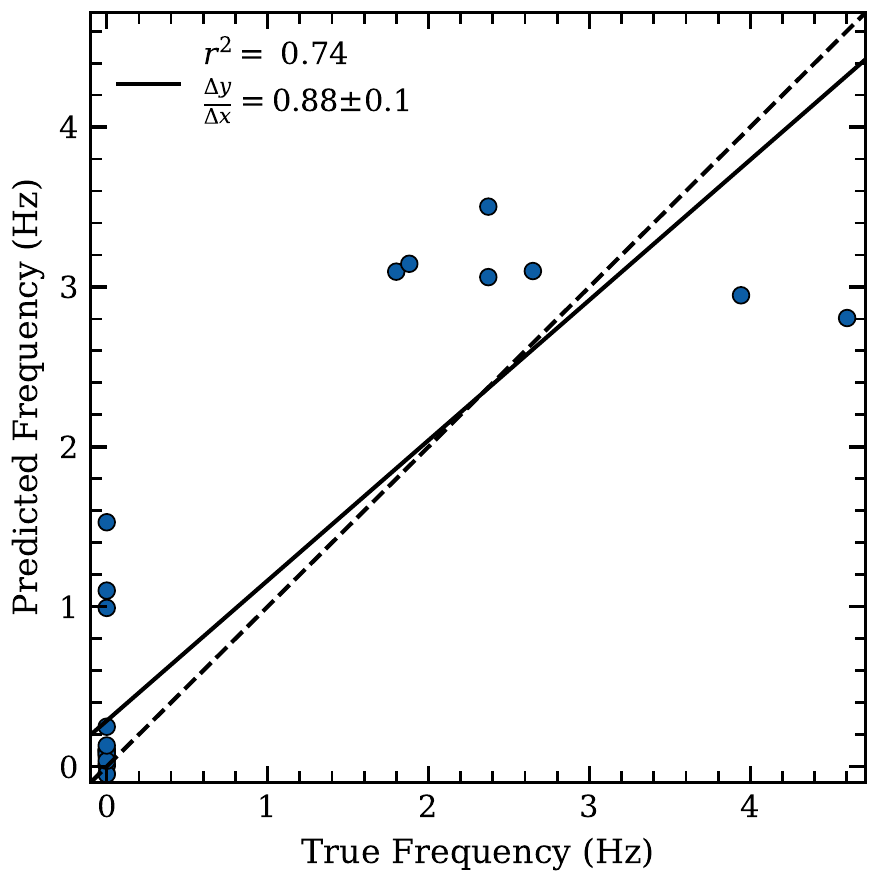}
    \includegraphics[width=0.3\textwidth]{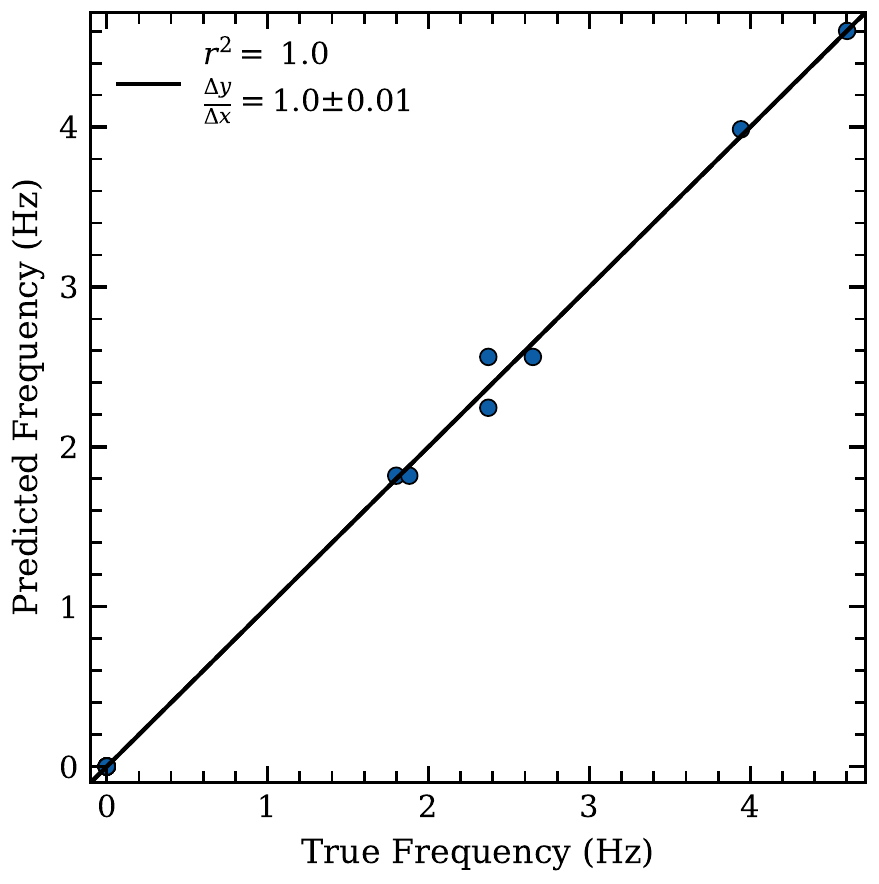}
    \includegraphics[width=0.3\textwidth]{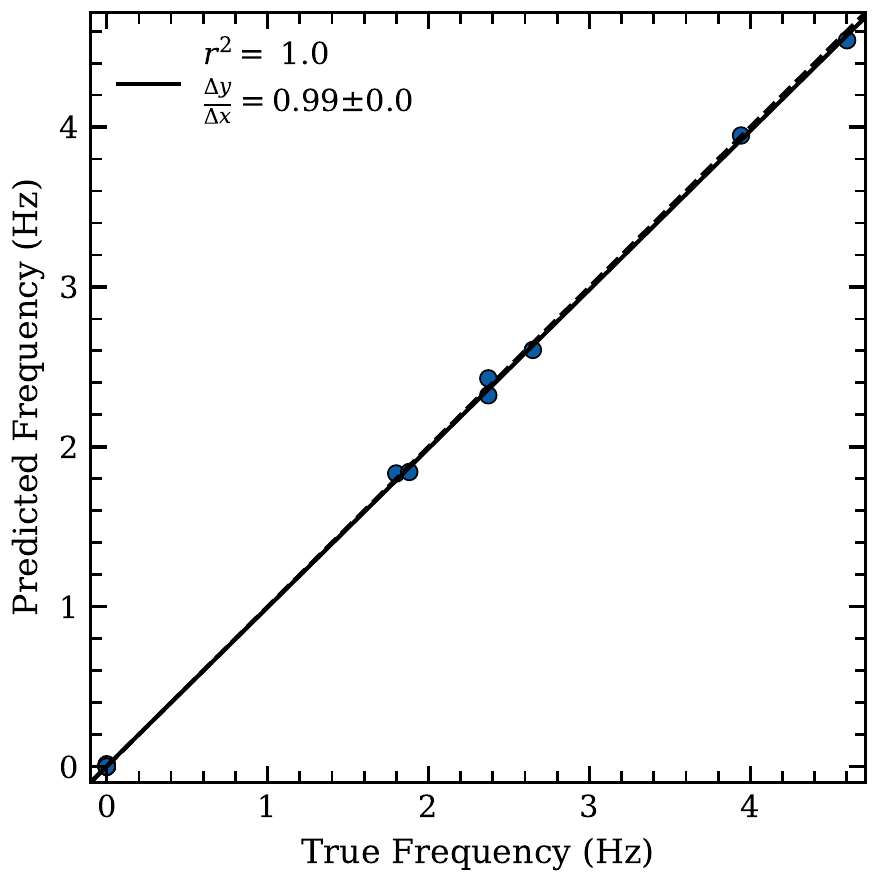}
    \includegraphics[width=0.3\textwidth]{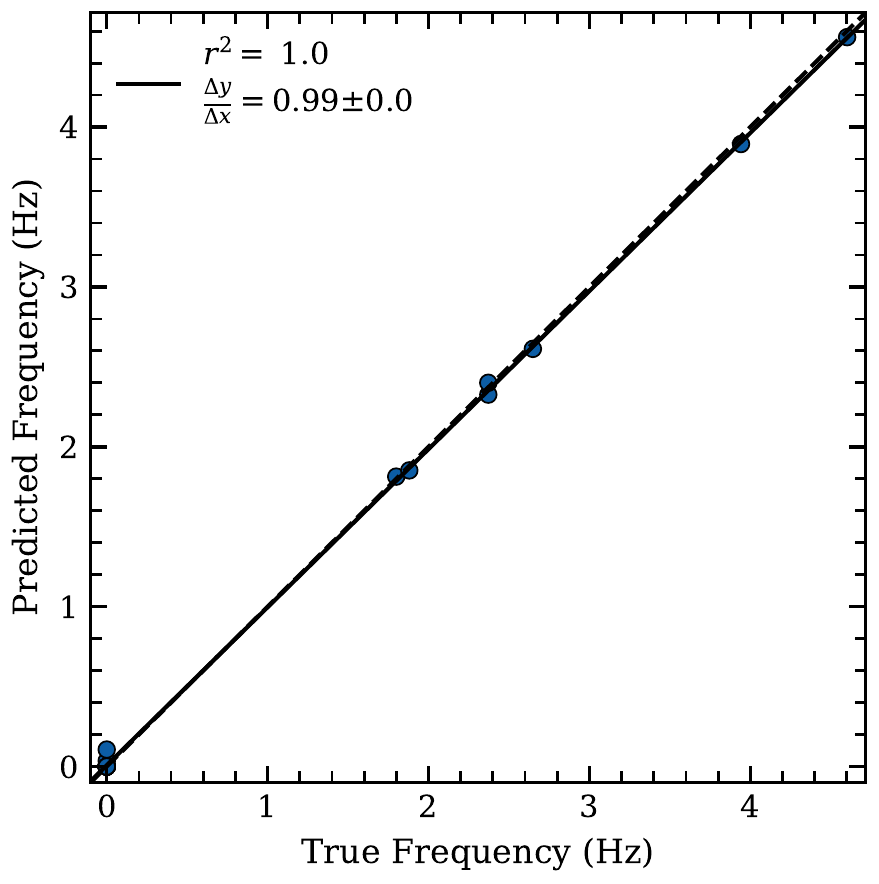} \\ 
     %% ALWAYS PUT SPECTRUM=TRUE ON BOTTOM BY ROW! %% 
   
    \caption{Same as Figure \ref{fig:results_regression_grs}, except for MAXI J1535-571 observations (processed feature input). The lesser number of points in these plots stems from both the smaller sample size of MAXI J1535-571 observations, as well as the clustering of values correctly predicted as zeros at the point $(0,0)$ where points cannot be seen individually in this plot). }
    \label{fig:results_regression_maxi_features}
\end{figure*}

\begin{figure*}
    \centering 
    \includegraphics[width=0.3\textwidth]{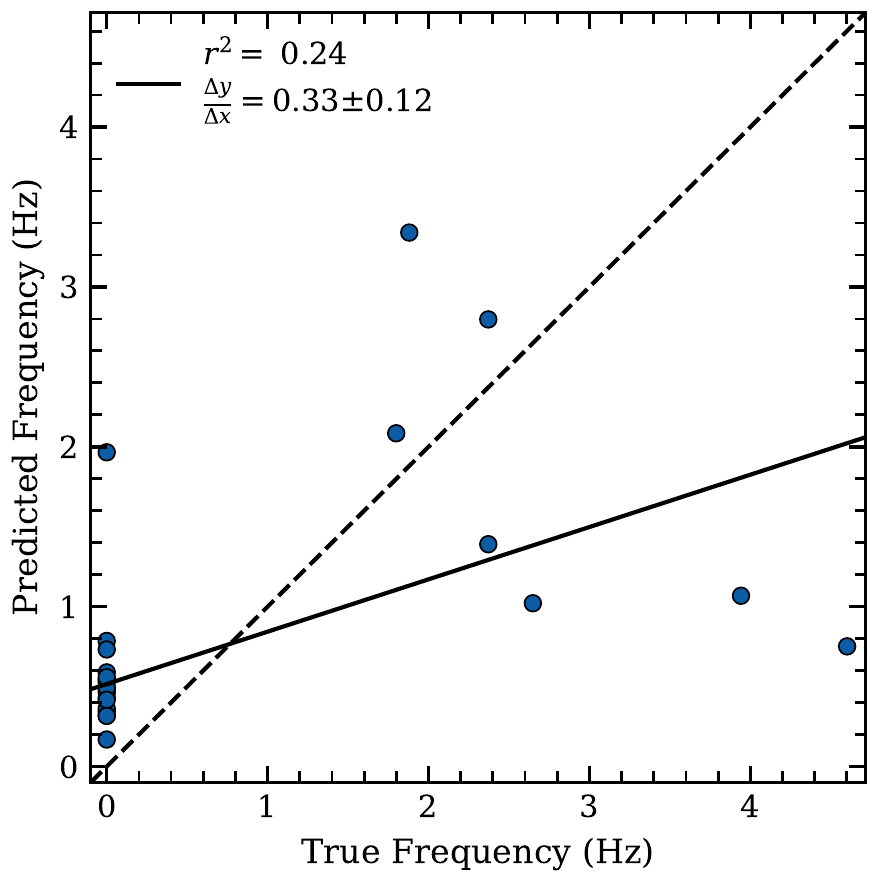}
    \includegraphics[width=0.3\textwidth]{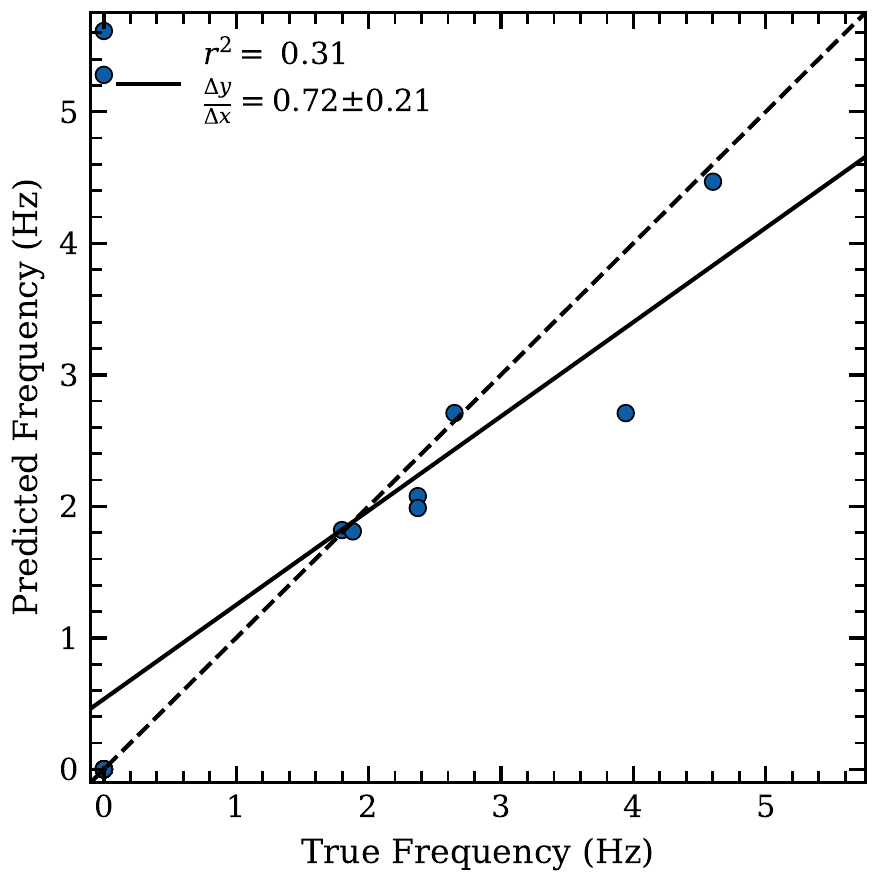}
    \includegraphics[width=0.3\textwidth]{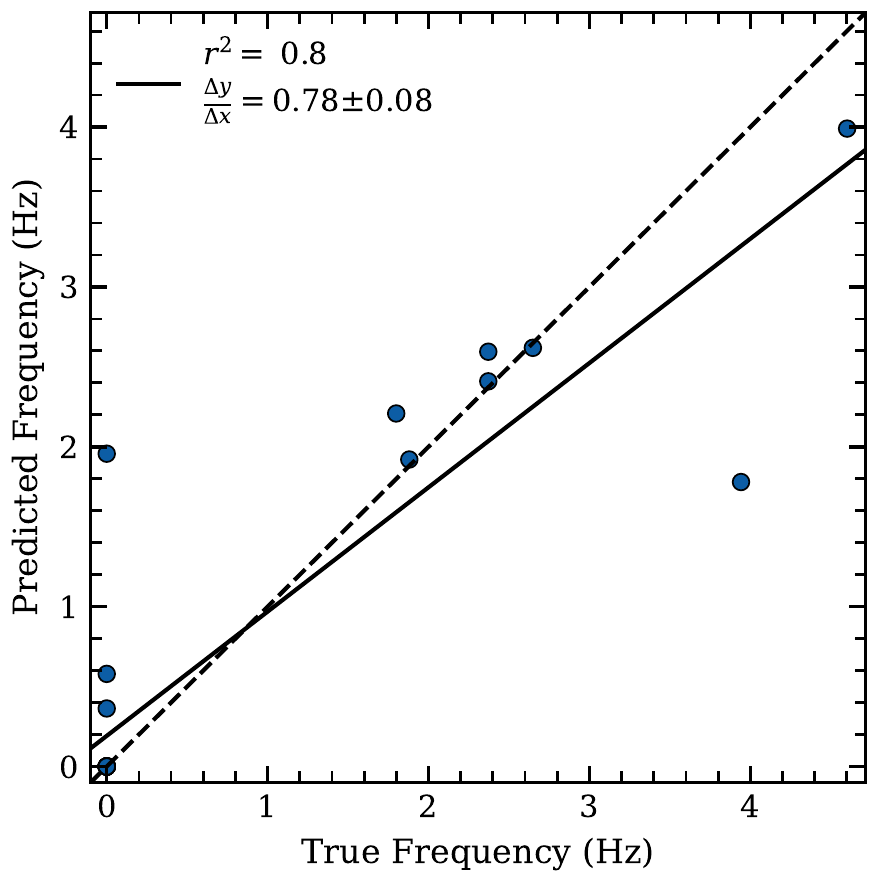}
    \includegraphics[width=0.3\textwidth]{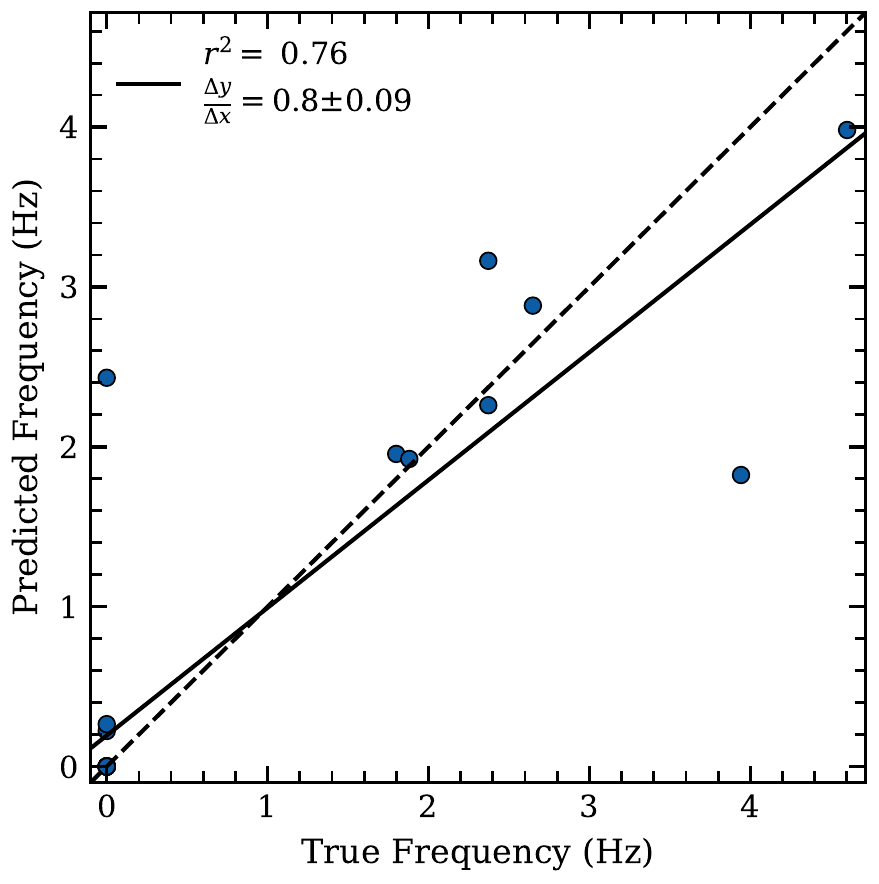}
    \caption{Same as Figure \ref{fig:results_regression_maxi_features}, except for MAXI J1535-571 observations (rebinned energy spectra as inputs). Note the increased dispersion and much less 1:1 relationships between true and predicted values for every model in the these plots compared to their equivalents in Figure \ref{fig:results_regression_maxi_features}}
    \label{fig:results_regression_maxi_spectrum}
\end{figure*}

\begin{figure*}
    \centering
    \includegraphics[width=0.24\textwidth]{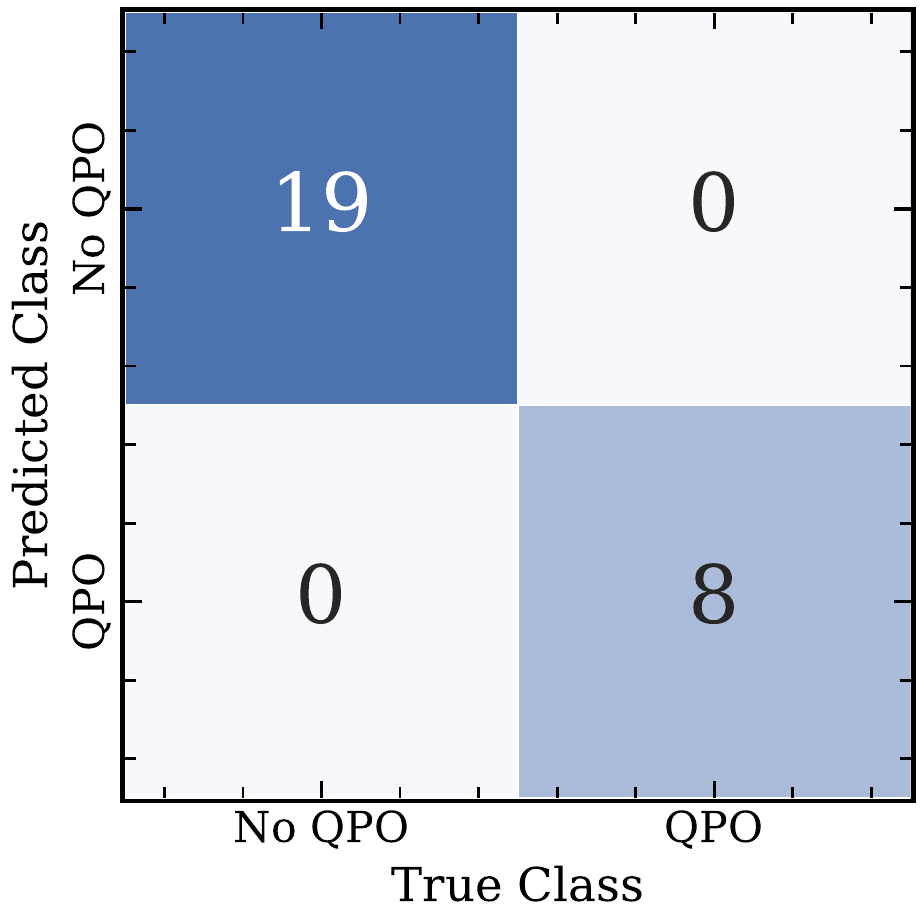}
    \includegraphics[width=0.24\textwidth]{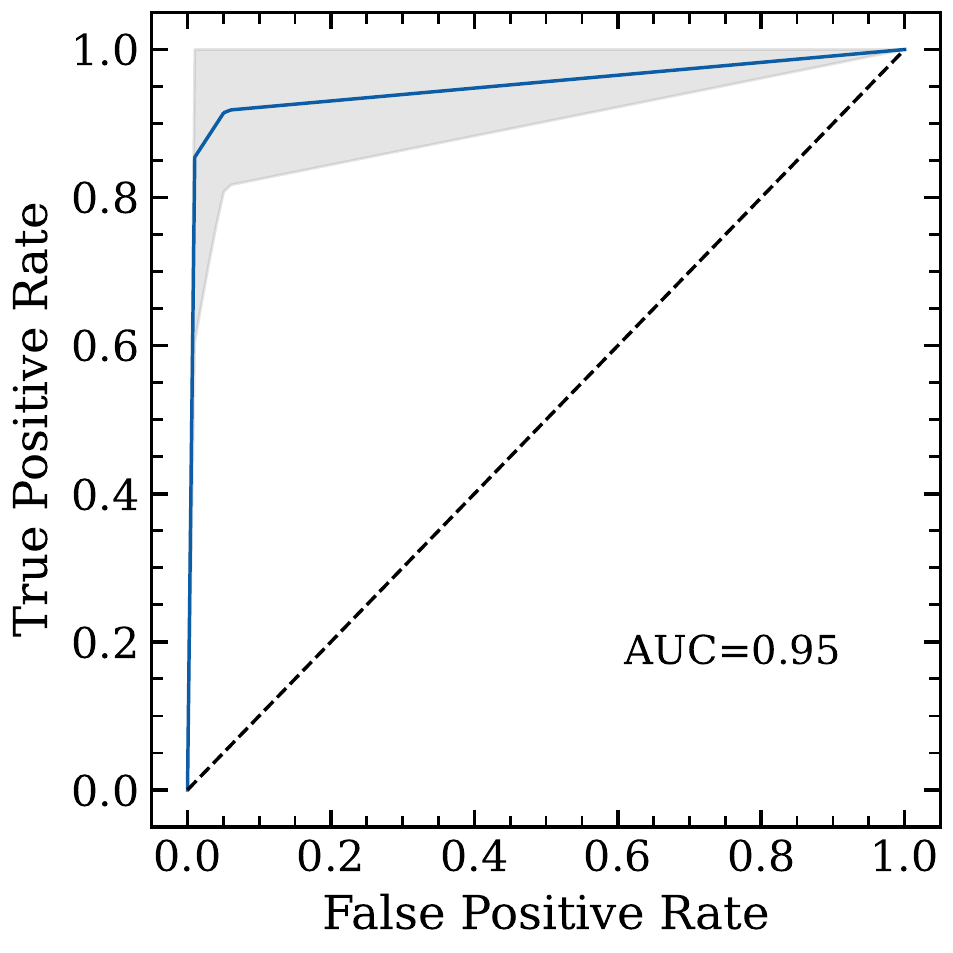}
    \includegraphics[width=0.24\textwidth]{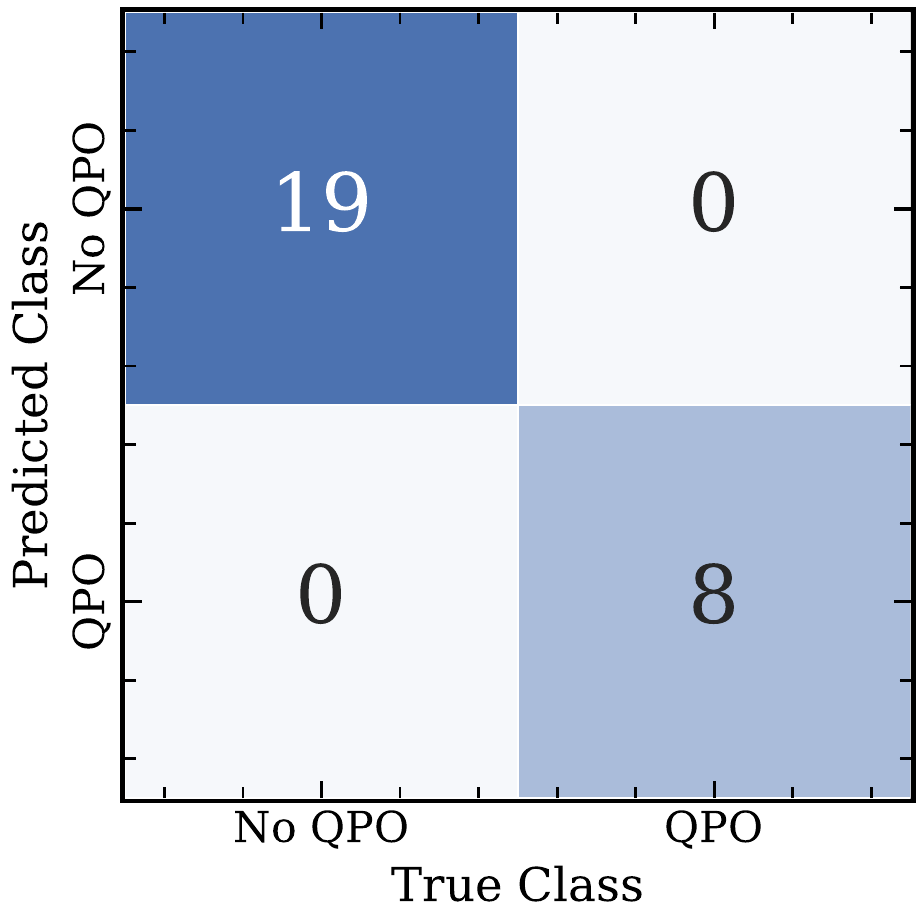}
    \includegraphics[width=0.24\textwidth]{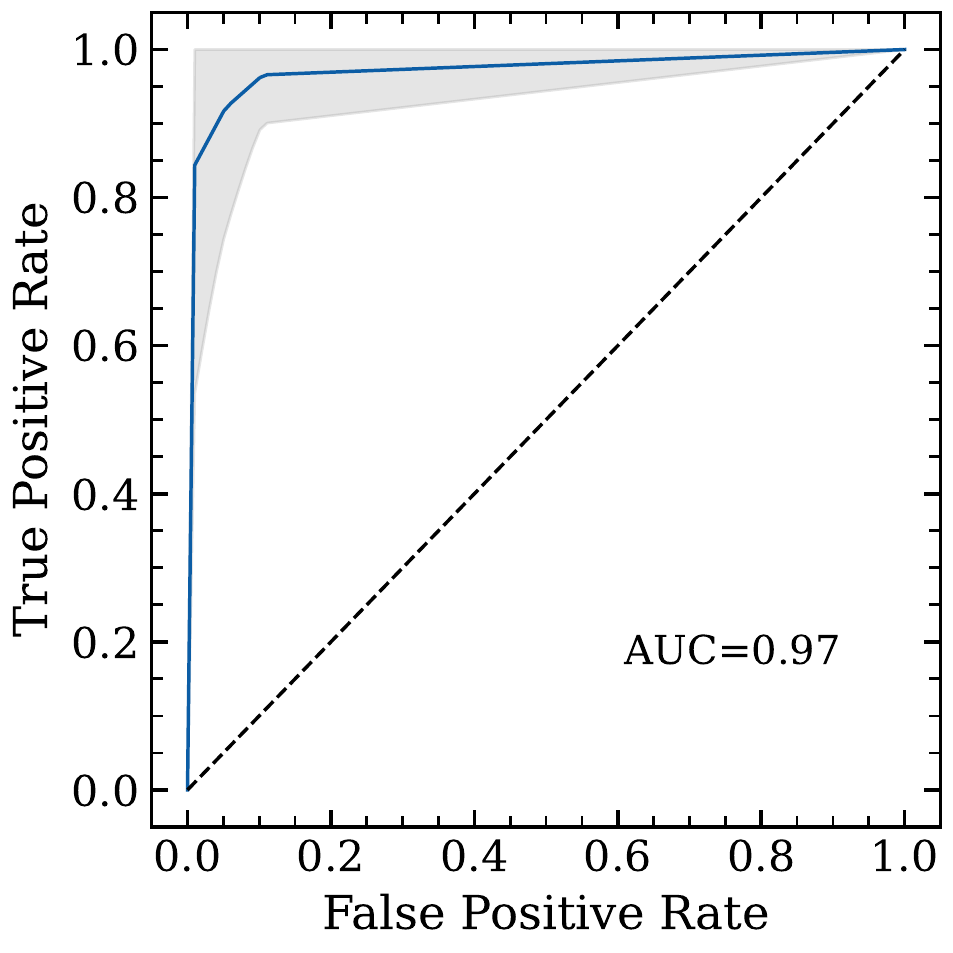}
    \includegraphics[width=0.24\textwidth]{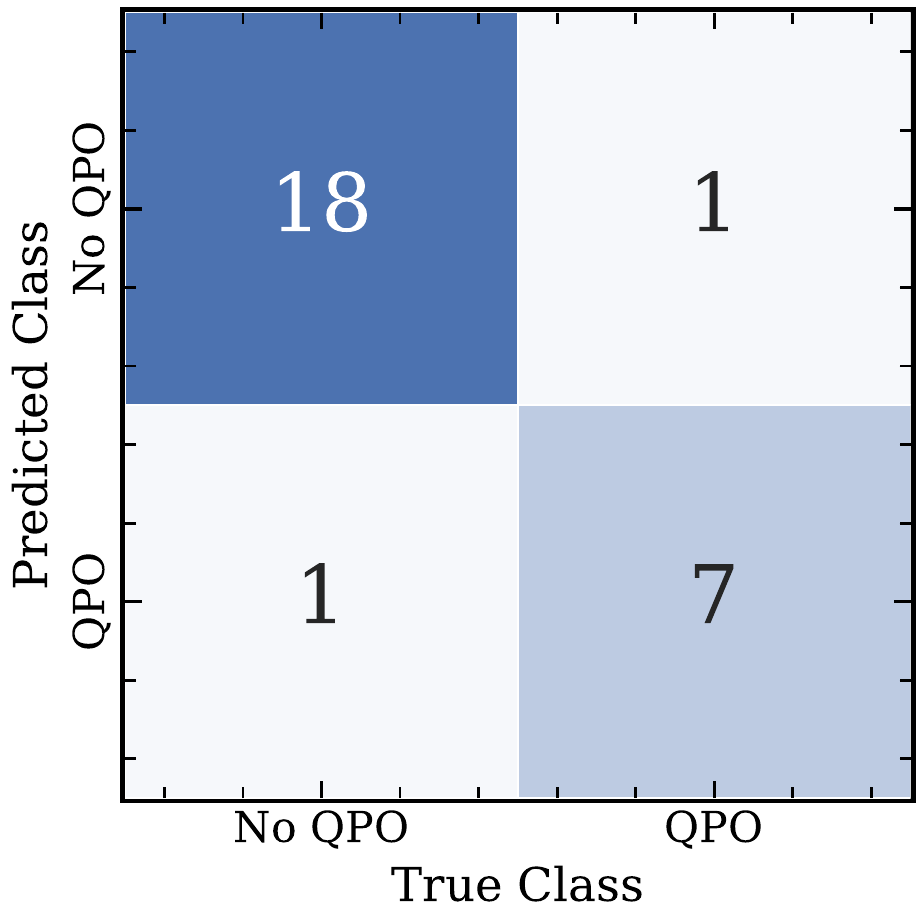}
    \includegraphics[width=0.24\textwidth]{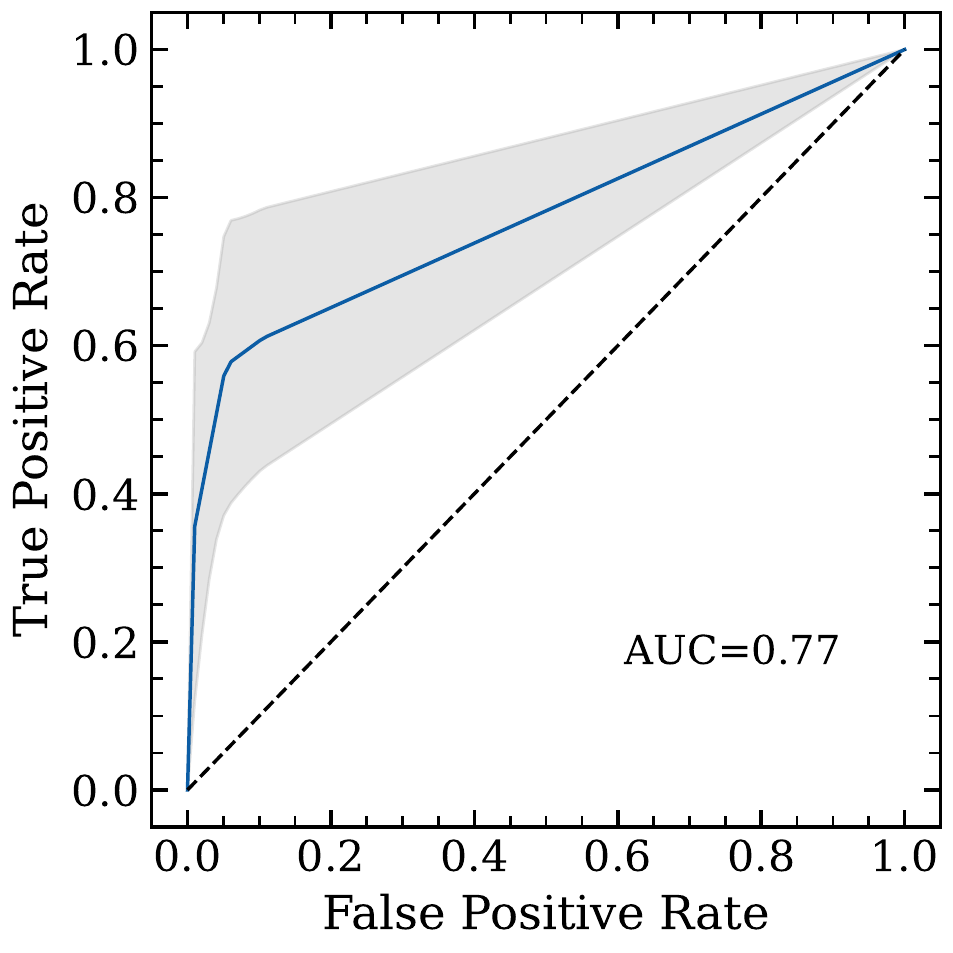}
    \includegraphics[width=0.24\textwidth]{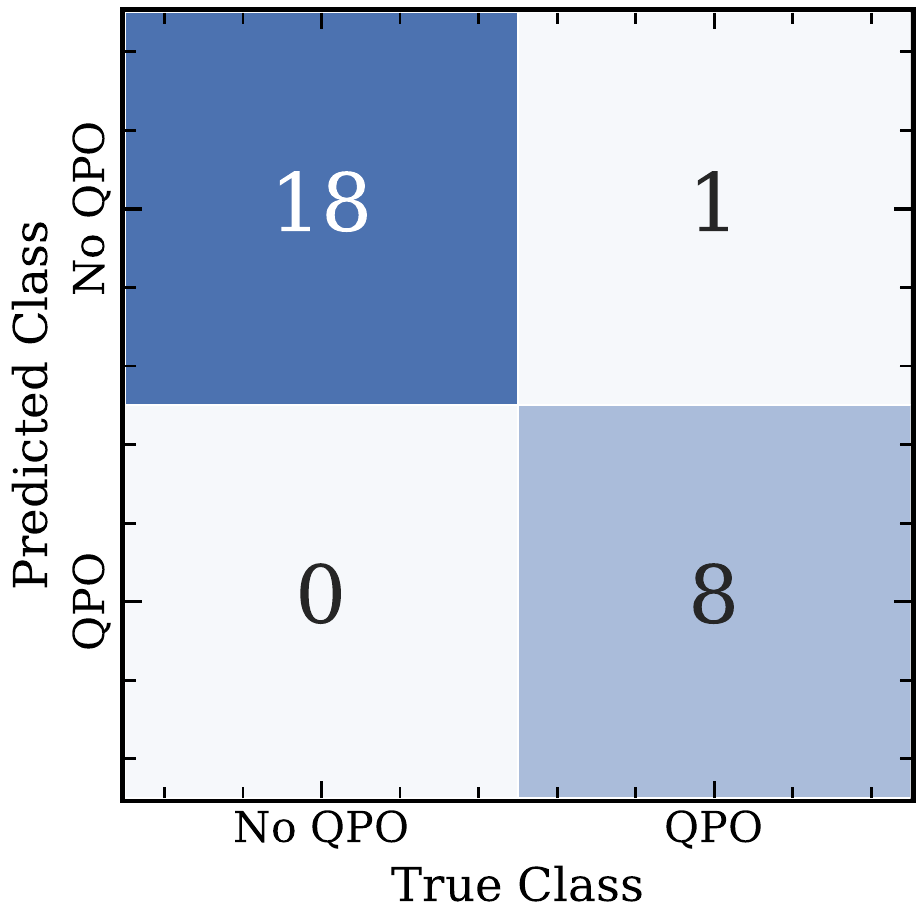}
    \includegraphics[width=0.24\textwidth]{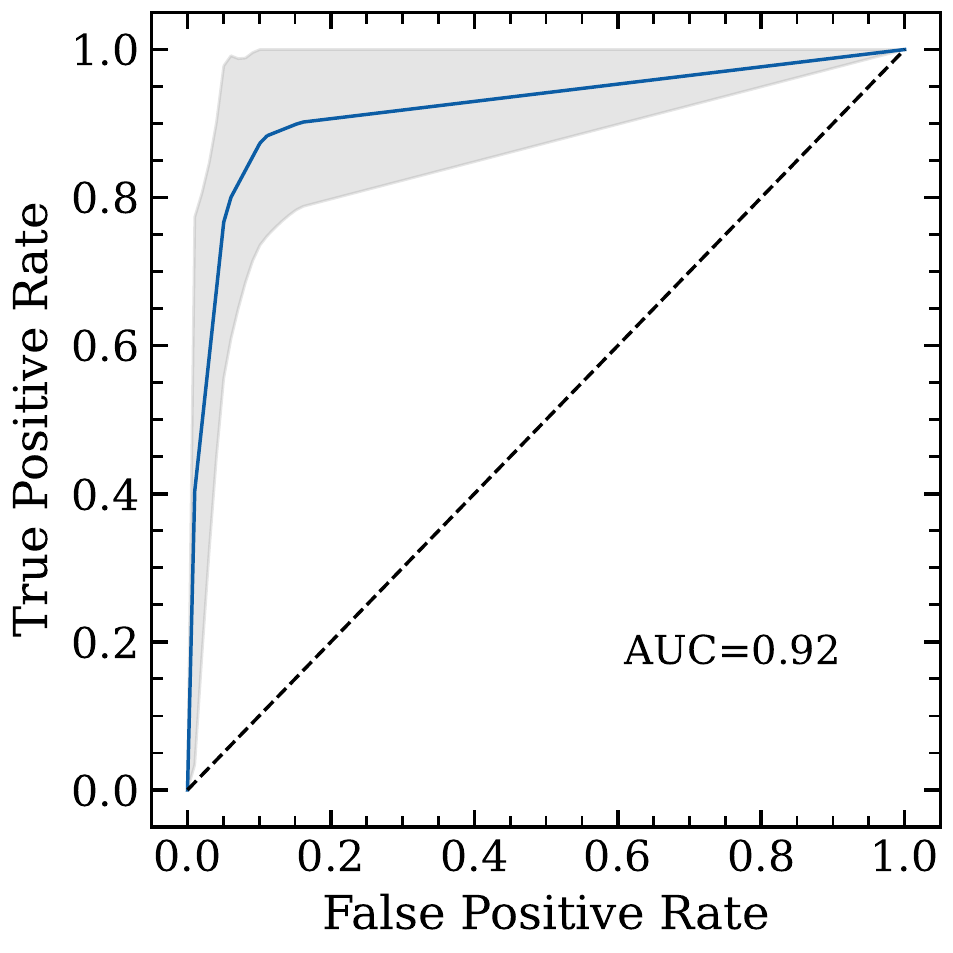}
    \caption{Confusion matrices and ROC Curves with labeled AUC values for MAXI J1535-571 binary classification cases. The left pairs correspond to logistic regression, whereas the right correspond to random forest. The confusion matrices are taken from the first tenth fold, whereas the ROC curves are averaged across all folds with $\pm1\sigma$ deviations denoted by the grey regions. The superior performance of the models working from processed inputs in the top row compared to their rebinned energy spectra input analogues in the bottom row is intriguing and discussed in more detail in Section \ref{sec:results}.}
    \label{fig:cm_and_roc_binary}
\end{figure*}

\begin{figure*}
    \centering
    \includegraphics[width=0.3\textwidth]{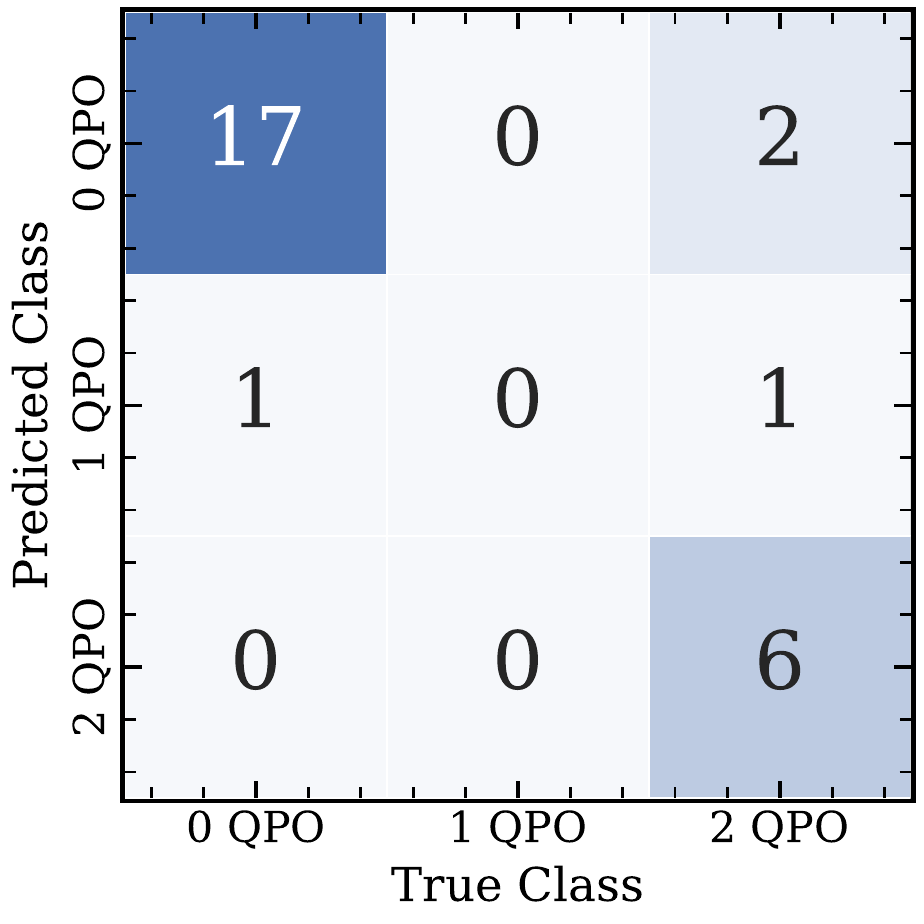}
    \includegraphics[width=0.3\textwidth]{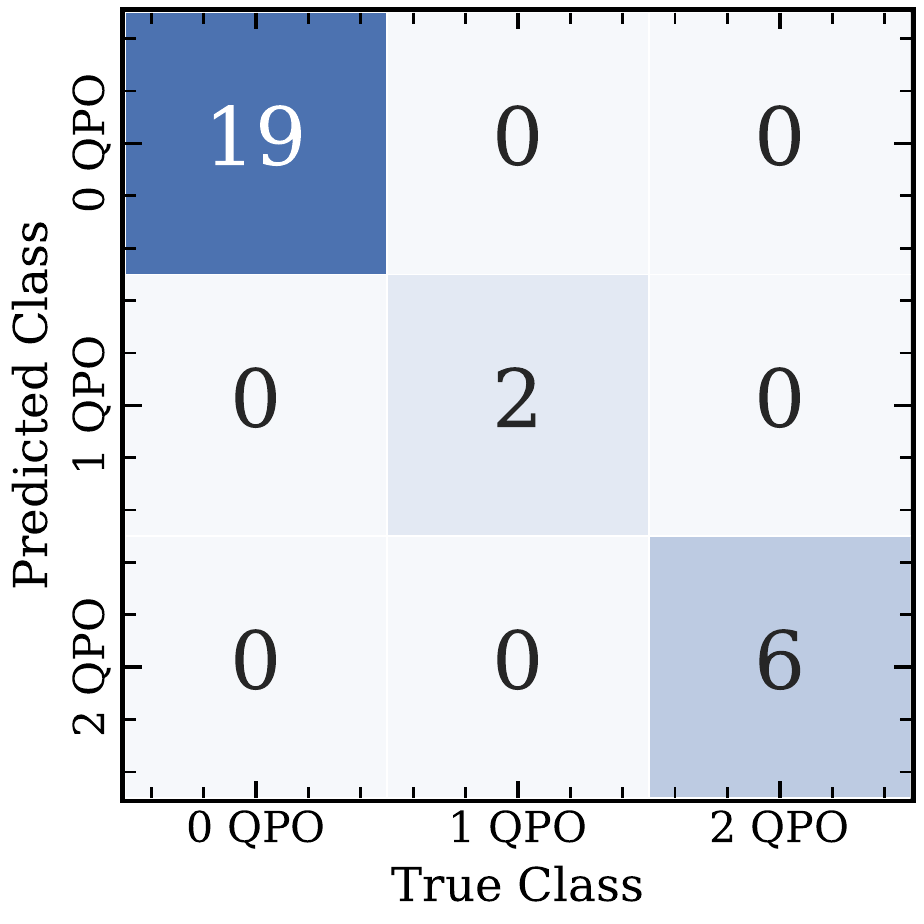} 
    \\
    \includegraphics[width=0.3\textwidth]{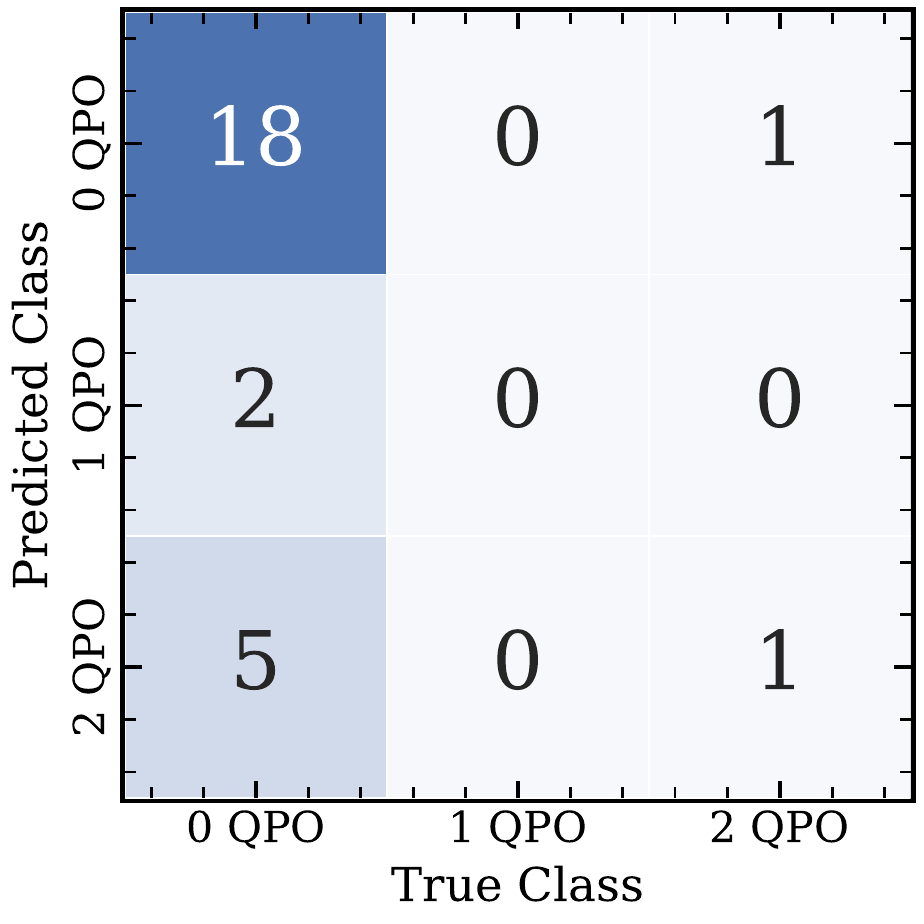}
    \includegraphics[width=0.3\textwidth]{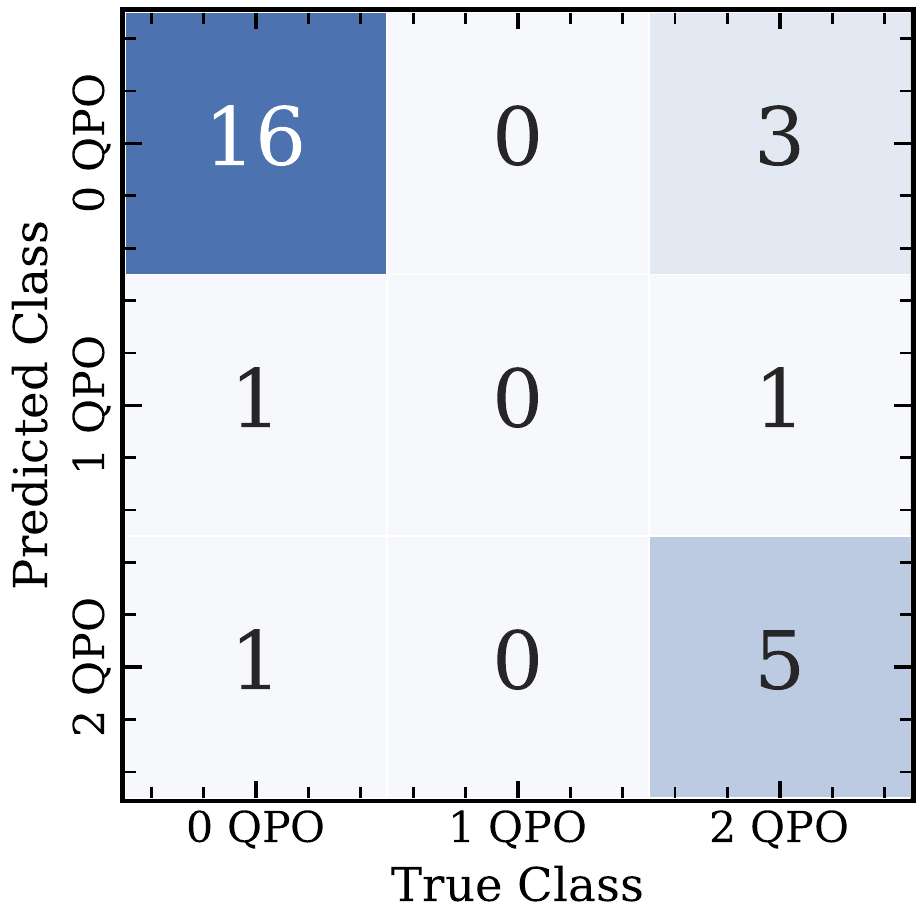}
    \caption{Confusion matrices for multiclass MAXI J1535-571 output, where the left column corresponds to logistic regression, the right column to random forest, the top row to processed input features, and the bottom row to rebinned energy spectra input features. Although only the accuracy of logistic regression decreases from binary to multinomial classification based on processed \texttt{XSPEC} input features, both models are significantly more inaccurate for the multinomial case based on energy spectra inputs compared to either binary case in Figure \ref{fig:cm_and_roc_binary}.}
    \label{fig:cm_multiclass}
\end{figure*}

\section{Discussion}\label{sec:discussion}

Now that we have demonstrated QPOs properties can be predicted—and in the following section show how features useful to these predictions can be analyzed—on the sources MAXI J1535-571 and GRS 1915+105 individually, we propose the next step would be to apply these methods in a future work on source-heterogeneous input data, a capability we intentionally incorporate into our QPOML library. To achieve this, it would be beneficial to construct a large standardized database of QPO and spectral data with a scope \textit{à la} \cite{blackcat}, for which the wealth of \textit{RXTE} observations will prove invaluable. Additionally, while increasing source sample size like this, it would also be fruitful to include neutron star LFQPOs and kHz QPOs in a followup study to generalize between sources, because unlike BH XRBs, NS XRBs are predominantly persistent and have significantly more observations with QPOs in archival \textit{RXTE} data in general \citep{mendez4U-1608,Migliari2003,belloniMendez2005,cenX-3QPOs}. That being said, the likely trade-off of using \textit{RXTE} data for these sources is that these QPOs will be predicted based on engineered XSPEC features instead of raw spectra given gain drift, as was the case with our analysis of GRS 1915+105 versus MAXI J1535-571. Another potential avenue for extending this work would involve exploring new input features to associate with QPOs, such as black hole spin, mass, inclination, jet properties, and QPO phase lags, and tracking the importance of variable features throughout outburst and accretion states to see if they evolve in tandem. Including scattering fraction as an input parameter promises interesting results as well, because QPO frequency and scattering fraction exhibit a correlation for sources like MAXI J1535-571 but an anti-correlation for other objects including GX 339-4, H1743-322 and XTE J1650-550 \citep{gargEnergyDependence}. Finally, how these non-parametric machine learning models interact with the polynomial/exponential versus sigmoidal relationship between frequency and power-law index for some black holes versus neutron stars \citep{titarchukSigmoidExp}, as well as how well models trained on distinct outbursts of certain objects perform for outbursts withheld from their training, would both also be of interest if these models are applied on samples that differ not only by source, but also by source type (BH or NS). Now, we turn to discussing feature importances in Section \ref{sec:feature_imps} and statistically compare the models we used throughout this work in Section \ref{sec:model_comparison}. 

\subsection{Feature Importances and Interpretation}\label{sec:feature_imps}

Feature importances refer to the relative attributed weights a model gives to different input features \citep{Saarela2021ComparisonOF}. In other words, they are measures for how helpful different features are for the model in making correct predictions, regardless of whether these predicted values are categorical or real-valued \citep{generalPermutations}. Before we discuss these, however, we will briefly describe our efforts to ensure the interpretability of our machine learning models. Interpretability is defined parsimoniously by \cite{tim-miller-interpretability} as the degree to which a human can understand the cause of a decision. Since most of our models are intrinsically complex (except for linear and logistic regression and decision trees), we seek \textit{post hoc} interpretability through feature importances \citep{post-hoc}. These values should not be interpreted as substitutes for other e.g. parametric importances, because they seek to explain how a machine learning model learns and interacts with its data. However, we believe that properly calculated feature importances may offer alternative helpful insight about the origins of QPOs, and we therefore take steps to avoid common pitfalls associated with these measures. For example, although it is common to discuss default impurity-based feature importances, this approach is flawed because it is both biased towards high-cardinality numerical input features, as well as computed on training set statistics, which means it may not accurately generalize to held-out data \citep{scikit-learn}. Additionally, although permutation importances are commonly put forward as a superior alternative, these suffer from multicollinearity, as in the process of permutating single features, an impactful feature could be erroneously ascribed as having little-to-no effect on model performance if it has high correlation with another feature \citep{Strobl2007,Nicodemus2010,hooker2019}. Therefore, we chose to to determine feature importances with the contemporary \texttt{TreeSHAP} algorithm as implemented in the Python package \texttt{shap} by \cite{SHAP2017}. This model extends game theoretic coalitional Shapley values to calculate SHapley Additive exPlanations (SHAP) in the presence of multicollinearity by incorporating conditional expected predictions \citep{shapleyvaluesoriginal,SHAP2017,molnar2022}. As hinted earlier and detailed in \cite{SHAP2017} and \cite{molnar2022}, an additional benefit of using tree based models is that through tree traversal and dynamic programming the computational cost for computing SHAP values is brought down from exponential time $\mathcal{O}(2^n)$ to   $\mathcal{O}(n^2)$ polynomial time. We calculate feature importances shown in Section \ref{sec:results} for each model $f$ by treating the model from the tenth fold in the first repetition as if they were taken from the test set, and averaging their $\phi_i(f,x)$ from Equation \ref{eq:shapley}, which represents the weighted average of differences in model performance when a feature $x$ out of $M$ simplified input features is present versus absent for all subsets $z'\subseteq x'$.

\begin{equation}\label{eq:shapley}
    \phi_i(f,x) = \sum_{z' \subseteq x'} \frac{|z'|!(M - |z'| -1)!}{M!} \left [ f_x(z') - f_x(z' \setminus i) \right ]
\end{equation}

\begin{figure*}
    \centering
    \includegraphics[width=0.35\textwidth]{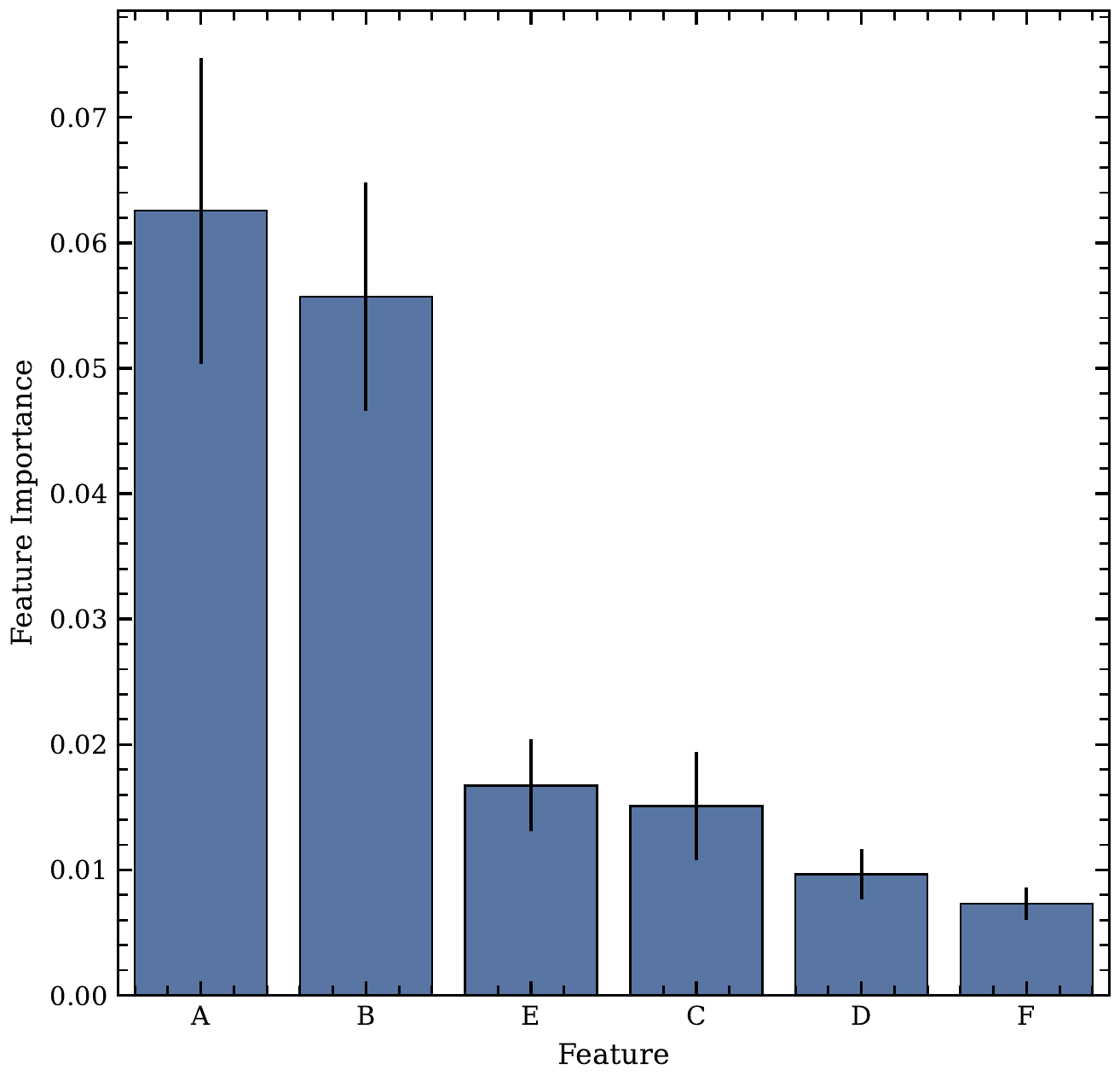}
    \includegraphics[width=0.35\textwidth]{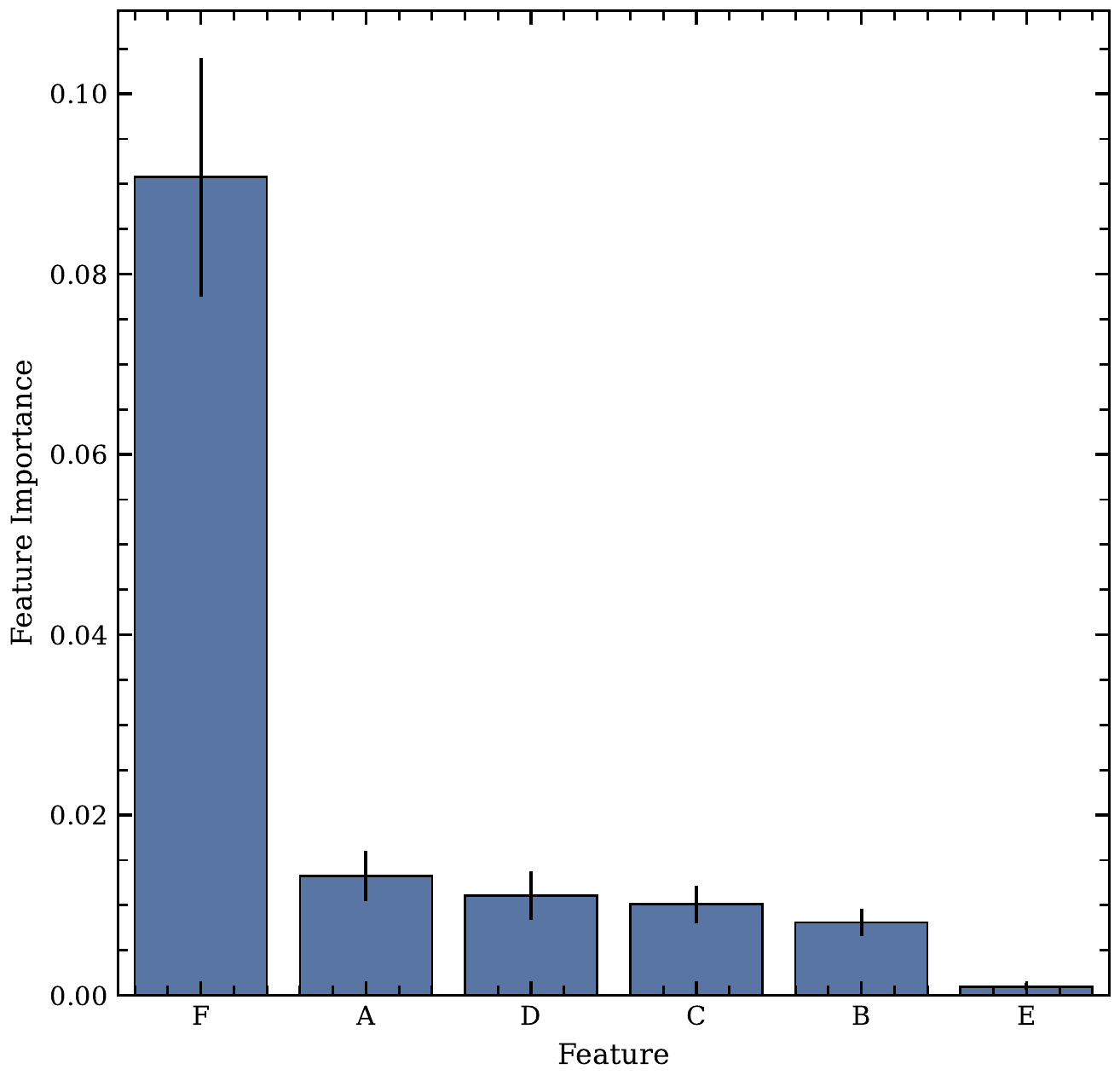}
    \includegraphics[width=0.35\textwidth]{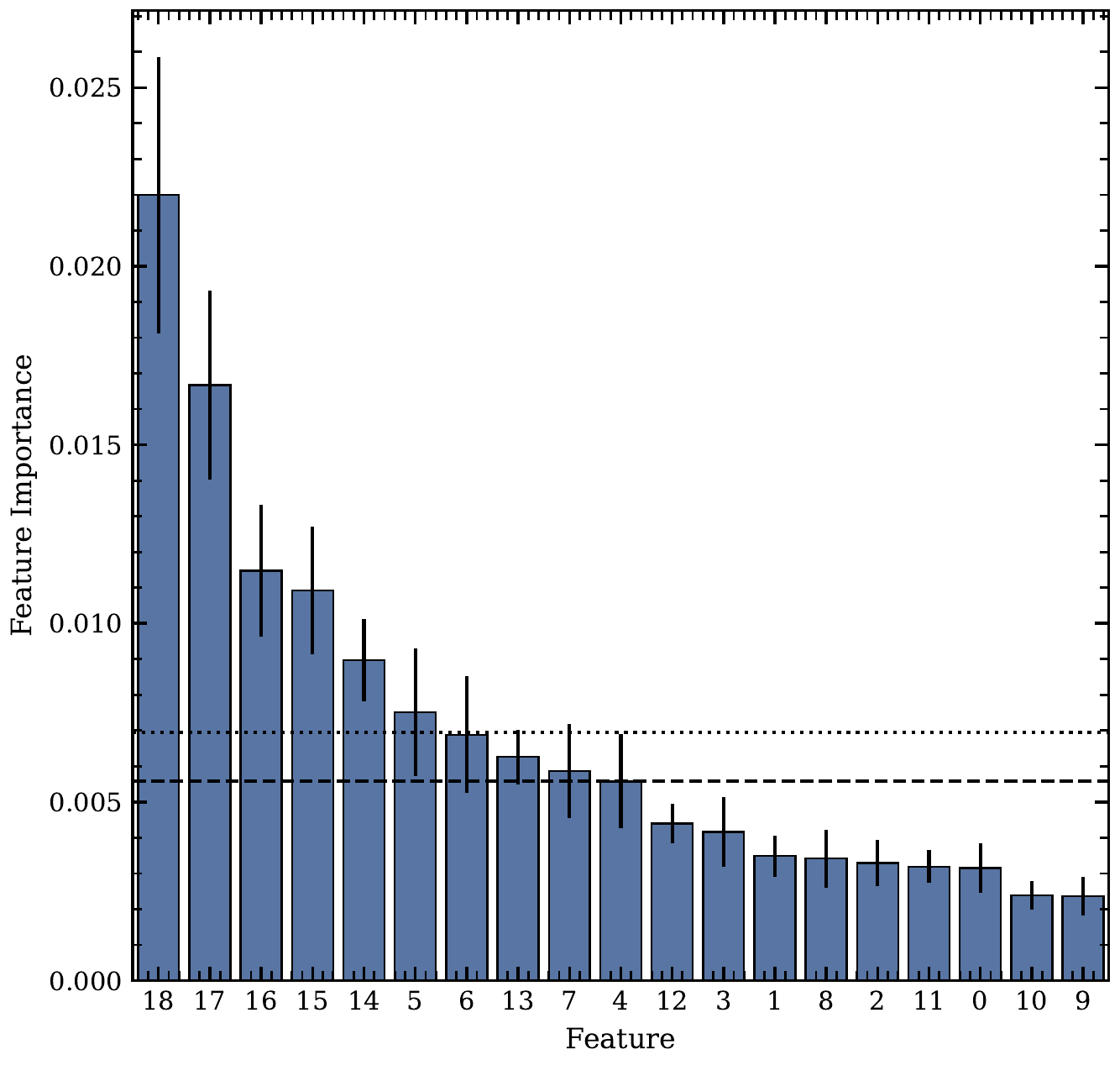}
    \caption{Tree-SHAP calculated average of absolute value SHAP feature importances for the most accurate predictive regression models for GRS 1915+105 engineered inputs (left, extra trees), MAXI J1535-571 engineered inputs (middle, extra trees), and MAXI J1535-571 energy spectra inputs (right, extra trees). The features denoted $A-F$ correspond to net count rate, hardness ratio, asymptotic power-law photon index, \texttt{nthcomp} normalization, inner-disk temperature, and \texttt{diskbb} normalization features, respectively. The error bars on each importance correspond to 99$\%$ confidence intervals on mean importances, the dashed line the median importance of all features, and the dotted line the mean of the same. Features corresponding to hard channel count rates are significantly more important than the median and mean feature importance, which is likely related to the higher energy origin of QPOs. An interesting difference between these plots and that for GRS 1915+105 in Figure \ref{fig:feature_importances_reg} is that Extra Trees primarily weights \texttt{diskbb} normalization for MAXI J1535-571 regression but splits primary importance for GRS 1915+105 between the net count rate and hardness ratio engineered inputs.}
    \label{fig:feature_importances_reg}
\end{figure*}

\begin{figure*}
    \centering
    \includegraphics[width=0.35\textwidth]{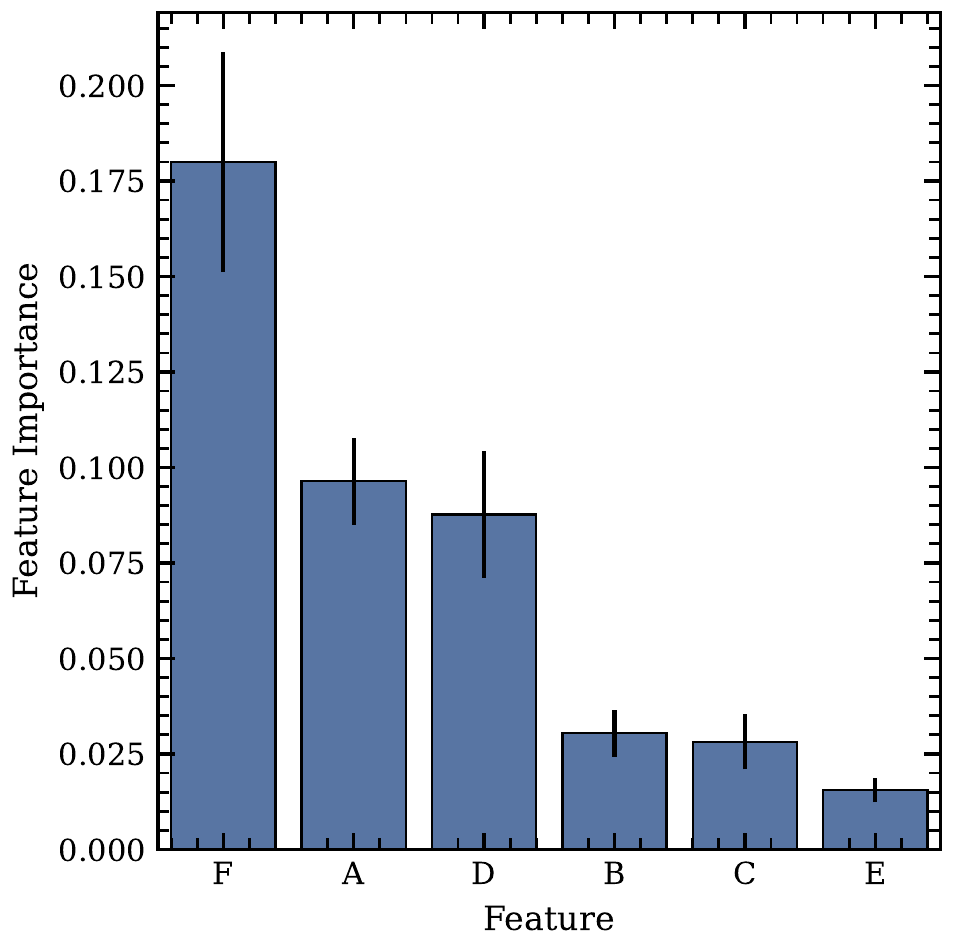}
    \includegraphics[width=0.35\textwidth]{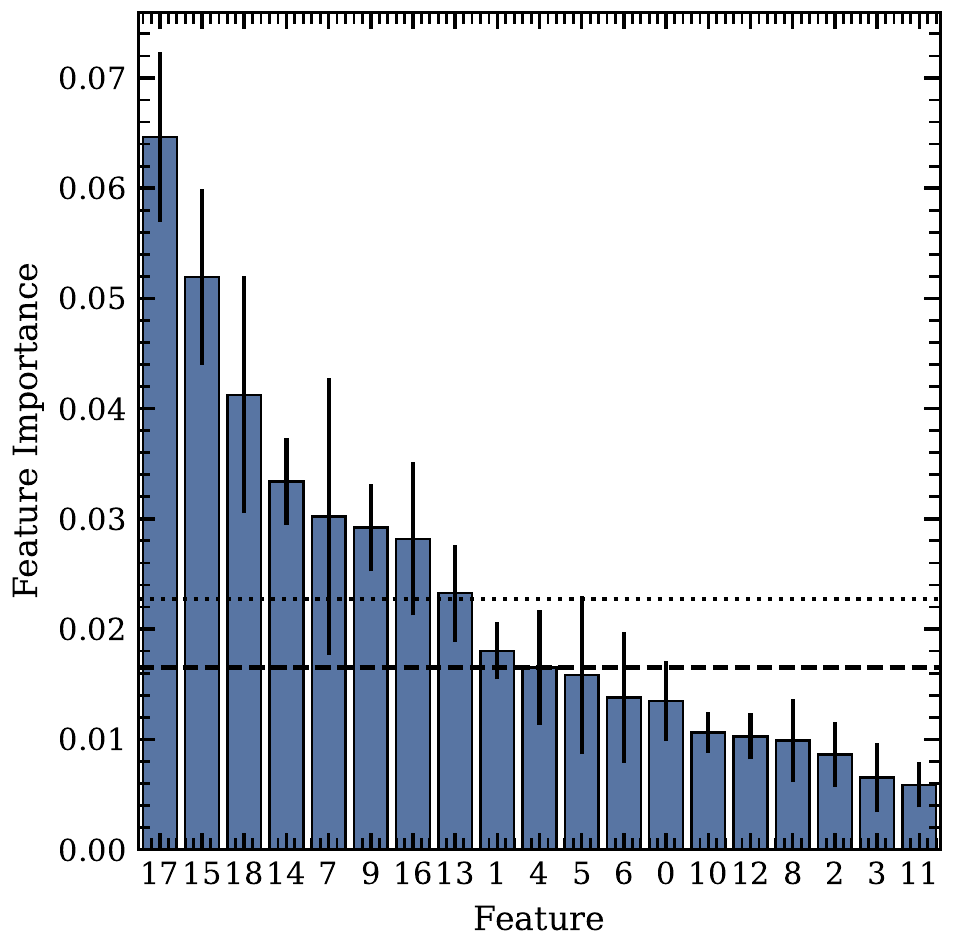}
    \caption{Similar to Figure \ref{fig:feature_importances_reg}, except for the best classification models for MAXI J1535-571 binary output based on engineered inputs (left, random forest), and energy spectral inputs (right, random forest). As seen for regression, hard energy channels similarly dominant feature importances for energy spectra input, yet, while \texttt{diskbb} normalization is still the most important processed feature for classification, more importance is attached here to net count rate and \texttt{nthcomp} normalization here than for regression on MAXI J1535-571.}
    \label{fig:feature_importances_class}
\end{figure*}

One of the most important things shown by Figure \ref{fig:feature_importances_reg} and \ref{fig:feature_importances_class} is that there are significant interesting differences between the feature importances attributed to the processed features for GRS 1915+105 and MAXI 1535-571, which may be related to the nuances of the process driving QPOs in these systems. For example, in GRS 1905+105, net count rate and hardness ratio are clearly the most important features, after which importance falls precipitously and remains uniformly modest for the rest, with this proportional decrease ranging from a factor of three for \texttt{nthcomp} asymptotic power law to six for \texttt{nthcomp} and \texttt{diskbb} normalization. Because we have used SHAP values for importance, we can rule out the un-importance of these features stemming from multicollinearity or training set artifacts, which means they could \textit{potentially} be related to curious physical related conditions. However, there is no ambiguity about the importance of net count rate and hardness, because an XRB outburst's q-shaped state-evolution in the hardness-intensity diagrams (HIDs) is known to also be indicative of changes in timing (e.g., QPO) properties as tracked in HIDs \citep{motta2015LotsofQPOs,mottaquickreview}. This is also in agreement with the findings of Figure 2 of \cite{evolvingPropertiesGRSCorona}, in which the QPO frequency of GRS 1915+105 is shown to vary with a somewhat inverse relationship with hardness ratio across mostly horizontal and vertical gradients in inner disk temperature and power law index, respectively. %Mention a motta paper for HID information and KNN style HID association. 

In contrast to GRS 1915+105, the feature importances for both the best regression and classification models on processed MAXI J1535-571 input features favor a single feature above all others: \texttt{diskbb} normalization (although in the case of classification, net count rate and \texttt{nthcomp} normalization are still significant for MAXI J1535-571). This quantity (ignoring relativistic and plasma corrections) approximately corresponds to the projected area of the inner-disk on the sky: $N_{\mathrm{disk}}=(\frac{R_{in}}{D_{10}})^2\cos(\theta)$, where $R_{in}$ is the apparent inner disk radius in km, $D_{10}$ is the distance to the source in 10 kpc units, and $\theta$ the angle of the disk \citep{XSPEC1999}. This prominent importance is intriguing because it implies a dependence between QPO presence and frequency on \texttt{diskbb} normalization and therefore inner disk radius. This is corroborated by \cite{gargEnergyDependence}, who find that QPO frequency correlates significantly with the inner disk radius for MAXI J1535-571 in data provided by \textit{AstroSat} according to the power law relationship $v_{\mathrm{QPO}}\propto \dot{M} R_{in}^p$, where $\dot{M}$ is mass-accretion rate \citep{astrosat}. However, \citep{gargEnergyDependence} do not find a clear relationship between \texttt{diskbb} normalization and QPO frequencies in the $\sim1.6-2.8$ Hz range. Overall, the similarity in feature importances for engineered features for regression and classification in MAXI J1535-571 shows that the same features that are  important in determining the parameterizations of QPOs are those important in determining their presence vs absence. Regarding the feature importances derived from the energy spectra, the highest energy channels are the most important for both regression and classification, with the five most important channel counts rates for each coming from the equivalent $[9.5-10), [9.0-9.5), [8.5-9.0), [8.0-8.5)$ and $[7.5-8.0)$ keV channels for regression and $[9.0-9.5), [9.5-10.0), [8.5-9.0), [8.0-8.5)$ and $[3.0-3.5)$ keV channels for classification. Notably, for both classification and regression only hard channels $\geq 3$ keV have importances significantly greater than the mean and median importances for all features in their respective sets at the $99\%$ confidence level. The fact that the high-energy spectral data is most informative of the QPOs is interesting and we speculate that this may be related to the fact that QPOs manifest more prominently at higher energies above the disk’s peak temperature. A broader perspective which generalizes these relationships to other BH systems is of high interest, but outside the scope of this work. Consequently, we are currently working on a comprehensive follow-up work, in which we will evaluate these models on data identically reprocessed for numerous black holes and neutron stars simultaneously. One additional difference between this preliminary work and that prospective one will be full inclusion of all LF QPO features for all sources (such as GRS 1915+105), because although focusing on the dominant frequency for QPOs in GRS 1915+105 served our purposes here, this would be a limitation in the future because such focus would not make it clear whether these trained forest methods would predict many false positives and false negatives for sources similar to GRS1915+105 that do include QBO-absent data, yet perform well nonetheless.

\subsection{Statistical Model Comparison} \label{sec:model_comparison}

As mentioned in Section \ref{sec:methods}, we included an ordinary least squares model as a benchmark for their utilization. As Figure \ref{fig:grs_average_performances}, Figure \ref{fig:results_regression_grs}, \ref{fig:results_regression_maxi_features}, and \ref{fig:results_regression_maxi_spectrum} demonstrate, each of our models outperform linear regression. In order to assess the significance of the improvements, we employ the \cite{nadeauandbengio} formulation of the frequentist Diebold-Mariano corrected paired t-test \citep{dieboldandmariano}, 

% Nadeu and Bengio: https://papers.nips.cc/paper/1661-inference-for-the-generalization-error.pdf

% Diebold-Mariano: http://www.est.uc3m.es/esp/nueva_docencia/comp_col_get/lade/tecnicas_prediccion/Practicas0708/Comparing%20Predictive%20Accuracy%20(Dielbold).pdf

\begin{equation}\label{eq:freq}
    t=\frac{\frac{1}{k \cdot r}\sum_{i=1}^{k}\sum_{j=1}^{r}x_{ij}}
{\sqrt{(\frac{1}{k \cdot r}+\frac{n_{\mathrm{test}}}{n_{\mathrm{train}}})\hat{\sigma}^2}}
\end{equation}

\noindent where $k=10$ and represents the number of k-fold validation folds, $r=10$ and equals the number of times we repeated the $k$-fold procedure, $x$ is the performance difference between two models, and $\hat{\sigma}^2$
represents the variance of these differences \citep{scikit-learn}. It is necessary to correct the $t-$values in this manner because the performances of the models are correlated with each fold upon which they are tested, as some folds may make it harder for one of, or all of, the models to generalize, whereas others make it easier, and thus the collective performance of the models varies. The results of these pairwise tests for all permutations of two models on both sources is shown in Table \ref{tab:pairwise}. 

We additionally implement the Bayesian \cite{benavoli} approach, which allows us to calculate the \textit{probability} that a given model is better than another, using the Student distribution formulated in Equation (\ref{eq:bay}): 

% Benavoli et al: http://www.jmlr.org/papers/volume18/16-305/16-305.pdf

\begin{equation}\label{eq:bay}
    \mathrm{St}(\mu;n-1,\overline{x},(\frac{1}{n}+\frac{n_{\mathrm{test}}}{n_{\mathrm{train}}}) %St = standard part function
\hat{\sigma}^2)
\end{equation}

\noindent where $n$ is the total number of samples, $\Bar{x}$ is the mean score difference, and $\hat{\sigma}^2$ is the \cite{nadeauandbengio} corrected variance in differences \citep{scikit-learn}. Both sets of these pairwise tests are also shown in Table \ref{tab:pairwise}. 

Based on these tests, it is clear that extra trees significantly out performs all other models, and interestingly, that each model that follows it in decreasing order of performance is significantly better than the remaining models following it, confirming the findings in Figure \ref{fig:grs_average_performances}. In fact, in all cases of regression (), the order of model performances is extra trees, random forest, decision tree, and finally, linear regression. This result is expected, with decision trees being more accurate than linear regression (because the former can leverage non-linear relationships between input features and QPOs), as well as for random forest to outperform individual decision trees (because random forests are ensemble aggregations of decision tree forests). The similar yet superior performance of extra trees in comparison to random forest is notable but not striking \citep{mathew2022optimized}, yet this improvement should be considered with the additional size of an extra trees model compared to a trained random forest counterpart (this difference ranges from larger in terms of leaf count) \citep{extratrees}. Nevertheless, based on these findings it is clear that these classical machine learning models have been able to fairly accurately optimize for individual sources. However, although extra trees may perform best in these individual source scenarios, it remains yet to be seen whether these classical models will be generalizable for accurate cross-source analyses (as proposed earlier) or if other models like neural networks will be required \citep{NIPS2017_10ce03a1}. Although it may seem reasonable to combine data from these two sources and evaluate the predictive performance of these models in such a source-heterogeneous space, this would not be appropriate because any the resultant feature importances would not communicate whether or not the input engineered or raw spectral features are being leveraged for intuition into the physical state of the objects, or if their importances just reflect the models picking up on artifacts from the data generation procedure. In other words, this could be considered a form of data leakage, considering differing instrumental sensitivities, QPOs identification methods for each source, etc. \citep{Hannun2021MeasuringDL,2022arXiv220903345Y}. Hence, this provides additional motivation for follow-up, in which energy and timing spectra from a single instrument are reprocessed in an identical manner for multiple objects to prevent instrumental artifacts from contaminating the findings potentially recoverable from such a source-heterogeneous data-set. 

%\subsection{Frequency Bias}\label{sec:freq_bias}

\begin{figure*}
    \centering
    \includegraphics[width=0.7\textwidth]{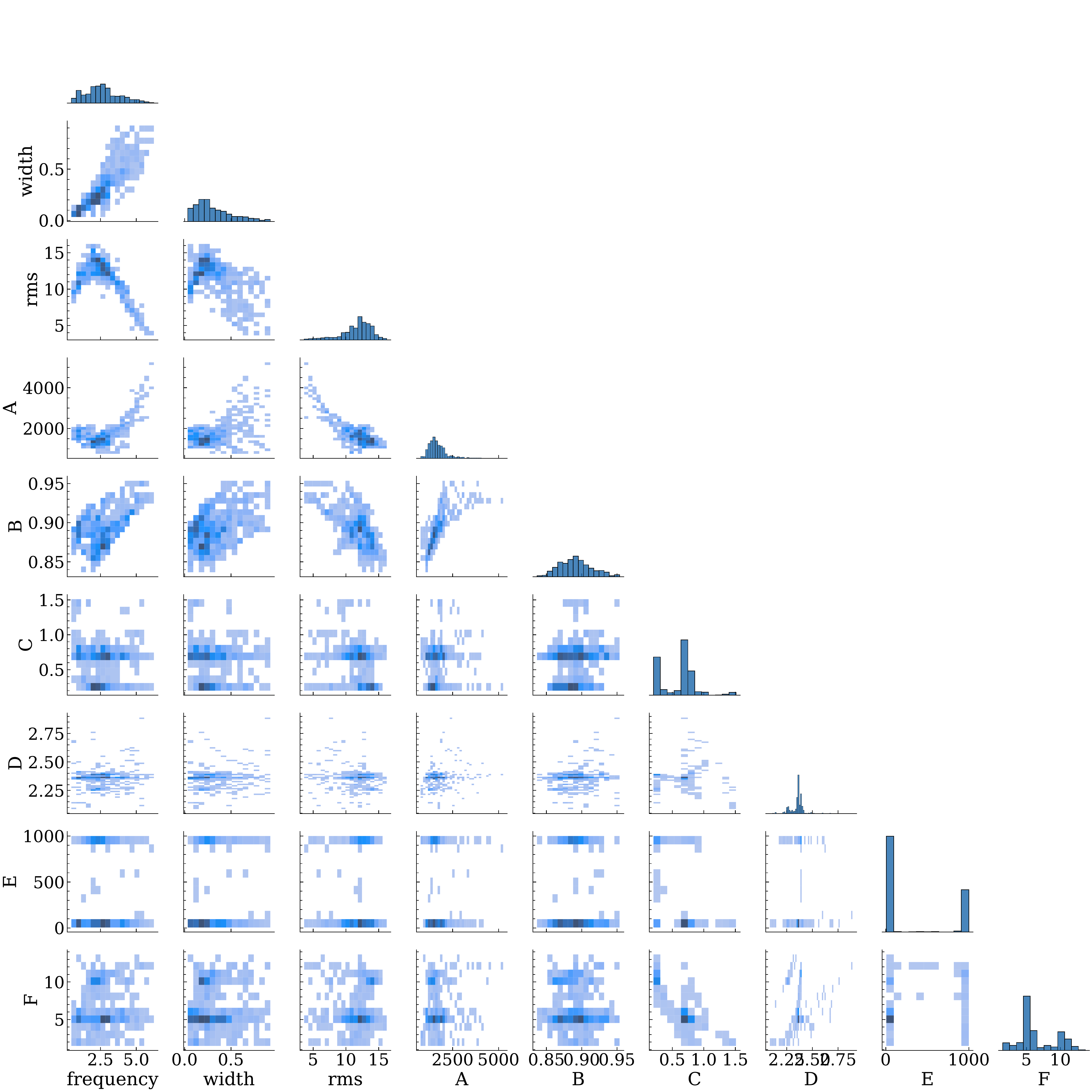}
    \caption{A pairplot displaying the pairwise relationships between engineered input and Lorentzian QPO output paramters for all GRS 1915+105 data. The letters in $A-F$ correspond to the net count rate, hardness ratio, asymptotic power-law photon index, \texttt{nthcomp} normalization, inner-disk temperature, and \texttt{diskbb} normalization features, respectively.}
    \label{fig:pairplot} 
\end{figure*}

\section{Conclusion} \label{sec:conclusion}

In this paper we have advanced novel approaches utilizing machine learning algorithms to link energy spectral properties (as both rebinned raw energy spectra and alternatively via engineered features derived from spectral fits) with the presence and properties of QPOs prominent in power-density spectra of two low-mass X-ray binary black hole systems. Specifically, we tested a selection of tree-based classical machine learning models using engineered features derived from energy spectra to predict QPO properties for fundamental QPOs in the black hole GRS 1915+105, and such derived features as well as raw rebinned energy spectra to characterize fundamental and harmonic QPOs in the black hole MAXI J1535-571. Additionally, we trained classification algorithms on the same data to predict the presence/absence of QPOs, as well as the multiclass QPO state of MAXI J1535-571 observations. We compared the performance of the machine learning models against each other, and found extra trees to perform best in all regression situations for both sources. Additionally, we compared every model against simplistic linear (regression) and logistic (classification) models as well, finding the machine learning models outperformed their linear counterpart in all regression cases, with linear regression notably struggling to correctly identify observations lacking QPOs. The main findings from this study are:

\begin{enumerate}
    \item All tested regression models yielded significantly better results on MAXI 1535-571 versus GRS 1915+105 data, despite the latter having 6x more data with QPOs and no issue with QPO absent observations. We attributed this to the multitude of unusual variability classes unique to GRS 1915+105 among \cite{MLStatesofGRS}.
    \item Kolmogorov-Smirnov tests on permutations of QPO parameter residuals showed that the best fitting regression model, Extra Trees, does not favor any particular QPO parameter and instead predicts for all with equal accuracy, including those for harmonics.
    \item Using rebinned raw spectral data as opposed to \texttt{XSPEC} derived features resulted in significantly worse performance for regression, binary classification, and multiclass classification on MAXI J1535-571 observations. 
    \item To enhance computational efficiency and ensure importance credibility, we calculated \texttt{TreeShap} feature importances immune to multicollinearity and found that for processed input features, extra trees determined the most significant features for GRS 1915+105 to be net count rate and hardness ratio, whereas the same model predicting for MAXI J1535-571 found \texttt{diskbb} normalization most important, which suggests a dependence on physical inner disk radius in this case. 
    \item We found almost all the rebinned channels which are the most important in determining the parameterizations of QPOs in regression are also those that are most important in determining their presence versus absence in classifying MAXI J1535-571 energy spectral data. Furthermore, for energy spectra, we found hard channels are the most important for both regression and classification, which aligns with the understanding of higher energy QPO manifestation above peak disk temperatures
    \item We have proposed future applications of these methods that range from extending the input feature space they are tested on (e.g. scattering fraction and inclination) to moving from single source to source/source-type heterogeneous samples to achieve our original goal of inter-object generalizations since in this paper we have introduced and laid the foundation for these methods on individual objects. 
\end{enumerate}

Finally, we based our work on our \texttt{QPOML} Python library, from input and output matrix construction and preprocessing, to hyperparameter tuning, model evaluation, and plot generation, which were all conveniently streamlined for application and both (i) executed as ``under-the-hood'' as possible while remaining user accessible; and (ii) easily extendable to any number of QPOs and any number of scalar observation features for any number of observations from any number of sources. This library is available on \href{https://github.com/thissop/QPOML}{GitHub}.

\section{Acknowledgements}

M.M. acknowledges support from the research program Athena with project number 184.034.002, which is (partly) financed by the Dutch Research Council (NUD). We also thank Virginia A. Cuneo for a helpful conversation early in this work, Michael Corcoran and Craig Gordon for assistance with some early technical issues. Finally, we thank Travis Austen with help recovering a significant amount of our work from a damaged virtual machine disk, and Brandon Barrios for Windows Subsystem for Linux advice. This work was made possible by the \textit{NICER} and \textit{RXTE} missions, as well as data from the High Energy Astrophysics Science Archive Research Center (HEARSARC) and NASA's Astrophysics Data System Bibliographic Services. This work has been advised by AstroAI. 

\section{Data Availability}
The data used for MAXI J1535-571 are available at the \textit{NICER} archive (https://heasarc.gsfc.nasa.gov/docs/nicer/nicer\_archive.html), and those for
GRS 1915+105 belong to their corresponding authors and are available at the following references \cite{GRSDATAPAPER} and \cite{zhangGRS2022}. The software used for energy spectral data analysis can be accessed from the HEASARC website
(https://heasarc.gsfc.nasa.gov/lheasoft/download.html). The \texttt{QPOML} code repository can be accesed via \href{https://github.com/thissop/QPOML}{GitHub}\\

\noindent \textit{Facilities:} \textit{NICER}, \textit{RXTE}

\noindent \textit{Software:} \textit{Software:} AstroPy \citep{astropy1,astropy2}, Keras \citep{keras}, Matplotlib \citep{Hunter2007}, NumPy \citep{numpy}, Pandas \citep{pandas}, SciencePlots \citep{SciencePlots}, SciPy \citep{Virtanen2020}, scikit-learn \citep{scikit-learn}, and seaborn \citep{Waskom2021}.

\section*{Appendix}
\setcounter{table}{0}
\renewcommand{\thetable}{A\arabic{table}}
\begin{table*}
    \caption{Pairwise fold corrected frequentist and Bayesian statistics for all regression model comparisons discussed in Section \ref{sec:discussion}, where GRS 1915+105 models are only tested on extracted (Processed) features, whereas MAXI J1535-571 models are tested on both Processed, as well as rebinned raw energy spectra features (Spectral). Abbreviations-wise, ET (Extra Trees), RF (Random Forest), and DT (Decision Tree). The $t$ values represent the fold-corrected Student's $t$ values of the differences of the average residual values for each model. These are accompanied by their corresponding $p$ values.}
    \label{tab:pairwise}
    \centering

    \begin{tabular}{lllrlrr}
\toprule
Source (Input Type) & First Model Name & Second Model Name &    t &            p &  \% Chance & \% Chance \\ & & & & & First Better & Second Better \\
\midrule
MAXI J1535-571 (Spectral) & & & & & & \\
&  ET & RF & 0.67 & 0.25 &                  74.74 &                   25.26 \\
&  ET & DT & 0.83 & 0.21 &                  79.60 &                   20.40 \\
&  ET & Linear & 5.73 & 0.00 &                 100.00 &                    0.00 \\
& RF & DT & 0.18 & 0.43 &                  57.12 &                   42.88 \\
& RF & Linear & 5.17 & 0.00 &                 100.00 &                    0.00 \\
& DT & Linear & 5.66 & 0.00 &                 100.00 &                    0.00 \\
MAXI J1535-571 (Processed) & & & & & & \\
&  ET & DT &  0.40 & 3.47e-01 &                  65.45 &                   34.55 \\
&  ET & RF &  0.60 & 2.76e-01 &                  72.59 &                   27.41 \\
&  ET & Linear & 11.21 & 9.35e-12 &                 100.00 &                    0.00 \\
& DT & RF &  0.15 & 4.40e-01 &                  56.00 &                   44.00 \\
& DT & Linear &  8.74 & 1.62e-09 &                 100.00 &                    0.00 \\
& RF & Linear &  9.73 & 1.86e-10 &                 100.00 &                    0.00 \\
GRS 1915+105 (Processed) & & & & & & \\
&  ET & RF &  1.25 & 1.07e-01 &                  89.37 &                   10.63 \\
&  ET & DT &  4.24 & 4.33e-05 &                 100.00 &                    0.00 \\
&  ET & Linear & 11.20 & 4.19e-16 &                 100.00 &                    0.00 \\
& RF & DT &  3.61 & 3.29e-04 &                  99.98 &                    0.02 \\
& RF & Linear &  9.04 & 9.27e-13 &                 100.00 &                    0.00 \\
& DT & Linear &  4.51 & 1.75e-05 &                 100.00 &                    0.00 \\
\bottomrule
\end{tabular}
\end{table*}


\begin{thebibliography}{}
\makeatletter
\relax
\def\mn@urlcharsother{\let\do\@makeother \do\$\do\&\do\#\do\^\do\_\do\%\do\~}
\def\mn@doi{\begingroup\mn@urlcharsother \@ifnextchar [ {\mn@doi@}
  {\mn@doi@[]}}
\def\mn@doi@[#1]#2{\def\@tempa{#1}\ifx\@tempa\@empty \href
  {http://dx.doi.org/#2} {doi:#2}\else \href {http://dx.doi.org/#2} {#1}\fi
  \endgroup}
\def\mn@eprint#1#2{\mn@eprint@#1:#2::\@nil}
\def\mn@eprint@arXiv#1{\href {http://arxiv.org/abs/#1} {{\tt arXiv:#1}}}
\def\mn@eprint@dblp#1{\href {http://dblp.uni-trier.de/rec/bibtex/#1.xml}
  {dblp:#1}}
\def\mn@eprint@#1:#2:#3:#4\@nil{\def\@tempa {#1}\def\@tempb {#2}\def\@tempc
  {#3}\ifx \@tempc \@empty \let \@tempc \@tempb \let \@tempb \@tempa \fi \ifx
  \@tempb \@empty \def\@tempb {arXiv}\fi \@ifundefined
  {mn@eprint@\@tempb}{\@tempb:\@tempc}{\expandafter \expandafter \csname
  mn@eprint@\@tempb\endcsname \expandafter{\@tempc}}}

\bibitem[\protect\citeauthoryear{Akaike}{Akaike}{1998}]{Akaike1998}
Akaike H.,  1998, Information Theory and an Extension of the Maximum Likelihood
  Principle.
Springer New York, New York, NY, pp 199--213,
  \mn@doi{10.1007/978-1-4612-1694-015}

\bibitem[\protect\citeauthoryear{Akanbi, Amiri  \& Fazeldehkordi}{Akanbi
  et~al.}{2015}]{Akanbi2015}
Akanbi O.~A.,  Amiri I.~S.,   Fazeldehkordi E.,  2015, in , A Machine-Learning
  Approach to Phishing Detection and Defense.
Elsevier, pp 45--54, \mn@doi{10.1016/b978-0-12-802927-5.00004-6}

\bibitem[\protect\citeauthoryear{Ampomah, Qin  \& Nyame}{Ampomah
  et~al.}{2020}]{evalTree-BasedEnsembles}
Ampomah E.~K.,  Qin Z.,   Nyame G.,  2020, \mn@doi [Information]
  {10.3390/info11060332}, 11

\bibitem[\protect\citeauthoryear{{Arnason}, {Barmby}  \& {Vulic}}{{Arnason}
  et~al.}{2020}]{arnason2020}
{Arnason} R.~M.,  {Barmby} P.,   {Vulic} N.,  2020, \mn@doi [\mnras]
  {10.1093/mnras/staa207}, \href
  {https://ui.adsabs.harvard.edu/abs/2020MNRAS.492.5075A} {492, 5075}

\bibitem[\protect\citeauthoryear{{Arnaud}, {Dorman}  \& {Gordon}}{{Arnaud}
  et~al.}{1999}]{XSPEC1999}
{Arnaud} K.,  {Dorman} B.,   {Gordon} C.,  1999, {XSPEC: An X-ray spectral
  fitting package}, Astrophysics Source Code Library, record ascl:9910.005
  (\mn@eprint {ascl} {9910.005})

\bibitem[\protect\citeauthoryear{{Astropy Collaboration} et~al.,}{{Astropy
  Collaboration} et~al.}{2013}]{astropy1}
{Astropy Collaboration} et~al., 2013, \mn@doi [\aap]
  {10.1051/0004-6361/201322068}, \href
  {https://ui.adsabs.harvard.edu/abs/2013A&A...558A..33A} {558, A33}

\bibitem[\protect\citeauthoryear{{Astropy Collaboration} et~al.,}{{Astropy
  Collaboration} et~al.}{2018}]{astropy2}
{Astropy Collaboration} et~al., 2018, \mn@doi [\aj] {10.3847/1538-3881/aabc4f},
  \href {https://ui.adsabs.harvard.edu/abs/2018AJ....156..123A} {156, 123}

\bibitem[\protect\citeauthoryear{{Belloni}, {Klein-Wolt}, {M{\'e}ndez}, {van
  der Klis}  \& {van Paradijs}}{{Belloni} et~al.}{2000}]{belloni2000}
{Belloni} T.,  {Klein-Wolt} M.,  {M{\'e}ndez} M.,  {van der Klis} M.,   {van
  Paradijs} J.,  2000, \aap, \href
  {http://adsabs.harvard.edu/abs/2000A%26A...355..271B} {355, 271}

\bibitem[\protect\citeauthoryear{{Belloni}, {Psaltis}  \& {van der
  Klis}}{{Belloni} et~al.}{2002}]{Belloni2002}
{Belloni} T.,  {Psaltis} D.,   {van der Klis} M.,  2002, \mn@doi [\apj]
  {10.1086/340290}, \href
  {https://ui.adsabs.harvard.edu/abs/2002ApJ...572..392B} {572, 392}

\bibitem[\protect\citeauthoryear{{Belloni}, {M{\'e}ndez}  \& {Homan}}{{Belloni}
  et~al.}{2005}]{belloniMendez2005}
{Belloni} T.,  {M{\'e}ndez} M.,   {Homan} J.,  2005, \mn@doi [\aap]
  {10.1051/0004-6361:20041377}, \href
  {https://ui.adsabs.harvard.edu/abs/2005A&A...437..209B} {437, 209}

\bibitem[\protect\citeauthoryear{{Belloni}, {Zhang}, {Kylafis}, {Reig}  \&
  {Altamirano}}{{Belloni} et~al.}{2020}]{belloni2020typeB}
{Belloni} T.~M.,  {Zhang} L.,  {Kylafis} N.~D.,  {Reig} P.,   {Altamirano} D.,
  2020, \mn@doi [\mnras] {10.1093/mnras/staa1843}, \href
  {https://ui.adsabs.harvard.edu/abs/2020MNRAS.496.4366B} {496, 4366}

\bibitem[\protect\citeauthoryear{{Benavoli}, {Corani}, {Demsar}  \&
  {Zaffalon}}{{Benavoli} et~al.}{2016}]{benavoli}
{Benavoli} A.,  {Corani} G.,  {Demsar} J.,   {Zaffalon} M.,  2016, arXiv
  e-prints, \href {https://ui.adsabs.harvard.edu/abs/2016arXiv160604316B} {p.
  arXiv:1606.04316}

\bibitem[\protect\citeauthoryear{Berkson}{Berkson}{1944}]{original-logistic}
Berkson J.,  1944, Journal of the American Statistical Association, 39, 357

\bibitem[\protect\citeauthoryear{{Bhargava}, {Belloni}, {Bhattacharya}  \&
  {Misra}}{{Bhargava} et~al.}{2019}]{Bhargava2019}
{Bhargava} Y.,  {Belloni} T.,  {Bhattacharya} D.,   {Misra} R.,  2019, \mn@doi
  [\mnras] {10.1093/mnras/stz1774}, \href
  {https://ui.adsabs.harvard.edu/abs/2019MNRAS.488..720B} {488, 720}

\bibitem[\protect\citeauthoryear{{Bildsten}}{{Bildsten}}{1998}]{thermonuclear}
{Bildsten} L.,  1998, in {Buccheri} R.,  {van Paradijs} J.,   {Alpar} A.,  eds,
   NATO Advanced Study Institute (ASI) Series C Vol. 515, The Many Faces of
  Neutron Stars.. p.~419 (\mn@eprint {arXiv} {astro-ph/9709094})

\bibitem[\protect\citeauthoryear{Boinee, Angelis  \& Foresti}{Boinee
  et~al.}{2008}]{metarandomforests}
Boinee P.,  Angelis A.~D.,   Foresti G.~L.,  2008, International Journal of
  Computer and Information Engineering, 2, 2246

\bibitem[\protect\citeauthoryear{Bouveyron, Celeux, Murphy  \&
  Raftery}{Bouveyron et~al.}{2019}]{bouveyron2019model}
Bouveyron C.,  Celeux G.,  Murphy T.,   Raftery A.,  2019, Model-Based
  Clustering and Classification for Data Science: With Applications in R.
Cambridge Series in Statistical and Probabilistic Mathematics, Cambridge
  University Press

\bibitem[\protect\citeauthoryear{Brefeld, Davis, Van~Haaren  \&
  Zimmermann}{Brefeld et~al.}{2020}]{brefeld2020machine}
Brefeld U.,  Davis J.,  Van~Haaren J.,   Zimmermann A.,  2020, Machine Learning
  and Data Mining for Sports Analytics: 7th International Workshop, MLSA 2020,
  Co-located with ECML/PKDD 2020, Ghent, Belgium, September 14--18, 2020,
  Proceedings.
Communications in Computer and Information Science, Springer International
  Publishing

\bibitem[\protect\citeauthoryear{Breiman}{Breiman}{1984}]{breiman1984original}
Breiman L.,  1984, Classification and Regression Trees.
(The Wadsworth statistics / probability series), Wadsworth International Group

\bibitem[\protect\citeauthoryear{Breiman}{Breiman}{1996}]{breiman1996bagging}
Breiman L.,  1996, Machine learning, 24, 123

\bibitem[\protect\citeauthoryear{Breiman}{Breiman}{2001}]{breiman2001random}
Breiman L.,  2001, Machine learning, 45, 5

\bibitem[\protect\citeauthoryear{Bruce \& Bruce}{Bruce \&
  Bruce}{2017}]{bruce2017practical}
Bruce P.,  Bruce A.,  2017, Practical Statistics for Data Scientists: 50
  Essential Concepts.
O'Reilly Media

\bibitem[\protect\citeauthoryear{Casari \& Zheng}{Casari \&
  Zheng}{2018}]{casari2018feature}
Casari A.,  Zheng A.,  2018, O'Reilly Media, Inc., p.~218

\bibitem[\protect\citeauthoryear{{Casella}, {Belloni}  \& {Stella}}{{Casella}
  et~al.}{2005}]{casellaABC}
{Casella} P.,  {Belloni} T.,   {Stella} L.,  2005, \mn@doi [\apj]
  {10.1086/431174}, \href
  {https://ui.adsabs.harvard.edu/abs/2005ApJ...629..403C} {629, 403}

\bibitem[\protect\citeauthoryear{{Castro Segura} et~al.,}{{Castro Segura}
  et~al.}{2022}]{neutronStarWind}
{Castro Segura} N.,  et~al., 2022, \mn@doi [\nat] {10.1038/s41586-021-04324-2},
  \href {https://ui.adsabs.harvard.edu/abs/2022Natur.603...52C} {603, 52}

\bibitem[\protect\citeauthoryear{{Castro-Tirado}, {Brandt}  \&
  {Lund}}{{Castro-Tirado} et~al.}{1992}]{grs-discovery}
{Castro-Tirado} A.~J.,  {Brandt} S.,   {Lund} N.,  1992, \iaucirc, \href
  {https://ui.adsabs.harvard.edu/abs/1992IAUC.5590....2C} {5590, 2}

\bibitem[\protect\citeauthoryear{Chollet}{Chollet}{2017}]{deepLearningPython}
Chollet F.,  2017, Deep Learning with Python.
Manning

\bibitem[\protect\citeauthoryear{Chollet et~al.}{Chollet et~al.}{2015}]{keras}
Chollet F.,  et~al., 2015, Keras, \url{https://keras.io}

\bibitem[\protect\citeauthoryear{{Chowdhury}, {Lin}, {Liaw}  \&
  {Kerby}}{{Chowdhury} et~al.}{2021}]{2021arXiv211102513C}
{Chowdhury} S.,  {Lin} Y.,  {Liaw} B.,   {Kerby} L.,  2021, arXiv e-prints,
  \href {https://ui.adsabs.harvard.edu/abs/2021arXiv211102513C} {p.
  arXiv:2111.02513}

\bibitem[\protect\citeauthoryear{{Corral-Santana, J. M.}, {Casares, J.},
  {Mu\~noz-Darias, T.}, {Bauer, F. E.}, {Mart\'{\i}nez-Pais, I. G.}  \&
  {Russell, D. M.}}{{Corral-Santana, J. M.} et~al.}{2016}]{blackcat}
{Corral-Santana, J. M.} {Casares, J.} {Mu\~noz-Darias, T.} {Bauer, F. E.}
  {Mart\'{\i}nez-Pais, I. G.}  {Russell, D. M.} 2016, \mn@doi [A\&A]
  {10.1051/0004-6361/201527130}, 587, A61

\bibitem[\protect\citeauthoryear{{C{\'u}neo} et~al.,}{{C{\'u}neo}
  et~al.}{2020}]{cuneo2020}
{C{\'u}neo} V.~A.,  et~al., 2020, \mn@doi [\mnras] {10.1093/mnras/staa1606},
  \href {https://ui.adsabs.harvard.edu/abs/2020MNRAS.496.1001C} {496, 1001}

\bibitem[\protect\citeauthoryear{Dangeti}{Dangeti}{2017}]{dangeti2017statisticsML}
Dangeti P.,  2017, Statistics for Machine Learning.
Packt Publishing

\bibitem[\protect\citeauthoryear{Diebold \& Mariano}{Diebold \&
  Mariano}{1995}]{dieboldandmariano}
Diebold F.~X.,  Mariano R.~S.,  1995, Journal of Business \& Economic
  Statistics, 13, 253

\bibitem[\protect\citeauthoryear{Dieleman, Willett  \& Dambre}{Dieleman
  et~al.}{2015}]{classifyGalaxies}
Dieleman S.,  Willett K.~W.,   Dambre J.,  2015, \mn@doi [Monthly Notices of
  the Royal Astronomical Society] {10.1093/mnras/stv632}, 450, 1441

\bibitem[\protect\citeauthoryear{{Done} \& {Gierli{\'n}ski}}{{Done} \&
  {Gierli{\'n}ski}}{2004}]{terrabytedone}
{Done} C.,  {Gierli{\'n}ski} M.,  2004, \mn@doi [Progress of Theoretical
  Physics Supplement] {10.1143/PTPS.155.9}, \href
  {https://ui.adsabs.harvard.edu/abs/2004PThPS.155....9D} {155, 9}

\bibitem[\protect\citeauthoryear{{Dong}, {Liu}, {Tuo}, {Steiner}, {Ge},
  {Garc{\'\i}a}  \& {Cao}}{{Dong} et~al.}{2022}]{Dong2022MAXIkTe}
{Dong} Y.,  {Liu} Z.,  {Tuo} Y.,  {Steiner} J.~F.,  {Ge} M.,  {Garc{\'\i}a}
  J.~A.,   {Cao} X.,  2022, \mn@doi [\mnras] {10.1093/mnras/stac1466}, \href
  {https://ui.adsabs.harvard.edu/abs/2022MNRAS.514.1422D} {514, 1422}

\bibitem[\protect\citeauthoryear{{Fabian}, {Rees}, {Stella}  \&
  {White}}{{Fabian} et~al.}{1989}]{ironlines1989}
{Fabian} A.~C.,  {Rees} M.~J.,  {Stella} L.,   {White} N.~E.,  1989, \mn@doi
  [\mnras] {10.1093/mnras/238.3.729}, \href
  {https://ui.adsabs.harvard.edu/abs/1989MNRAS.238..729F} {238, 729}

\bibitem[\protect\citeauthoryear{{Fisher}, {Rudin}  \& {Dominici}}{{Fisher}
  et~al.}{2018}]{generalPermutations}
{Fisher} A.,  {Rudin} C.,   {Dominici} F.,  2018, arXiv e-prints, \href
  {https://ui.adsabs.harvard.edu/abs/2018arXiv180101489F} {p. arXiv:1801.01489}

\bibitem[\protect\citeauthoryear{Floares, Ferisgan, Onita, Ciuparu, Calin  \&
  Manolache}{Floares et~al.}{2017}]{floares2017smallest}
Floares A.,  Ferisgan M.,  Onita D.,  Ciuparu A.,  Calin G.,   Manolache F.,
  2017, Int J Oncol Cancer Ther, 2, 13

\bibitem[\protect\citeauthoryear{{Fragile}, {Straub}  \& {Blaes}}{{Fragile}
  et~al.}{2016}]{typeCmodel2016}
{Fragile} P.~C.,  {Straub} O.,   {Blaes} O.,  2016, \mn@doi [\mnras]
  {10.1093/mnras/stw1428}, \href
  {https://ui.adsabs.harvard.edu/abs/2016MNRAS.461.1356F} {461, 1356}

\bibitem[\protect\citeauthoryear{Fudenberg \& Liang}{Fudenberg \&
  Liang}{2020}]{mlandtheory}
Fudenberg D.,  Liang A.,  2020, SIGecom Exch., 18, 4–11

\bibitem[\protect\citeauthoryear{{Galeev}, {Rosner}  \& {Vaiana}}{{Galeev}
  et~al.}{1979}]{corona1979}
{Galeev} A.~A.,  {Rosner} R.,   {Vaiana} G.~S.,  1979, \mn@doi [\apj]
  {10.1086/156957}, \href
  {https://ui.adsabs.harvard.edu/abs/1979ApJ...229..318G} {229, 318}

\bibitem[\protect\citeauthoryear{{Gallo}, {Fender}  \& {Kaiser}}{{Gallo}
  et~al.}{2005}]{gallo2005}
{Gallo} E.,  {Fender} R.,   {Kaiser} C.,  2005, in {Burderi} L.,  {Antonelli}
  L.~A.,  {D'Antona} F.,  {di Salvo} T.,  {Israel} G.~L.,  {Piersanti} L.,
  {Tornamb{\`e}} A.,   {Straniero} O.,  eds,  American Institute of Physics
  Conference Series Vol. 797, Interacting Binaries: Accretion, Evolution, and
  Outcomes. pp 189--196 (\mn@eprint {arXiv} {astro-ph/0501374}),
  \mn@doi{10.1063/1.2130232}

\bibitem[\protect\citeauthoryear{{Gao} et~al.,}{{Gao}
  et~al.}{2017}]{globaltypeB}
{Gao} H.~Q.,  et~al., 2017, \mn@doi [\mnras] {10.1093/mnras/stw3146}, \href
  {https://ui.adsabs.harvard.edu/abs/2017MNRAS.466..564G} {466, 564}

\bibitem[\protect\citeauthoryear{{Garc{\'\i}a}, {M{\'e}ndez}, {Karpouzas},
  {Belloni}, {Zhang}  \& {Altamirano}}{{Garc{\'\i}a}
  et~al.}{2021}]{garciaMendez2021}
{Garc{\'\i}a} F.,  {M{\'e}ndez} M.,  {Karpouzas} K.,  {Belloni} T.,  {Zhang}
  L.,   {Altamirano} D.,  2021, \mn@doi [\mnras] {10.1093/mnras/staa3944},
  \href {https://ui.adsabs.harvard.edu/abs/2021MNRAS.501.3173G} {501, 3173}

\bibitem[\protect\citeauthoryear{{Garc{\'\i}a}, {Karpouzas}, {M{\'e}ndez},
  {Zhang}, {Zhang}, {Belloni}  \& {Altamirano}}{{Garc{\'\i}a}
  et~al.}{2022a}]{garciaGRS2022MNRAS}
{Garc{\'\i}a} F.,  {Karpouzas} K.,  {M{\'e}ndez} M.,  {Zhang} L.,  {Zhang} Y.,
  {Belloni} T.,   {Altamirano} D.,  2022a, \mn@doi [\mnras]
  {10.1093/mnras/stac1202}, \href
  {https://ui.adsabs.harvard.edu/abs/2022MNRAS.513.4196G} {513, 4196}

\bibitem[\protect\citeauthoryear{{Garc{\'\i}a}, {Karpouzas}, {M{\'e}ndez},
  {Zhang}, {Zhang}, {Belloni}  \& {Altamirano}}{{Garc{\'\i}a}
  et~al.}{2022b}]{evolvingPropertiesGRSCorona}
{Garc{\'\i}a} F.,  {Karpouzas} K.,  {M{\'e}ndez} M.,  {Zhang} L.,  {Zhang} Y.,
  {Belloni} T.,   {Altamirano} D.,  2022b, \mn@doi [\mnras]
  {10.1093/mnras/stac1202}, \href
  {https://ui.adsabs.harvard.edu/abs/2022MNRAS.513.4196G} {513, 4196}

\bibitem[\protect\citeauthoryear{{Gardenier} \& {Uttley}}{{Gardenier} \&
  {Uttley}}{2018}]{gardenier2018}
{Gardenier} D.~W.,  {Uttley} P.,  2018, \mn@doi [\mnras]
  {10.1093/mnras/sty2524}, \href
  {https://ui.adsabs.harvard.edu/abs/2018MNRAS.481.3761G} {481, 3761}

\bibitem[\protect\citeauthoryear{{Garg}, {Misra}  \& {Sen}}{{Garg}
  et~al.}{2022}]{gargEnergyDependence}
{Garg} A.,  {Misra} R.,   {Sen} S.,  2022, \mn@doi [\mnras]
  {10.1093/mnras/stac1490}, \href
  {https://ui.adsabs.harvard.edu/abs/2022MNRAS.514.3285G} {514, 3285}

\bibitem[\protect\citeauthoryear{Garrett}{Garrett}{2021}]{SciencePlots}
Garrett J.~D.,  2021, \mn@doi [] {10.5281/zenodo.4106649}

\bibitem[\protect\citeauthoryear{{Gendreau}, {Arzoumanian}  \&
  {Okajima}}{{Gendreau} et~al.}{2012}]{NICER}
{Gendreau} K.~C.,  {Arzoumanian} Z.,   {Okajima} T.,  2012, in Space Telescopes
  and Instrumentation 2012: Ultraviolet to Gamma Ray. p. 844313,
  \mn@doi{10.1117/12.926396}

\bibitem[\protect\citeauthoryear{Geurts, Ernst  \& Wehenkel}{Geurts
  et~al.}{2006}]{extratrees}
Geurts P.,  Ernst D.,   Wehenkel L.,  2006, \mn@doi [Mach. Learn.]
  {10.1007/s10994-006-6226-1}, 63, 3–42

\bibitem[\protect\citeauthoryear{Giacconi, Gursky, Paolini  \& Rossi}{Giacconi
  et~al.}{1962}]{scox-1}
Giacconi R.,  Gursky H.,  Paolini F.~R.,   Rossi B.~B.,  1962, \mn@doi [Phys.
  Rev. Lett.] {10.1103/PhysRevLett.9.439}, 9, 439

\bibitem[\protect\citeauthoryear{Gilmore}{Gilmore}{2004}]{gilmore2004}
Gilmore G.,  2004, \mn@doi [Science] {10.1126/science.1100370}, 304, 1915

\bibitem[\protect\citeauthoryear{Goodfellow, Bengio  \& Courville}{Goodfellow
  et~al.}{2016}]{goodfellow2016deep}
Goodfellow I.,  Bengio Y.,   Courville A.,  2016, Deep Learning.
Adaptive Computation and Machine Learning series, MIT Press, \url
  {https://books.google.com/books?id=omivDQAAQBAJ}

\bibitem[\protect\citeauthoryear{{Greiner}}{{Greiner}}{2003}]{Greiner2003}
{Greiner} J.,  2003, in {van den Heuvel} E.~P.,  {Kaper} L.,  {Rol} E.,
  {Wijers} R. A.~M.~J.,  eds,  Astronomical Society of the Pacific Conference
  Series Vol. 308, From X-ray Binaries to Gamma-Ray Bursts: Jan van Paradijs
  Memorial Symposium. p.~111

\bibitem[\protect\citeauthoryear{{Greiner}, {Cuby}, {McCaughrean},
  {Castro-Tirado}  \& {Mennickent}}{{Greiner} et~al.}{2001}]{GRSCompanion}
{Greiner} J.,  {Cuby} J.~G.,  {McCaughrean} M.~J.,  {Castro-Tirado} A.~J.,
  {Mennickent} R.~E.,  2001, \mn@doi [\aap] {10.1051/0004-6361:20010771}, \href
  {https://ui.adsabs.harvard.edu/abs/2001A&A...373L..37G} {373, L37}

\bibitem[\protect\citeauthoryear{Han, Kamber  \& Pei}{Han
  et~al.}{2012}]{Han2012}
Han J.,  Kamber M.,   Pei J.,  2012, in , Data Mining.
Elsevier, pp 83--124, \mn@doi{10.1016/b978-0-12-381479-1.00003-4}

\bibitem[\protect\citeauthoryear{{Hannikainen} et~al.,}{{Hannikainen}
  et~al.}{2005}]{Hannikainen2005}
{Hannikainen} D.~C.,  et~al., 2005, \mn@doi [\aap]
  {10.1051/0004-6361:20042250}, \href
  {https://ui.adsabs.harvard.edu/abs/2005A&A...435..995H} {435, 995}

\bibitem[\protect\citeauthoryear{Hannun, Guo  \& van~der Maaten}{Hannun
  et~al.}{2021}]{Hannun2021MeasuringDL}
Hannun A.~Y.,  Guo C.,   van~der Maaten L.,  2021, in Conference on Uncertainty
  in Artificial Intelligence.

\bibitem[\protect\citeauthoryear{Harris et~al.,}{Harris et~al.}{2020}]{numpy}
Harris C.~R.,  et~al., 2020, \mn@doi [Nature] {10.1038/s41586-020-2649-2}, 585,
  357

\bibitem[\protect\citeauthoryear{{Homan} \& {Belloni}}{{Homan} \&
  {Belloni}}{2005}]{homanBelloniQPOstates}
{Homan} J.,  {Belloni} T.,  2005, \mn@doi [\apss] {10.1007/s10509-005-1197-4},
  \href {https://ui.adsabs.harvard.edu/abs/2005Ap&SS.300..107H} {300, 107}

\bibitem[\protect\citeauthoryear{{Hooker}, {Mentch}  \& {Zhou}}{{Hooker}
  et~al.}{2019}]{hooker2019}
{Hooker} G.,  {Mentch} L.,   {Zhou} S.,  2019, arXiv e-prints, \href
  {https://ui.adsabs.harvard.edu/abs/2019arXiv190503151H} {p. arXiv:1905.03151}

\bibitem[\protect\citeauthoryear{Hunter}{Hunter}{2007}]{Hunter2007}
Hunter J.~D.,  2007, \mn@doi [Computing in Science \& Engineering]
  {10.1109/MCSE.2007.55}, 9, 90

\bibitem[\protect\citeauthoryear{{Huppenkothen}, {Heil}, {Hogg}  \&
  {Mueller}}{{Huppenkothen} et~al.}{2017a}]{grsML2017}
{Huppenkothen} D.,  {Heil} L.~M.,  {Hogg} D.~W.,   {Mueller} A.,  2017a,
  \mn@doi [\mnras] {10.1093/mnras/stw3190}, \href
  {https://ui.adsabs.harvard.edu/abs/2017MNRAS.466.2364H} {466, 2364}

\bibitem[\protect\citeauthoryear{{Huppenkothen}, {Heil}, {Hogg}  \&
  {Mueller}}{{Huppenkothen} et~al.}{2017b}]{MLStatesofGRS}
{Huppenkothen} D.,  {Heil} L.~M.,  {Hogg} D.~W.,   {Mueller} A.,  2017b,
  \mn@doi [\mnras] {10.1093/mnras/stw3190}, \href
  {https://ui.adsabs.harvard.edu/abs/2017MNRAS.466.2364H} {466, 2364}

\bibitem[\protect\citeauthoryear{Ingram \& Motta}{Ingram \&
  Motta}{2019}]{ingram2019}
Ingram A.,  Motta S.~E.,  2019, New Astronomy Reviews

\bibitem[\protect\citeauthoryear{{Ingram}, {Done}  \& {Fragile}}{{Ingram}
  et~al.}{2009}]{precessionandlense}
{Ingram} A.,  {Done} C.,   {Fragile} P.~C.,  2009, \mn@doi [\mnras]
  {10.1111/j.1745-3933.2009.00693.x}, \href
  {https://ui.adsabs.harvard.edu/abs/2009MNRAS.397L.101I} {397, L101}

\bibitem[\protect\citeauthoryear{{Ivezi{\'c}}, {Connolly}, {VanderPlas}  \&
  {Gray}}{{Ivezi{\'c}} et~al.}{2014}]{2014MLinAstro}
{Ivezi{\'c}} {\v{Z}}.,  {Connolly} A.~J.,  {VanderPlas} J.~T.,   {Gray} A.,
  2014, {Statistics, Data Mining, and Machine Learning in Astronomy: A
  Practical Python Guide for the Analysis of Survey Data},
  \mn@doi{10.1515/9781400848911.
}

\bibitem[\protect\citeauthoryear{{Jonker}, {van der Klis}  \&
  {Wijnands}}{{Jonker} et~al.}{1999}]{oneHZqpo}
{Jonker} P.~G.,  {van der Klis} M.,   {Wijnands} R.,  1999, \mn@doi [\apjl]
  {10.1086/311840}, \href
  {https://ui.adsabs.harvard.edu/abs/1999ApJ...511L..41J} {511, L41}

\bibitem[\protect\citeauthoryear{KS-}{KS-}{2008}]{KS-test-review}
 2008, Kolmogorov--Smirnov Test.
Springer New York, New York, NY, pp 283--287,
  \mn@doi{10.1007/978-0-387-32833-1_214}

\bibitem[\protect\citeauthoryear{Kalinin \& Foster}{Kalinin \&
  Foster}{2020}]{kalinin2020handbook}
Kalinin S.,  Foster I.,  2020, Handbook On Big Data And Machine Learning In The
  Physical Sciences (In 2 Volumes).
World Scientific Series On Emerging Technologies, World Scientific Publishing
  Company

\bibitem[\protect\citeauthoryear{Kandanaarachchi, Mu{\~{n}}oz, Hyndman  \&
  Smith-Miles}{Kandanaarachchi et~al.}{2019}]{Kandanaarachchi2019}
Kandanaarachchi S.,  Mu{\~{n}}oz M.~A.,  Hyndman R.~J.,   Smith-Miles K.,
  2019, \mn@doi [Data Mining and Knowledge Discovery]
  {10.1007/s10618-019-00661-z}, 34, 309

\bibitem[\protect\citeauthoryear{{Kato}}{{Kato}}{2005}]{hectohertzKato2005}
{Kato} S.,  2005, \mn@doi [\pasj] {10.1093/pasj/57.3.L17}, \href
  {https://ui.adsabs.harvard.edu/abs/2005PASJ...57L..17K} {57, L17}

\bibitem[\protect\citeauthoryear{{Kato} \& {Fukue}}{{Kato} \&
  {Fukue}}{1980}]{corrugation}
{Kato} S.,  {Fukue} J.,  1980, \pasj, \href
  {https://ui.adsabs.harvard.edu/abs/1980PASJ...32..377K} {32, 377}

\bibitem[\protect\citeauthoryear{Kline}{Kline}{1998}]{kline1998principles}
Kline R.,  1998, Principles and Practice of Structural Equation Modeling.
Methodology in the Social Sciences, Guilford Publications

\bibitem[\protect\citeauthoryear{{Kojima} et~al.,}{{Kojima}
  et~al.}{2020}]{discoverGalaxies}
{Kojima} T.,  et~al., 2020, \mn@doi [\apj] {10.3847/1538-4357/aba047}, \href
  {https://ui.adsabs.harvard.edu/abs/2020ApJ...898..142K} {898, 142}

\bibitem[\protect\citeauthoryear{{Koljonen} \& {Hovatta}}{{Koljonen} \&
  {Hovatta}}{2021}]{Koljonen2021}
{Koljonen} K.~I.~I.,  {Hovatta} T.,  2021, \mn@doi [\aap]
  {10.1051/0004-6361/202039581}, \href
  {https://ui.adsabs.harvard.edu/abs/2021A&A...647A.173K} {647, A173}

\bibitem[\protect\citeauthoryear{Kremer, Stensbo-Smidt, Gieseke, Pedersen  \&
  Igel}{Kremer et~al.}{2017}]{bigUNIbigDATA}
Kremer J.,  Stensbo-Smidt K.,  Gieseke F.,  Pedersen K.,   Igel C.,  2017,
  \mn@doi [IEEE Intelligent Systems] {10.1109/MIS.2017.40}, 32, 16

\bibitem[\protect\citeauthoryear{{Kubota}, {Tanaka}, {Makishima}, {Ueda},
  {Dotani}, {Inoue}  \& {Yamaoka}}{{Kubota} et~al.}{1998}]{Kubota1998}
{Kubota} A.,  {Tanaka} Y.,  {Makishima} K.,  {Ueda} Y.,  {Dotani} T.,  {Inoue}
  H.,   {Yamaoka} K.,  1998, \mn@doi [\pasj] {10.1093/pasj/50.6.667}, \href
  {https://ui.adsabs.harvard.edu/abs/1998PASJ...50..667K} {50, 667}

\bibitem[\protect\citeauthoryear{Kuhn \& Johnson}{Kuhn \&
  Johnson}{2019}]{kuhn2019applied}
Kuhn M.,  Johnson K.,  2019, Applied Predictive Modeling.
Springer New York

\bibitem[\protect\citeauthoryear{Lakshminarayanan}{Lakshminarayanan}{2016}]{lakshminarayanan2016decision}
Lakshminarayanan B.,  2016, PhD thesis, UCL (University College London)

\bibitem[\protect\citeauthoryear{{Leahy}, {Elsner}  \& {Weisskopf}}{{Leahy}
  et~al.}{1983}]{leahy1983norm}
{Leahy} D.~A.,  {Elsner} R.~F.,   {Weisskopf} M.~C.,  1983, \mn@doi [\apj]
  {10.1086/161288}, 272, 256

\bibitem[\protect\citeauthoryear{{Li}, {Zheng}, {Wang}  \& {Wang}}{{Li}
  et~al.}{2020}]{solarflareML}
{Li} X.,  {Zheng} Y.,  {Wang} X.,   {Wang} L.,  2020, \mn@doi [\apj]
  {10.3847/1538-4357/ab6d04}, \href
  {https://ui.adsabs.harvard.edu/abs/2020ApJ...891...10L} {891, 10}

\bibitem[\protect\citeauthoryear{Lieberman \& Morris}{Lieberman \&
  Morris}{2014}]{multicollinearityclass}
Lieberman M.,  Morris J.,  2014, 40, 5

\bibitem[\protect\citeauthoryear{{Liu}, {van Paradijs}  \& {van den
  Heuvel}}{{Liu} et~al.}{2007}]{lmxbCatalog}
{Liu} Q.~Z.,  {van Paradijs} J.,   {van den Heuvel} E.~P.~J.,  2007, \mn@doi
  [\aap] {10.1051/0004-6361:20077303}, \href
  {https://ui.adsabs.harvard.edu/abs/2007A&A...469..807L} {469, 807}

\bibitem[\protect\citeauthoryear{{Liu}, {Liu}, {Bambi}  \& {Ji}}{{Liu}
  et~al.}{2022}]{HXMT-MAXI-SPIN}
{Liu} Q.,  {Liu} H.,  {Bambi} C.,   {Ji} L.,  2022, \mn@doi [\mnras]
  {10.1093/mnras/stac616}, \href
  {https://ui.adsabs.harvard.edu/abs/2022MNRAS.512.2082L} {512, 2082}

\bibitem[\protect\citeauthoryear{{Lones}}{{Lones}}{2021}]{avoidMLpitfalls}
{Lones} M.~A.,  2021, arXiv e-prints, \href
  {https://ui.adsabs.harvard.edu/abs/2021arXiv210802497L} {p. arXiv:2108.02497}

\bibitem[\protect\citeauthoryear{{Lundberg} \& {Lee}}{{Lundberg} \&
  {Lee}}{2017}]{SHAP2017}
{Lundberg} S.,  {Lee} S.-I.,  2017, arXiv e-prints, \href
  {https://ui.adsabs.harvard.edu/abs/2017arXiv170507874L} {p. arXiv:1705.07874}

\bibitem[\protect\citeauthoryear{Ma \& He}{Ma \& He}{2013}]{ma2013imbalanced}
Ma Y.,  He H.,  2013, Imbalanced Learning: Foundations, Algorithms, and
  Applications.
Wiley

\bibitem[\protect\citeauthoryear{Massey}{Massey}{1951}]{KS-test}
Massey F.~J.,  1951, Journal of the American Statistical Association, 46, 68

\bibitem[\protect\citeauthoryear{Mathew}{Mathew}{2022}]{mathew2022optimized}
Mathew T.~E.,  2022, Journal of Theoretical and Applied Information Technology,
  100

\bibitem[\protect\citeauthoryear{{McClintock} \& {Remillard}}{{McClintock} \&
  {Remillard}}{2006}]{McClintockRemillard2006}
{McClintock} J.~E.,  {Remillard} R.~A.,  2006, in , Vol.~39, Compact stellar
  X-ray sources.
pp 157--213

\bibitem[\protect\citeauthoryear{{M{\'e}ndez} \& {Belloni}}{{M{\'e}ndez} \&
  {Belloni}}{2021}]{mendezBelloni2021}
{M{\'e}ndez} M.,  {Belloni} T.~M.,  2021, in {Belloni} T.~M.,  {M{\'e}ndez} M.,
    {Zhang} C.,  eds,  Astrophysics and Space Science Library Vol. 461,
  Astrophysics and Space Science Library. pp 263--331 (\mn@eprint {arXiv}
  {2010.08291}), \mn@doi{10.1007/978-3-662-62110-3_6}

\bibitem[\protect\citeauthoryear{{M{\'e}ndez}, {van der Klis}, {Ford},
  {Wijnands}  \& {van Paradijs}}{{M{\'e}ndez} et~al.}{1999}]{mendez4U-1608}
{M{\'e}ndez} M.,  {van der Klis} M.,  {Ford} E.~C.,  {Wijnands} R.,   {van
  Paradijs} J.,  1999, \mn@doi [\apjl] {10.1086/311836}, \href
  {https://ui.adsabs.harvard.edu/abs/1999ApJ...511L..49M} {511, L49}

\bibitem[\protect\citeauthoryear{{M{\'e}ndez}, {Altamirano}, {Belloni}  \&
  {Sanna}}{{M{\'e}ndez} et~al.}{2013}]{MendezHFQPOs2013}
{M{\'e}ndez} M.,  {Altamirano} D.,  {Belloni} T.,   {Sanna} A.,  2013, \mn@doi
  [\mnras] {10.1093/mnras/stt1431}, \href
  {https://ui.adsabs.harvard.edu/abs/2013MNRAS.435.2132M} {435, 2132}

\bibitem[\protect\citeauthoryear{{M{\'e}ndez}, {Karpouzas}, {Garc{\'\i}a},
  {Zhang}, {Zhang}, {Belloni}  \& {Altamirano}}{{M{\'e}ndez}
  et~al.}{2022}]{mendez2022couplingNATURE}
{M{\'e}ndez} M.,  {Karpouzas} K.,  {Garc{\'\i}a} F.,  {Zhang} L.,  {Zhang} Y.,
  {Belloni} T.~M.,   {Altamirano} D.,  2022, \mn@doi [Nature Astronomy]
  {10.1038/s41550-022-01617-y}, \href
  {https://ui.adsabs.harvard.edu/abs/2022NatAs...6..577M} {6, 577}

\bibitem[\protect\citeauthoryear{{Migliari}, {van der Klis}  \&
  {Fender}}{{Migliari} et~al.}{2003}]{Migliari2003}
{Migliari} S.,  {van der Klis} M.,   {Fender} R.~P.,  2003, \mn@doi [\mnras]
  {10.1046/j.1365-8711.2003.07186.x}, \href
  {https://ui.adsabs.harvard.edu/abs/2003MNRAS.345L..35M} {345, L35}

\bibitem[\protect\citeauthoryear{{Miller}}{{Miller}}{2017}]{tim-miller-interpretability}
{Miller} T.,  2017, arXiv e-prints, \href
  {https://ui.adsabs.harvard.edu/abs/2017arXiv170607269M} {p. arXiv:1706.07269}

\bibitem[\protect\citeauthoryear{Miller et~al.,}{Miller
  et~al.}{2018}]{miller2018}
Miller J.~M.,  et~al., 2018, \mn@doi [The Astrophysical Journal]
  {10.3847/2041-8213/aacc61}, 860, L28

\bibitem[\protect\citeauthoryear{{Mirabel} \& {Rodr{\'\i}guez}}{{Mirabel} \&
  {Rodr{\'\i}guez}}{1994}]{grsjetinclination}
{Mirabel} I.~F.,  {Rodr{\'\i}guez} L.~F.,  1994, \mn@doi [\nat]
  {10.1038/371046a0}, \href
  {https://ui.adsabs.harvard.edu/abs/1994Natur.371...46M} {371, 46}

\bibitem[\protect\citeauthoryear{{Mitsuda} et~al.,}{{Mitsuda}
  et~al.}{1984}]{Mitsuda1984}
{Mitsuda} K.,  et~al., 1984, \pasj, \href
  {https://ui.adsabs.harvard.edu/abs/1984PASJ...36..741M} {36, 741}

\bibitem[\protect\citeauthoryear{Molnar}{Molnar}{2022}]{molnar2022}
Molnar C.,  2022, Interpretable Machine Learning, 2 edn

\bibitem[\protect\citeauthoryear{{Molteni}, {Sponholz}  \&
  {Chakrabarti}}{{Molteni} et~al.}{1996}]{propagatingoscillatoryshock}
{Molteni} D.,  {Sponholz} H.,   {Chakrabarti} S.~K.,  1996, \mn@doi [\apj]
  {10.1086/176775}, \href
  {https://ui.adsabs.harvard.edu/abs/1996ApJ...457..805M} {457, 805}

\bibitem[\protect\citeauthoryear{{Motta}}{{Motta}}{2016}]{mottaquickreview}
{Motta} S.~E.,  2016, \mn@doi [Astronomische Nachrichten]
  {10.1002/asna.201612320}, \href
  {https://ui.adsabs.harvard.edu/abs/2016AN....337..398M} {337, 398}

\bibitem[\protect\citeauthoryear{Motta, Muñoz-Darias, Casella, Belloni  \&
  Homan}{Motta et~al.}{2011}]{motta2011}
Motta S.,  Muñoz-Darias T.,  Casella P.,  Belloni T.,   Homan J.,  2011,
  \mn@doi [Monthly Notices of the Royal Astronomical Society]
  {10.1111/j.1365-2966.2011.19566.x}, 418, 2292

\bibitem[\protect\citeauthoryear{{Motta}, {Casella}, {Henze},
  {Mu{\~n}oz-Darias}, {Sanna}, {Fender}  \& {Belloni}}{{Motta}
  et~al.}{2015}]{motta2015LotsofQPOs}
{Motta} S.~E.,  {Casella} P.,  {Henze} M.,  {Mu{\~n}oz-Darias} T.,  {Sanna} A.,
   {Fender} R.,   {Belloni} T.,  2015, \mn@doi [\mnras]
  {10.1093/mnras/stu2579}, \href
  {https://ui.adsabs.harvard.edu/abs/2015MNRAS.447.2059M} {447, 2059}

\bibitem[\protect\citeauthoryear{Mundfrom, Smith  \& Kay}{Mundfrom
  et~al.}{2018}]{multicollinearityregression}
Mundfrom D.,  Smith M.,   Kay L.,  2018, \mn@doi [General Linear Model Journal]
  {10.31523/glmj.044001.003}, 44, 24

\bibitem[\protect\citeauthoryear{Nadeau \& Bengio}{Nadeau \&
  Bengio}{2004}]{nadeauandbengio}
Nadeau C.,  Bengio Y.,  2004, Machine Learning, 52, 239

\bibitem[\protect\citeauthoryear{{Nakahira} et~al.,}{{Nakahira}
  et~al.}{2018}]{Nakahira2018}
{Nakahira} S.,  et~al., 2018, \mn@doi [\pasj] {10.1093/pasj/psy093}, \href
  {https://ui.adsabs.harvard.edu/abs/2018PASJ...70...95N} {70, 95}

\bibitem[\protect\citeauthoryear{{Negoro} et~al.,}{{Negoro}
  et~al.}{2017a}]{maxiATELdiscovery}
{Negoro} H.,  et~al., 2017a, The Astronomer's Telegram, 10699, 1

\bibitem[\protect\citeauthoryear{{Negoro} et~al.,}{{Negoro}
  et~al.}{2017b}]{MAXIisBHATel}
{Negoro} H.,  et~al., 2017b, The Astronomer's Telegram, \href
  {https://ui.adsabs.harvard.edu/abs/2017ATel10708....1N} {10708, 1}

\bibitem[\protect\citeauthoryear{{Neilsen}}{{Neilsen}}{2013}]{bhwinds}
{Neilsen} J.,  2013, \mn@doi [Advances in Space Research]
  {10.1016/j.asr.2013.04.021}, \href
  {https://ui.adsabs.harvard.edu/abs/2013AdSpR..52..732N} {52, 732}

\bibitem[\protect\citeauthoryear{Neyshabur, Bhojanapalli, Mcallester  \&
  Srebro}{Neyshabur et~al.}{2017}]{NIPS2017_10ce03a1}
Neyshabur B.,  Bhojanapalli S.,  Mcallester D.,   Srebro N.,  2017, in Guyon
  I.,  Luxburg U.~V.,  Bengio S.,  Wallach H.,  Fergus R.,  Vishwanathan S.,
  Garnett R.,  eds,  Vol. 30, Advances in Neural Information Processing
  Systems. Curran Associates, Inc., \url
  {https://proceedings.neurips.cc/paper/2017/file/10ce03a1ed01077e3e289f3e53c72813-Paper.pdf}

\bibitem[\protect\citeauthoryear{Nicodemus, Malley, Strobl  \&
  Ziegler}{Nicodemus et~al.}{2010}]{Nicodemus2010}
Nicodemus K.~K.,  Malley J.~D.,  Strobl C.,   Ziegler A.,  2010, \mn@doi [{BMC}
  Bioinformatics] {10.1186/1471-2105-11-110}, 11

\bibitem[\protect\citeauthoryear{{Nowak}, {Wilms}  \& {Dove}}{{Nowak}
  et~al.}{1999}]{quality1999}
{Nowak} M.~A.,  {Wilms} J.,   {Dove} J.~B.,  1999, \mn@doi [\apj]
  {10.1086/307189}, \href
  {https://ui.adsabs.harvard.edu/abs/1999ApJ...517..355N} {517, 355}

\bibitem[\protect\citeauthoryear{Olson \& Delen}{Olson \&
  Delen}{2008}]{olson2008advanced}
Olson D.,  Delen D.,  2008, Advanced Data Mining Techniques.
Springer Berlin Heidelberg

\bibitem[\protect\citeauthoryear{{Orwat-Kapola}, {Bird}, {Hill}, {Altamirano}
  \& {Huppenkothen}}{{Orwat-Kapola} et~al.}{2022}]{grsML2022}
{Orwat-Kapola} J.~K.,  {Bird} A.~J.,  {Hill} A.~B.,  {Altamirano} D.,
  {Huppenkothen} D.,  2022, \mn@doi [\mnras] {10.1093/mnras/stab3043}, \href
  {https://ui.adsabs.harvard.edu/abs/2022MNRAS.509.1269O} {509, 1269}

\bibitem[\protect\citeauthoryear{{Parikh}, {Russell}, {Wijnands},
  {Miller-Jones}, {Sivakoff}  \& {Tetarenko}}{{Parikh}
  et~al.}{2019}]{hysteresis-MAXI}
{Parikh} A.~S.,  {Russell} T.~D.,  {Wijnands} R.,  {Miller-Jones} J.~C.~A.,
  {Sivakoff} G.~R.,   {Tetarenko} A.~J.,  2019, \mn@doi [\apjl]
  {10.3847/2041-8213/ab2636}, \href
  {https://ui.adsabs.harvard.edu/abs/2019ApJ...878L..28P} {878, L28}

\bibitem[\protect\citeauthoryear{Pattnaik, Sharma, Alabarta, Altamirano,
  Chakraborty, Kembhavi, Méndez  \& Orwat-Kapola}{Pattnaik
  et~al.}{2020}]{Pattnaik2020}
Pattnaik R.,  Sharma K.,  Alabarta K.,  Altamirano D.,  Chakraborty M.,
  Kembhavi A.,  Méndez M.,   Orwat-Kapola J.~K.,  2020, \mn@doi [Monthly
  Notices of the Royal Astronomical Society] {10.1093/mnras/staa3899}, 501,
  3457

\bibitem[\protect\citeauthoryear{{Pattnaik}, {Sharma}, {Alabarta},
  {Altamirano}, {Chakraborty}, {Kembhavi}, {M{\'e}ndez}  \&
  {Orwat-Kapola}}{{Pattnaik} et~al.}{2021}]{Pattnaik2021}
{Pattnaik} R.,  {Sharma} K.,  {Alabarta} K.,  {Altamirano} D.,  {Chakraborty}
  M.,  {Kembhavi} A.,  {M{\'e}ndez} M.,   {Orwat-Kapola} J.~K.,  2021, \mn@doi
  [\mnras] {10.1093/mnras/staa3899}, \href
  {https://ui.adsabs.harvard.edu/abs/2021MNRAS.501.3457P} {501, 3457}

\bibitem[\protect\citeauthoryear{{Pearson}, {Palafox}  \& {Griffith}}{{Pearson}
  et~al.}{2018}]{mlExos}
{Pearson} K.~A.,  {Palafox} L.,   {Griffith} C.~A.,  2018, \mn@doi [\mnras]
  {10.1093/mnras/stx2761}, \href
  {https://ui.adsabs.harvard.edu/abs/2018MNRAS.474..478P} {474, 478}

\bibitem[\protect\citeauthoryear{Pedregosa et~al.,}{Pedregosa
  et~al.}{2011}]{scikit-learn}
Pedregosa F.,  et~al., 2011, Journal of Machine Learning Research, 12, 2825

\bibitem[\protect\citeauthoryear{{Raichur} \& {Paul}}{{Raichur} \&
  {Paul}}{2008}]{cenX-3QPOs}
{Raichur} H.,  {Paul} B.,  2008, \mn@doi [\apj] {10.1086/591037}, \href
  {https://ui.adsabs.harvard.edu/abs/2008ApJ...685.1109R} {685, 1109}

\bibitem[\protect\citeauthoryear{{Rao}, {Singh}  \& {Bhattacharya}}{{Rao}
  et~al.}{2016}]{astrosat}
{Rao} A.~R.,  {Singh} K.~P.,   {Bhattacharya} D.,  2016, arXiv e-prints, \href
  {https://ui.adsabs.harvard.edu/abs/2016arXiv160806051R} {p. arXiv:1608.06051}

\bibitem[\protect\citeauthoryear{Raudys \& Jain}{Raudys \&
  Jain}{1991}]{smallsample1991}
Raudys S.,  Jain A.,  1991, \mn@doi [IEEE Transactions on Pattern Analysis and
  Machine Intelligence] {10.1109/34.75512}, 13, 252

\bibitem[\protect\citeauthoryear{{Reid}, {McClintock}, {Steiner}, {Steeghs},
  {Remillard}, {Dhawan}  \& {Narayan}}{{Reid} et~al.}{2014}]{GRSDISTANCE}
{Reid} M.~J.,  {McClintock} J.~E.,  {Steiner} J.~F.,  {Steeghs} D.,
  {Remillard} R.~A.,  {Dhawan} V.,   {Narayan} R.,  2014, \mn@doi [\apj]
  {10.1088/0004-637X/796/1/2}, 796, 2

\bibitem[\protect\citeauthoryear{{Remillard}, {McClintock}, {Orosz}  \&
  {Levine}}{{Remillard} et~al.}{2006}]{Remillard2006}
{Remillard} R.~A.,  {McClintock} J.~E.,  {Orosz} J.~A.,   {Levine} A.~M.,
  2006, \mn@doi [\apj] {10.1086/498556}, \href
  {http://adsabs.harvard.edu/abs/2006ApJ...637.1002R} {637, 1002}

\bibitem[\protect\citeauthoryear{{Remillard} et~al.,}{{Remillard}
  et~al.}{2022}]{Remillard_3C50}
{Remillard} R.~A.,  et~al., 2022, \mn@doi [\aj] {10.3847/1538-3881/ac4ae6},
  \href {https://ui.adsabs.harvard.edu/abs/2022AJ....163..130R} {163, 130}

\bibitem[\protect\citeauthoryear{{Revnivtsev}, {Churazov}, {Gilfanov}  \&
  {Sunyaev}}{{Revnivtsev} et~al.}{2001}]{Revnivtsev2001}
{Revnivtsev} M.,  {Churazov} E.,  {Gilfanov} M.,   {Sunyaev} R.,  2001, \mn@doi
  [\aap] {10.1051/0004-6361:20010434}, \href
  {https://ui.adsabs.harvard.edu/abs/2001A&A...372..138R} {372, 138}

\bibitem[\protect\citeauthoryear{{Richards} et~al.,}{{Richards}
  et~al.}{2011}]{mlVariableStars2011}
{Richards} J.~W.,  et~al., 2011, \mn@doi [\apj] {10.1088/0004-637X/733/1/10},
  \href {https://ui.adsabs.harvard.edu/abs/2011ApJ...733...10R} {733, 10}

\bibitem[\protect\citeauthoryear{Rodríguez, Rodríguez-Rodríguez  \&
  Woo}{Rodríguez et~al.}{2022}]{astroMLarticlenotPython}
Rodríguez J.-V.,  Rodríguez-Rodríguez I.,   Woo W.~L.,  2022, \mn@doi [WIREs
  Data Mining and Knowledge Discovery] {https://doi.org/10.1002/widm.1476}, 12,
  e1476

\bibitem[\protect\citeauthoryear{{Ross} \& {Fabian}}{{Ross} \&
  {Fabian}}{2005}]{x-rayreflectionmodels2005}
{Ross} R.~R.,  {Fabian} A.~C.,  2005, \mn@doi [\mnras]
  {10.1111/j.1365-2966.2005.08797.x}, \href
  {https://ui.adsabs.harvard.edu/abs/2005MNRAS.358..211R} {358, 211}

\bibitem[\protect\citeauthoryear{Saarela \& Jauhiainen}{Saarela \&
  Jauhiainen}{2021}]{Saarela2021ComparisonOF}
Saarela M.,  Jauhiainen S.,  2021, SN Applied Sciences, 3, 1

\bibitem[\protect\citeauthoryear{{Schlegel}}{{Schlegel}}{1995}]{Schlegel1995}
{Schlegel} E.~M.,  1995, \mn@doi [Reports on Progress in Physics]
  {10.1088/0034-4885/58/11/001}, \href
  {https://ui.adsabs.harvard.edu/abs/1995RPPh...58.1375S} {58, 1375}

\bibitem[\protect\citeauthoryear{{Schmidt} et~al.,}{{Schmidt}
  et~al.}{2021}]{MLgravWave}
{Schmidt} S.,  et~al., 2021, \mn@doi [\prd] {10.1103/PhysRevD.103.043020},
  \href {https://ui.adsabs.harvard.edu/abs/2021PhRvD.103d3020S} {103, 043020}

\bibitem[\protect\citeauthoryear{{Shakura} \& {Sunyaev}}{{Shakura} \&
  {Sunyaev}}{1973}]{SS73}
{Shakura} N.~I.,  {Sunyaev} R.~A.,  1973, \aap, \href
  {http://adsabs.harvard.edu/abs/1973A%26A....24..337S} {24, 337}

\bibitem[\protect\citeauthoryear{Shapley}{Shapley}{1952}]{shapleyvaluesoriginal}
Shapley L.~S.,  1952, A Value for N-Person Games.
RAND Corporation, Santa Monica, CA, \mn@doi{10.7249/P0295}

\bibitem[\protect\citeauthoryear{Sheather}{Sheather}{2008}]{Sheather2008-mc}
Sheather S.~J.,  2008, A modern approach to regression with {R}, 2009 edn.
Springer Texts in Statistics, Springer, New York, NY

\bibitem[\protect\citeauthoryear{Singh, Thakur  \& Sharma}{Singh
  et~al.}{2016}]{supervisedreview}
Singh A.,  Thakur N.,   Sharma A.,  2016, in 2016 3rd International Conference
  on Computing for Sustainable Global Development (INDIACom). pp 1310--1315

\bibitem[\protect\citeauthoryear{{Sreehari} \& {Nandi}}{{Sreehari} \&
  {Nandi}}{2021}]{Sreehari2021}
{Sreehari} H.,  {Nandi} A.,  2021, \mn@doi [\mnras] {10.1093/mnras/stab151},
  \href {https://ui.adsabs.harvard.edu/abs/2021MNRAS.502.1334S} {502, 1334}

\bibitem[\protect\citeauthoryear{{Sreehari}, {Nandi}, {Das}, {Agrawal},
  {Mandal}, {Ramadevi}  \& {Katoch}}{{Sreehari}
  et~al.}{2020}]{astrosatviewofGRS}
{Sreehari} H.,  {Nandi} A.,  {Das} S.,  {Agrawal} V.~K.,  {Mandal} S.,
  {Ramadevi} M.~C.,   {Katoch} T.,  2020, \mn@doi [\mnras]
  {10.1093/mnras/staa3135}, \href
  {https://ui.adsabs.harvard.edu/abs/2020MNRAS.499.5891S} {499, 5891}

\bibitem[\protect\citeauthoryear{{Sridhar}, {Bhattacharyya}, {Chandra}  \&
  {Antia}}{{Sridhar} et~al.}{2019}]{MAXI-distance}
{Sridhar} N.,  {Bhattacharyya} S.,  {Chandra} S.,   {Antia} H.~M.,  2019,
  \mn@doi [\mnras] {10.1093/mnras/stz1476}, \href
  {https://ui.adsabs.harvard.edu/abs/2019MNRAS.487.4221S} {487, 4221}

\bibitem[\protect\citeauthoryear{{Stella} \& {Vietri}}{{Stella} \&
  {Vietri}}{1998}]{lense-thirring-original}
{Stella} L.,  {Vietri} M.,  1998, \mn@doi [\apjl] {10.1086/311075}, \href
  {https://ui.adsabs.harvard.edu/abs/1998ApJ...492L..59S} {492, L59}

\bibitem[\protect\citeauthoryear{{Stella} \& {Vietri}}{{Stella} \&
  {Vietri}}{1999}]{stellaVietriISCO}
{Stella} L.,  {Vietri} M.,  1999, \mn@doi [\prl] {10.1103/PhysRevLett.82.17},
  \href {https://ui.adsabs.harvard.edu/abs/1999PhRvL..82...17S} {82, 17}

\bibitem[\protect\citeauthoryear{Strobl, Boulesteix, Zeileis  \&
  Hothorn}{Strobl et~al.}{2007}]{Strobl2007}
Strobl C.,  Boulesteix A.-L.,  Zeileis A.,   Hothorn T.,  2007, \mn@doi [{BMC}
  Bioinformatics] {10.1186/1471-2105-8-25}, 8

\bibitem[\protect\citeauthoryear{Strobl, Boulesteix, Kneib, Augustin  \&
  Zeileis}{Strobl et~al.}{2008}]{Strobl2008}
Strobl C.,  Boulesteix A.-L.,  Kneib T.,  Augustin T.,   Zeileis A.,  2008,
  \mn@doi [{BMC} Bioinformatics] {10.1186/1471-2105-9-307}, 9

\bibitem[\protect\citeauthoryear{{Taam}, {Chen}  \& {Swank}}{{Taam}
  et~al.}{1996}]{Taam1996}
{Taam} R.~E.,  {Chen} X.,   {Swank} J.~H.,  1996, in American Astronomical
  Society Meeting Abstracts. p. 35.08

\bibitem[\protect\citeauthoryear{{Tagger} \& {Pellat}}{{Tagger} \&
  {Pellat}}{1999}]{accretion-ejection}
{Tagger} M.,  {Pellat} R.,  1999, \aap, \href
  {https://ui.adsabs.harvard.edu/abs/1999A&A...349.1003T} {349, 1003}

\bibitem[\protect\citeauthoryear{{Tauris} \& {van den Heuvel}}{{Tauris} \& {van
  den Heuvel}}{2006}]{tauris2006}
{Tauris} T.~M.,  {van den Heuvel} E.~P.~J.,  2006, in , Vol.~39, Compact
  stellar X-ray sources.
pp 623--665

\bibitem[\protect\citeauthoryear{Thomas}{Thomas}{1952}]{thomas1952country}
Thomas D.,  1952, In Country Sleep: And Other Poems.
James Laughlin

\bibitem[\protect\citeauthoryear{{Titarchuk} \& {Shaposhnikov}}{{Titarchuk} \&
  {Shaposhnikov}}{2005}]{titarchukSigmoidExp}
{Titarchuk} L.,  {Shaposhnikov} N.,  2005, \mn@doi [\apj] {10.1086/429986},
  \href {https://ui.adsabs.harvard.edu/abs/2005ApJ...626..298T} {626, 298}

\bibitem[\protect\citeauthoryear{Truss \& Done}{Truss \&
  Done}{2006}]{Truss2006}
Truss M.~R.,  Done C.,  2006, Monthly Notices of the Royal Astronomical
  Society: Letters, 368

\bibitem[\protect\citeauthoryear{Vanwinckelen \& Blockeel}{Vanwinckelen \&
  Blockeel}{2012}]{repeatedkfold}
Vanwinckelen G.,  Blockeel H.,  2012.

\bibitem[\protect\citeauthoryear{{Verner}, {Ferland}, {Korista}  \&
  {Yakovlev}}{{Verner} et~al.}{1996}]{VernerCrossSections}
{Verner} D.~A.,  {Ferland} G.~J.,  {Korista} K.~T.,   {Yakovlev} D.~G.,  1996,
  \mn@doi [\apj] {10.1086/177435}, \href
  {https://ui.adsabs.harvard.edu/abs/1996ApJ...465..487V} {465, 487}

\bibitem[\protect\citeauthoryear{Vieira \& Digiampietri}{Vieira \&
  Digiampietri}{2022}]{post-hoc}
Vieira C.~P.,  Digiampietri L.~A.,  2022. SBSI.
Association for Computing Machinery, New York, NY, USA,
  \mn@doi{10.1145/3535511.3535512}

\bibitem[\protect\citeauthoryear{{Virtanen} et~al.,}{{Virtanen}
  et~al.}{2020}]{Virtanen2020}
{Virtanen} P.,  et~al., 2020, \mn@doi [Nature Methods]
  {10.1038/s41592-019-0686-2}, \href
  {https://ui.adsabs.harvard.edu/abs/2020NatMe..17..261V} {17, 261}

\bibitem[\protect\citeauthoryear{{Wang}}{{Wang}}{2016}]{wang2016QPOreview}
{Wang} J.,  2016, \mn@doi [International Journal of Astronomy and Astrophysics]
  {10.4236/ijaa.2016.61006}, \href
  {https://ui.adsabs.harvard.edu/abs/2016IJAA....6...82W} {6, 82}

\bibitem[\protect\citeauthoryear{Waskom}{Waskom}{2021}]{Waskom2021}
Waskom M.~L.,  2021, \mn@doi [Journal of Open Source Software]
  {10.21105/joss.03021}, 6, 3021

\bibitem[\protect\citeauthoryear{{W}es {M}c{K}inney}{{W}es
  {M}c{K}inney}{2010}]{pandas}
{W}es {M}c{K}inney 2010, in {S}t\'efan van~der {W}alt {J}arrod {M}illman eds,
  {P}roceedings of the 9th {P}ython in {S}cience {C}onference. pp 56 -- 61,
  \mn@doi{10.25080/Majora-92bf1922-00a}

\bibitem[\protect\citeauthoryear{{White} \& {Holt}}{{White} \&
  {Holt}}{1982}]{coronae1982}
{White} N.~E.,  {Holt} S.~S.,  1982, \mn@doi [\apj] {10.1086/159991}, \href
  {https://ui.adsabs.harvard.edu/abs/1982ApJ...257..318W} {257, 318}

\bibitem[\protect\citeauthoryear{{Wilms}, {Allen}  \& {McCray}}{{Wilms}
  et~al.}{2000}]{Wilms2000}
{Wilms} J.,  {Allen} A.,   {McCray} R.,  2000, \mn@doi [\apj] {10.1086/317016},
  \href {http://adsabs.harvard.edu/abs/2000ApJ...542..914W} {542, 914}

\bibitem[\protect\citeauthoryear{Wolpert}{Wolpert}{2002}]{NoFreeLunch}
Wolpert D.~H.,  2002, The Supervised Learning No-Free-Lunch Theorems.
Springer London, London, pp 25--42, \mn@doi{10.1007/978-1-4471-0123-93}

\bibitem[\protect\citeauthoryear{{Xu}, {Shi}, {Tsang}, {Ong}, {Gong}  \&
  {Shen}}{{Xu} et~al.}{2019}]{multioutputreview}
{Xu} D.,  {Shi} Y.,  {Tsang} I.~W.,  {Ong} Y.-S.,  {Gong} C.,   {Shen} X.,
  2019, arXiv e-prints, \href
  {https://ui.adsabs.harvard.edu/abs/2019arXiv190100248X} {p. arXiv:1901.00248}

\bibitem[\protect\citeauthoryear{{Yang}, {Brower-Sinning}, {Lewis}  \&
  {K{\"a}stner}}{{Yang} et~al.}{2022a}]{2022arXiv220903345Y}
{Yang} C.,  {Brower-Sinning} R.~A.,  {Lewis} G.~A.,   {K{\"a}stner} C.,  2022a,
  \mn@doi [arXiv e-prints] {10.48550/arXiv.2209.03345}, \href
  {https://ui.adsabs.harvard.edu/abs/2022arXiv220903345Y} {p. arXiv:2209.03345}

\bibitem[\protect\citeauthoryear{{Yang}, {Hare}, {Kargaltsev}, {Volkov}, {Chen}
   \& {Rangelov}}{{Yang} et~al.}{2022b}]{yang2022}
{Yang} H.,  {Hare} J.,  {Kargaltsev} O.,  {Volkov} I.,  {Chen} S.,   {Rangelov}
  B.,  2022b, \mn@doi [\apj] {10.3847/1538-4357/ac952b}, \href
  {https://ui.adsabs.harvard.edu/abs/2022ApJ...941..104Y} {941, 104}

\bibitem[\protect\citeauthoryear{{Yasodhara}, {Asgarian}, {Huang}  \&
  {Sobhani}}{{Yasodhara} et~al.}{2021}]{featureimportancestrees}
{Yasodhara} A.,  {Asgarian} A.,  {Huang} D.,   {Sobhani} P.,  2021, arXiv
  e-prints, \href {https://ui.adsabs.harvard.edu/abs/2021arXiv211000086Y} {p.
  arXiv:2110.00086}

\bibitem[\protect\citeauthoryear{{Zdziarski}, {Johnson}  \&
  {Magdziarz}}{{Zdziarski} et~al.}{1996}]{Zdziarski1996}
{Zdziarski} A.~A.,  {Johnson} W.~N.,   {Magdziarz} P.,  1996, \mn@doi [\mnras]
  {10.1093/mnras/283.1.193}, \href
  {https://ui.adsabs.harvard.edu/abs/1996MNRAS.283..193Z} {283, 193}

\bibitem[\protect\citeauthoryear{{Zhang}, {Jahoda}, {Swank}, {Morgan}  \&
  {Giles}}{{Zhang} et~al.}{1995}]{zhangPDSsubtraction}
{Zhang} W.,  {Jahoda} K.,  {Swank} J.~H.,  {Morgan} E.~H.,   {Giles} A.~B.,
  1995, \mn@doi [\apj] {10.1086/176111}, 449, 930

\bibitem[\protect\citeauthoryear{{Zhang} et~al.,}{{Zhang}
  et~al.}{2020}]{GRSDATAPAPER}
{Zhang} L.,  et~al., 2020, \mn@doi [\mnras] {10.1093/mnras/staa797}, 494, 1375

\bibitem[\protect\citeauthoryear{{Zhang}, {M{\'e}ndez}, {Garc{\'\i}a},
  {Karpouzas}, {Zhang}, {Liu}, {Belloni}  \& {Altamirano}}{{Zhang}
  et~al.}{2022}]{zhangGRS2022}
{Zhang} Y.,  {M{\'e}ndez} M.,  {Garc{\'\i}a} F.,  {Karpouzas} K.,  {Zhang} L.,
  {Liu} H.,  {Belloni} T.~M.,   {Altamirano} D.,  2022, \mn@doi [\mnras]
  {10.1093/mnras/stac1050}, \href
  {https://ui.adsabs.harvard.edu/abs/2022MNRAS.514.2891Z} {514, 2891}

\bibitem[\protect\citeauthoryear{{Zhu} et~al.,}{{Zhu}
  et~al.}{2014}]{detectPulsars}
{Zhu} W.~W.,  et~al., 2014, \mn@doi [\apj] {10.1088/0004-637X/781/2/117}, \href
  {https://ui.adsabs.harvard.edu/abs/2014ApJ...781..117Z} {781, 117}

\bibitem[\protect\citeauthoryear{{{\.Z}ycki}, {Done}  \& {Smith}}{{{\.Z}ycki}
  et~al.}{1999}]{Zycki1999}
{{\.Z}ycki} P.~T.,  {Done} C.,   {Smith} D.~A.,  1999, \mn@doi [\mnras]
  {10.1046/j.1365-8711.1999.02885.x}, \href
  {https://ui.adsabs.harvard.edu/abs/1999MNRAS.309..561Z} {309, 561}

\bibitem[\protect\citeauthoryear{{de Beurs}, {Islam}, {Gopalan}  \&
  {Vrtilek}}{{de Beurs} et~al.}{2022}]{deBeurs2022}
{de Beurs} Z.~L.,  {Islam} N.,  {Gopalan} G.,   {Vrtilek} S.~D.,  2022, \mn@doi
  [\apj] {10.3847/1538-4357/ac6184}, \href
  {https://ui.adsabs.harvard.edu/abs/2022ApJ...933..116D} {933, 116}

\bibitem[\protect\citeauthoryear{van~de Schoot \& Mio{\v{c}}evi{\'c}}{van~de
  Schoot \& Mio{\v{c}}evi{\'c}}{2020}]{van2020small}
van~de Schoot R.,  Mio{\v{c}}evi{\'c} M.,  2020, Small Sample Size Solutions: A
  Guide for Applied Researchers and Practitioners.
European Association of Methodology Series, Taylor \& Francis

\bibitem[\protect\citeauthoryear{{van den Eijnden}, {Degenaar}, {Russell},
  {Wijnands}, {Miller-Jones}, {Sivakoff}  \& {Hern{\'a}ndez Santisteban}}{{van
  den Eijnden} et~al.}{2018}]{neutronstarjet}
{van den Eijnden} J.,  {Degenaar} N.,  {Russell} T.~D.,  {Wijnands} R.,
  {Miller-Jones} J.~C.~A.,  {Sivakoff} G.~R.,   {Hern{\'a}ndez Santisteban}
  J.~V.,  2018, \mn@doi [\nat] {10.1038/s41586-018-0524-1}, \href
  {https://ui.adsabs.harvard.edu/abs/2018Natur.562..233V} {562, 233}

\bibitem[\protect\citeauthoryear{{van der Klis}}{{van der
  Klis}}{2006}]{vanDerKlis2006}
{van der Klis} M.,  2006, in , Vol.~39, Compact stellar X-ray sources.
pp 39--112

\makeatother
\end{thebibliography}
\end{document}